\documentclass[aps,pra,reprint,groupedaddress,showkeys,floatfix,superscriptaddress,longbibliography,nofootinbib]{revtex4-2} 
\usepackage{graphicx}
\usepackage{amsmath}
\usepackage{amssymb}
\usepackage{babel}
\usepackage{hyperref}
\usepackage{siunitx} 
\usepackage{tcolorbox}
\tcbuselibrary{theorems}

\newtcbtheorem{formaldef}{Aspect}{
 colback=green!5,
 colframe=green!35!black,
 fonttitle=\bfseries,
 separator sign none
}{def}

\begin{document}

\title{Material-Based Intelligence: \\ Self-organizing, Autonomous and Adaptive Cognition Embodied in Physical Substrates}

\author{Vladimir A. Baulin}
\email{vbaulin@activeinference.institute}
\affiliation{Active Inference Institute, Crescent City, California, 95531, USA} 
\affiliation{Universitat Rovira i Virgili, Tarragona, Spain} 

\author{Rudolf M. Füchslin}
\affiliation{Zurich University of Applied Sciences, Zurich, Switzerland}
\affiliation{European Centre for Living Technology (ECLT) Ca Bottacin, Dorsoduro 3911, Calle Crosera, 30123 Venice, Italy}

\author{Achille Giacometti}
\affiliation{Dipartimento di Scienze Molecolari e Nanosistemi, Universita Ca Foscari Venezia, Via Torino 155, 30172 Venezia, Italy}
\affiliation{European Centre for Living Technology (ECLT) Ca Bottacin, Dorsoduro 3911, Calle Crosera, 30123 Venice, Italy}

\author{Helmut Hauser}
\affiliation{School of Engineering Mathematics and Technology, University of Bristol, Bristol, UK}

\author{Marco Werner}
\affiliation{Division Theory of Polymers, Leibniz-Institut fur Polymerforschung Dresden e.V., Dresden, Germany}

\date{\today}

\begin{abstract}
The design of intelligent materials often draws parallels with the complex adaptive behaviors of biological organisms, where robust functionality stems from sophisticated hierarchical organization and emergent long-distance coordination among a myriad local components. Current synthetic materials, despite integrating advanced sensors and actuators, predominantly demonstrate only simple, pre-programmed stimulus-response functionalities, falling short of robustly autonomous intelligent behavior. These systems typically execute tasks determined by rigid design or external control, fundamentally lacking the intricate internal feedback loops, dynamic adaptation, self-generated learning, and genuine self-determination characteristic of biological agents.
This perspective proposes a fundamentally different approach focusing on architectures where material-based intelligence is not pre-designed ,but arises spontaneously from self-organization harnessing far-from-equilibrium dynamics. Such approach leverage minimal physical models, intrinsically embedding information-theoretic control within the material’s own physics and its seamless coupling with the environment. This work explores interdisciplinary concepts from material physics, chemistry, biology, and computation, identifying concrete pathways toward developing materials that not only react, but actively perceive, adapt, learn, self-correct, and potentially self-construct, moving \textit{beyond biomimicry} to cultivate fully synthetic, self-evolving systems without external conrol. This framework outlines the fundamental requirements for, and constraints upon, future architectures where complex, goal-directed functionalities emerge synergistically from integrated local processes, distinguishing material-based intelligence from traditional hardware-software divisions. This demands that concepts of high-level goals and robust, replicable memory mechanisms are encoded and enacted through the material's inherent dynamics, inherently blurring the distinction between system output and process.
\end{abstract}

\maketitle

\section{Introduction}
\label{sec:intro}

The ambition to create materials that inherently process information, learn from experience, and react adaptively to internal states or external environmental changes, i.e. exhibiting what we term material-based intelligence (MBI), represents a significant challenge across material science, physics, engineering, and artificial intelligence \cite{bryant_physical_2023, wool_self-healing_2008, kaspar_rise_2021}.
Such ambition draws inspiration from biology where complex functionality from molecular recognition to organism-like behaviour arises from intricate, multi-scale interactions within intrinsically "intelligent" matter \cite{mcmillen_collective_2024, kaspar_rise_2021, wegst_bioinspired_2015, levin_multiscale_2024}. These biological systems demonstrate levels of autonomy, adaptive complexity, and problem-solving capability that current synthetic materials typically struggle to achieve \cite{pfeifer_self-organization_2007}. However, emergent "intelligent" behaviour in MBI may not necessarily follow the biological analogies.

Similarly, a useful distinction can be drawn between this emerging paradigm of MBI and traditional machine-based intelligence, where one can distinguish between the algorithm to be executed (the software) and the physical substrate on which the algorithm is executed (the hardware in the traditional sense of a computing machine). This separation relies on the fact that algorithms can be formulated as stepwise discrete manipulations of bit strings. This twofold discretisation (with respect to values and time) enables establishing computations on a dynamical physical device without having to account for the details of said dynamics. This "coarse-graining" or projection of an underlying complex physical dynamics onto one that can serve as an interface between abstract considerations and the physical world bears a large number of advantages, among them the fact that programs are to some degree independent of computers. However, this independence comes at a price. 

An algorithm is an abstract construct, distinct from any physical process. Therefore, emulating physical behavior requires that the governing laws and interactions be explicitly programmed. Explicit physical simulations appear to constitute only a subset of all computations. However, they underlie many applications, particularly of systems that cope with tasks in the real world (control of a chemical reactor or a robot). Furthermore, nature is inherently parallel; algorithms can be made to run in parallel, but the flow of information between parallel threads has to be organized. Conventional computiong systems must first de-physicalize sensory input, i.e. transduce physical sensory input into an abstract, digital representation for computation. Subsequently, the digital output requires re-physicalizition, i.e. conversion back into a physical action via actuators. MBI, however, may circumvent this intermediate layer of abstraction, integrating sensing, computation, and actuation directly within the material.

A characteristic of conventional computing architectures, typically features a clear architectural separation between core functions \cite{yao_fully_2020, yang_bicoss_2022}. This is exemplified, for example, in contemporary computers, by the pervasive von Neumann architecture with its continued data transfer between physically distinct memory and processing units, that inherently suppressed from exploiting the full, often subtle and rich, physical dynamics available within the material itself \cite{yao_fully_2020, yang_bicoss_2022, mead_neuromorphic_1990}. This deliberate decoupling from physically intrinsic laws and functions creates significant computational costs, particularly for tasks involving large datasets or the simulation of complex physical dynamics \cite{mead_neuromorphic_1990, kaspar_rise_2021}. We emphasize that these computational costs come together with corresponding high energy consumption, which puts limitations on miniaturization and/or the potential for the implementation of small, independent information processing systems\footnote{Note: the CPU itself may be small, but the devices providing it with energy or cooling it aren't}. 

Today, this traditional paradigm is also predominantly employed for simulating conceptual models of neural interaction in the form of artificial neural networks (ANNs). This approach, however, forces a brain-inspired model onto a fundamentally brain-unlike architecture. The brain itself shows "deep structural differences" from a von Neumann machine, an observation noted by von Neumann himself \cite{vonNeumann2012computer} and later established as a foundational principle of neuromorphic engineering by Carver Mead \cite{mead_neuromorphic_1990}. The brain's architecture avoids the "von Neumann bottleneck" by co-locating memory and processing at the synaptic level, operating with massive parallelism and extraordinary energy efficiency—consuming mere watts while performing tasks that require megawatts in supercomputers \cite{mead_neuromorphic_1990}. Modern paradigms like memcomputing explicitly seek to overcome this limitation by designing hardware where the same physical elements both store and process information, moving closer to the brain's integrated structure \cite{di_ventra_parallel_2013, kuncic_emergent_2018}.
Stored-program computer concepts like the von-Neumann architecture were the solution~\cite{vonNeumann2012computer} to overcome the limited accuracy and reproducibility of computation results by the earlier analog computers.
The fact that (some aspects of) inherently non-v. Neumann neural structures can be emulated by a v. Neumann architecture proves the versatility and universality of this architecture. While its versatility has enabled the recent, impressive progress in artificial intelligence, to sustain the "emulation" today requires dedicated computing farms and large energy supplies \cite{kindig_ai_2024}. Here, both feed-forward- and backpropagation steps need iterative read-processing-write cycles across the entire memory, where the communication between memory and processor acts as a bottleneck. Moreover, since stored-program computers operate (ideally) fully deterministic, artificial neural networks implemented therein make explicit use of (pseudo-)randomness when initializing networks weights, during training, or targeting statistically distributed outputs - more generally: when mimicking the statistical nature of nature-inspired embodied intelligence.

Conversely, material-based intelligence explores an alternative route by intentionally blurring or entirely eliminating the hardware-software distinction \cite{harrison_mind_2022}, seeking to overcome the outlined traditional limitations through the co-location and integration of functions like memory, and computation directly within the physical substrate \cite{di_ventra_parallel_2013, mcevoy_materials_2015}. This approach leverages inherent physical laws and material properties as computational resources \cite{milkowski_morphological_2018}, and may allow to integrate additional features in place that enable for sensing, actuation, and communication. The material's structure, its intrinsic dynamics, and its direct interactions with the environment embody the 'program' rather than merely executing external instructions on a passive substrate \cite{mead_neuromorphic_1990}. This concept resonates strongly with morphological computation, where an agent’s physical form actively contributes to information processing and control in the context of robotics \cite{pfeifer_morphological_2006, pfeifer_self-organization_2007, hauser_towards_2011, hauser_towards_2011-1,fuchslin2013morphological, milkowski_morphological_2018, mengaldo_concise_2022, zhao_exploring_2024}. For instance, to promote a soft robot's complex interactions with the environment or stabilize it's locomotion, the programming of morphological-~\cite{mertan_no-brainer_2025}, topological-~\cite{li_computational_2024}, or chemical reaction networks can be exploited. 
This effectively reduces the need for explicit, step-by-step algorithmic computation typically performed by a separate processing unit, and thereby \textit{outsources control to physical and / or chemical processing}. A physical embedding, which will be discussed in some more detail in Sec. \ref{sec:defining_mbi}, allows a computational material system to be deeply coupled with its environment, influencing and being influenced by it, enabling a richer computational dialogue than systems reliant on pre-defined, low-bandwidth input/output channels \cite{zhao_exploring_2024, vihmar_how_2023}. Biological systems often exemplify this profound integration: high-level neural control might select a dynamic regime (\textit{e.g.}, a walking gait, which can be viewed as an attractor in state space), while the intricate physics of the musculoskeletal system handles low-level stabilization and adaptation to minor environmental perturbations \cite{pfeifer_self-organization_2007, hauser_leveraging_2023}. Single-cell organisms are an excellent example of integration of memory, sensing, actuation, and computation. The training of complex motor skills in biology inherently involves adapting both neural pathways and the body’s mechanical response characteristics \cite{hoffmann_trade-offs_2014}.

Contemporary research is advancing towards MBI along three distinct but convergent vectors, each targeting a core pillar of embodied cognition. The first vector pursues \textbf{embodied action and morphological computation}, primarily through the field of soft robotics, where compliant materials and clever mechanical design are used to achieve adaptable locomotion and manipulation with minimal central control \cite{rus_design_2015, lu_bioinspired_2018, hu_small-scale_2018}. The second vector focuses on creating \textbf{embodied memory}, engineering materials that can store information directly in their physical state variables. This is achieved through mechanisms like programmed shape changes in advanced polymers \cite{xiao_artificial_2020, zhao_phase_2021, xia_dynamic_2022} or by harnessing the persistent, non-equilibrium conformations of polymer chains themselves \cite{reiter_memorizing_2020}. The third vector aims to realize \textbf{embodied information processing}, developing substrates that perform computation intrinsically through their local physical laws. This is exemplified by microscale neuromorphic devices, such as nanowire networks \cite{kuncic_emergent_2018, loeffler_neuromorphic_2023} and iontronic channels \cite{kamsma_brain-inspired_2024}, which demonstrate rudimentary learning and decision-making functions.

These efforts, while promising, often exhibit fragmented characteristics and they are far from the integrated, autonomous nature of biological cognition. While implementations using active polymers \cite{lu_bioinspired_2018, soto_programmable_2023} or memristive networks \cite{loeffler_neuromorphic_2023, tanaka_molecular_2018, di_ventra_parallel_2013} may show some decision-making~\cite{feng_memristive_2025}, actuation~\cite{ionov:l:2015}, or memory capabilities~\cite{chen:n:2021,xia:am:2021,zhang:aaem:2025}, they frequently rely on external programming or control \cite{wang_robo-matter_2024, alapan_shape-encoded_2019}, respond primarily to specific, predefined inputs \cite{calvino_microcapsule-containing_2018, bayat_self-indicating_2024}, and possess limited internal adaptability or capacity for autonomous self-improvement \cite{jiao_mechanical_2023, luo_highly_2023, bordiga_automated_2024, sabelhaus_-situ_2022}. This points to the need to move beyond merely combining predefined functionalities within a single material under external control and to explore architectures emphasizing autonomy and self-regulation emerging directly from local component interactions. While exploiting physical dynamics (morphological computation) is powerful \cite{pfeifer_morphological_2006}, achieving higher intelligence likely requires integrating this with internal feedback, adaptive memory that is actively shaped by the system's experience, and intrinsic information processing. For example, in "Natural Induction," a physical network's structure is modified by the stress of sub-optimal states, creating a memory that biases future behavior towards better solutions without external reward \cite{buckley_natural_2024}. This form of goal-directed memory, where the material learns from its own history to improve performance, represents a critical step beyond simple responsivity.

A defining feature of Material-Based Intelligence is that the physical dimensions and timescales of the system are not merely engineering constraints, but are fundamentally constitutive of the computation itself. This is a direct consequence of embedding the algorithm into the physics of the substrate. For instance, the time required for a computation is often limited by the physical propagation speed of information through the material, which could be the speed of a chemical reaction front \cite{kim_polymeric_2015}, the diffusion of ions \cite{kamsma_brain-inspired_2024}, or the propagation of a mechanical wave \cite{bordiga_automated_2024}. Similarly, the minimal physical size of the system is often dictated by the complexity of the problem it must solve; a material designed to solve a maze must, in some sense, be large enough to represent the maze's state space \cite{kramar_encoding_2021}.

This stands in stark contrast to, for example, conventional computing. While a conventional system also has physical limits (e.g., transistor size, clock speed), these parameters are engineered to be as universal and problem-agnostic as possible. The time a conventional computer takes to solve a problem is an abstract measure of algorithmic steps, not the physical duration of a process like diffusion across a specific distance. For an MBI system, the computation time \textit{is} that physical duration. The 'algorithm's' time step is a physical relaxation time. Therefore, in MBI, the relationship between the task, the time required, and the physical size of the system is not an indirect engineering consideration, but a direct, physically determined, and often inseparable property of the intelligent matter itself.

Defining the minimal size of the system requires considering both its capacity for memorizing and learning. By a number of $n$ constituents with state variables $x$, a total number of distinct states that can be memorized is defined by the $n$-dimensional integral over the state space with reference to the accuracy $\Delta x$ at which states can be distinguished (falling back to $2^n$ in case of binary discrete memory entities). In a dynamical system, in principle, memory can be distributed in time in terms of temporal states, $x(t)$. Learning capacity is a complementary measure that defines the ability to translate between memorized states and coordinated responses or associated memory patterns. It requires to recognize or produce correlated memory states, and depends on an efficient mediation of between information and its abstracted and ordered representations. Hence, while the memory requirement scales the minimal system size according to the information density, learning capacity depends foremost on the coordination between memory entities. Inspired from a connectionist's view in context of biological neural network: If we define a cognitive process as being the result of a number $n_c$ of communication steps between a set of constituents at a distance $d$ in space, a rough estimate for the processing time can be given by $n_c$[$\tau_d$+$\tau_{intern}$], where $\tau_d$ is the propagation time.

The prevalent theme in many current “smart” materials remains responsivity: a property changes predictably upon a specific input \cite{bayat_self-indicating_2024}, like shape changes with electric fields in nematic liquid crystals \cite{yao_nematic_2022, kos_nematic_2022}, stiffness changes with temperature in phase-change composites \cite{chen_enormous-stiffness-changing_2022, zhao_phase_2021}, or optical signals from mechanical stress in self-reporting polymers \cite{hu_self-reporting_2025}. These systems react, often passively, e.g. through simple thresholds, based on external triggers. They mirror still the conventional computing approach by reducing general computation to simple building blocks (logic gates in computers and similarly simple modules in materials) and thus they generally lack: (1) complex local computation beyond simple predefined thresholds \cite{calvino_microcapsule-containing_2018, milana_morphological_2022}; (2) robust flexibility via dynamic self-organization instead of relying solely on fixed, engineered geometries \cite{ke_three-dimensional_2012, scheidegger_modelling_2020}; (3) autonomous self-improvement without external control or continuous retraining loops \cite{bordiga_automated_2024, sabelhaus_-situ_2022, rauba_self-healing_2024}; and (4) an active memory that shapes future computations, rather than merely recording past events as a final output \cite{xiao_artificial_2020, reiter_memorizing_2020, lee_shape_2022}.

Thus, the transition towards MBI exhibiting a broad characteristics of computation involves fundamental changes along several interdependent dimensions. First, we have to move from relying only on centralized, global feedback coupled with external computation, e.g. mechanical logic gates \cite{jiao_mechanical_2023-1, adamatzky_computing_2011}, physical networks trained with external back propagation \cite{wright_deep_2022, li_training_2024} or by physical reservoirs where only a software-based readout layer is trained \cite{fernando_pattern_2003, nakajima_information_2015}, to intricate local feedback mechanisms enabling material-intrinsic computation, thereby, opening the potential for exploiting the intrinsic parallelism of nature. 

Second, we must move beyond pre-programmed optimization for fixed tasks. A hallmark of true MBI is the capacity for \textit{intrinsic adaptation}, where the material itself reconfigures its response to solve novel or unforeseen environmental challenges without external reprogramming. We point out, however, that the line between "pre-programmed" and "adaptive" is a matter of perspective. For example, a chemotactic oil droplet is not programmed with an explicit algorithm, yet its self-propulsion emerges from a deep interplay between its internal chemistry and external physical gradients, allowing it to navigate complex environments in a highly flexible manner \cite{hanczyc_chemical_2010, cejkova_dynamics_2014, horibe_mode_2011}.

Third, the hallmark of conventional hardware-software separation should be replaced by a design that exploits the specific physical dynamics of the system. This deep embodiment often comes at the expense of the universality found in abstract computation, creating specialized, 'arational' systems whose intelligence is bound to their physical form \cite{fuchslin_ai_2022}. A key design dimension in this paradigm is the degree of determinism in the material's response. This dimension creates a spectrum of cognitive strategies. At one end, highly deterministic and reliable dynamics are ideal for robust control and logic, a principle we explore further in the context of \textbf{multistable and ordered systems (Sec. \ref{sec:strategies_for_cognition}A)}. At the other end, systems with high susceptibility to stimulus patterns and inherent stochasticity are better suited for generalization and creativity, a strategy exemplified by systems operating near \textbf{criticality (Sec. \ref{sec:strategies_for_cognition}C)}. This spectrum also forces us to refine our notion of agency: a material system can be causally \textit{responsible} for its output without being \textit{accountable} in a symbolic or moral sense, a distinction crucial for defining the goals of autonomous MBI.

Fourth and finally, the discrete modes of operation where feedback, learning, and training are distinct from the operational mode must be replaced by a continuous simultaneity and integration of information processing and learning. This complements towards an integration in space \textit{and} time domain. For MBI to emerge, systems must, consequently, exhibit characteristics where: (i) the environment is a deeply intertwined participant through rich physical interaction, with information flow directly influencing, and being influenced by, the material's body \cite{zhao_exploring_2024, harrison_mind_2022, vihmar_how_2023}; (ii) the behavior must be governed by cycles of \textit{intrinsic physical feedback}, where the material's internal state and its actions cyclically influence each other across multiple timescales, enabling autonomous self-regulation \cite{buckley_natural_2024, huang_self-regulation_2018}; and (iii) information is not merely transferred, but processed and interpreted autonomously by the material itself to modulate its future behavior based on emergent goals and its history, drastically reducing reliance on centralized, external computation \cite{kramar_encoding_2021, srinivasa_criticality_2015}.

To achieve this shift from simple responsivity to genuine intelligence, a system must incorporate its own history into its present dynamics. This is not a matter of recording the full past but of maintaining a \textit{fading memory} where recent events are weighted more heavily than distant ones \cite{nakajima_information_2015}. Crucially, this memory must be \textit{active}, not passive. A passive memory, like a scratch on a surface or the ruptured microcapsules in self-reporting polymers \cite{calvino_microcapsule-containing_2018, hu_self-reporting_2025}, merely records an event. An \textit{active memory}, in contrast, is a dynamically maintained physical state that actively participates in and shapes the system's future computations and actions. It is an inseparable, evolving component of the material's ongoing processes.

Excellent examples of active memory are found across different substrates. In biological systems, the methylation level of bacterial chemotaxis receptors serves as a short-term memory of past ligand concentrations, actively tuning the cell's present sensitivity \cite{wadhams_making_2004}. In synthetic materials, the history-dependent ion distribution in an iontronic memristor actively modulates its current conductance, enabling synaptic-like plasticity \cite{kamsma_brain-inspired_2024}. In theoretical models, the adapted natural lengths of viscoelastic elements in a physical network act as a memory of previously found solutions, actively biasing the energy landscape to make future problem-solving more efficient \cite{buckley_natural_2024}. Integrating such active memory is the key to implementing the complex local (and emergent long-distance) feedback loops necessary for genuine self-organization and adaptation. 

This article addresses several pathways central to this paradigm shift: First, can the complex notion of material-based intelligence be captured within clear, operational definitions? Secondly, what are the minimal physical requirements and intrinsic characteristics necessary for a material system to exhibit genuinely intelligent behavior, transcending mere programmed responsivity towards autonomy, learning, and self-organization? Thirdly, how can complex computation be fundamentally realized directly within soft and active matter systems? And finally, what theoretical and experimental pathways show promise for realizing such advanced intelligent materials, particularly emphasizing mechanisms grounded in non-equilibrium physics and inspired by biological principles, and how can these elusive qualities be experimentally verified and characterized?

\section{Manifestations of Material-Based Intelligence}
\label{sec:defining_mbi}

\begin{figure*}[htbp] % Use figure* for full-width figure in two-column layout
\centering
\includegraphics[width=0.8\textwidth]{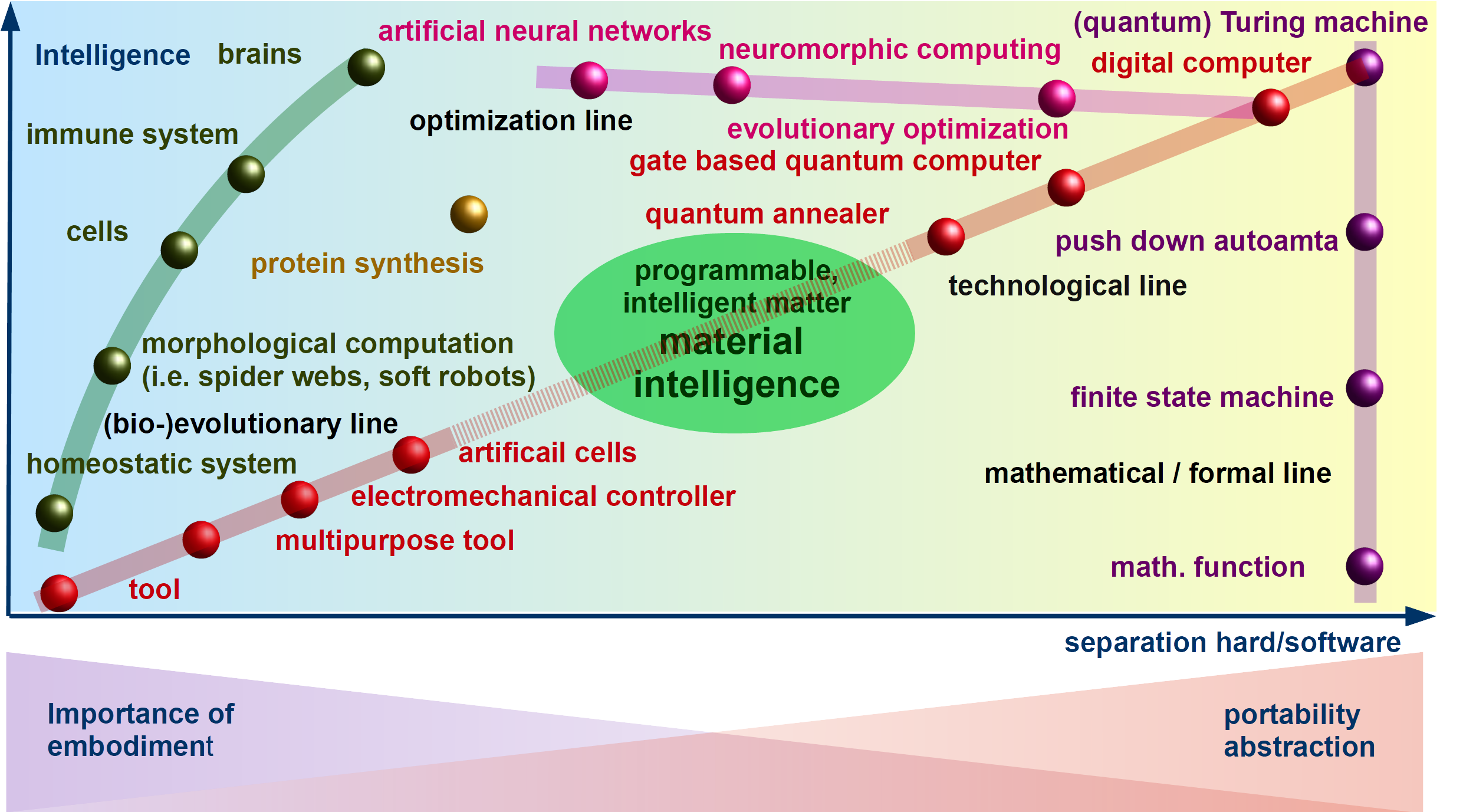} 
 \caption{\textbf{The Landscape of Intelligent Systems and the vast land for Material-Based Intelligence.} This schematic maps different forms of intelligence onto a design space defined by two key axes: the degree of hardware-software separation (horizontal) and the breadth of problem-solving capability (vertical). Distinct lines of development traverse this landscape. The \textbf{mathematical/formal Line} (purple) illustrates increasing computational power within a paradigm of high abstraction and hardware independence. The \textbf{(bio-)evolutionary Line} (green) shows how natural systems achieved sophisticated cognitive capabilities (\textit{e.g.}, cells, immune systems, brains) while maintaining a deep integration of information processing with their physical bodies. This line defines a trajectory deep into the high-integration, high-functionality quadrant. The \textbf{technological Line} (red) and a branch, the \textbf{optimization line} (magenta) charts the development from embodied tools to electro-mechanical systems, before making a significant leap to abstract digital computation, thereby leaving a conceptual "gap" in the development of highly capable yet physically integrated systems. The jump from analogue controllers to machines capable of various degrees of formal computation is wide and consequently, the line connecting them is dashed. \textbf{Material-Based Intelligence (MBI)} is positioned to fill this gap, aiming to engineer systems that exhibit complex, autonomous functions directly through their physical structure and dynamics. This paradigm sits at the intersection of biological inspiration and a new technological frontier, leveraging the principles of embodiment over pure abstraction. There are outliers not covered by lines in this diagram. As a prominent example, we noted protein synthesis. Although it relies deeply on the details of physics and chemistry, the process can produce a combinatorially rich number of outcomes in a programmable manner.}
\label{fig:spectrum}
\end{figure*}

We regard the distinction between machine-based and material-based intelligence not as a mutually exclusive choice but as the respective ends of a spectrum characterized by the degree to which one can clearly distinguish between hard- and software. Before we discuss this spectrum (illustrated in Fig. \ref{fig:spectrum}), we exemplify machine- and material-based intelligence with respect to a conventional "device", i.e. a controlled system that uses the results of computations for some purpose, e.g. the control of actuators and sensors or other means to interact with the physical world.

%%%%%%%%%%%%%% FIG 1 %%%%%%%%%%%%%%%%%
\begin{figure*}[htbp] % Use figure* for full-width figure in two-column layout
\centering
\includegraphics[width=0.8\textwidth]{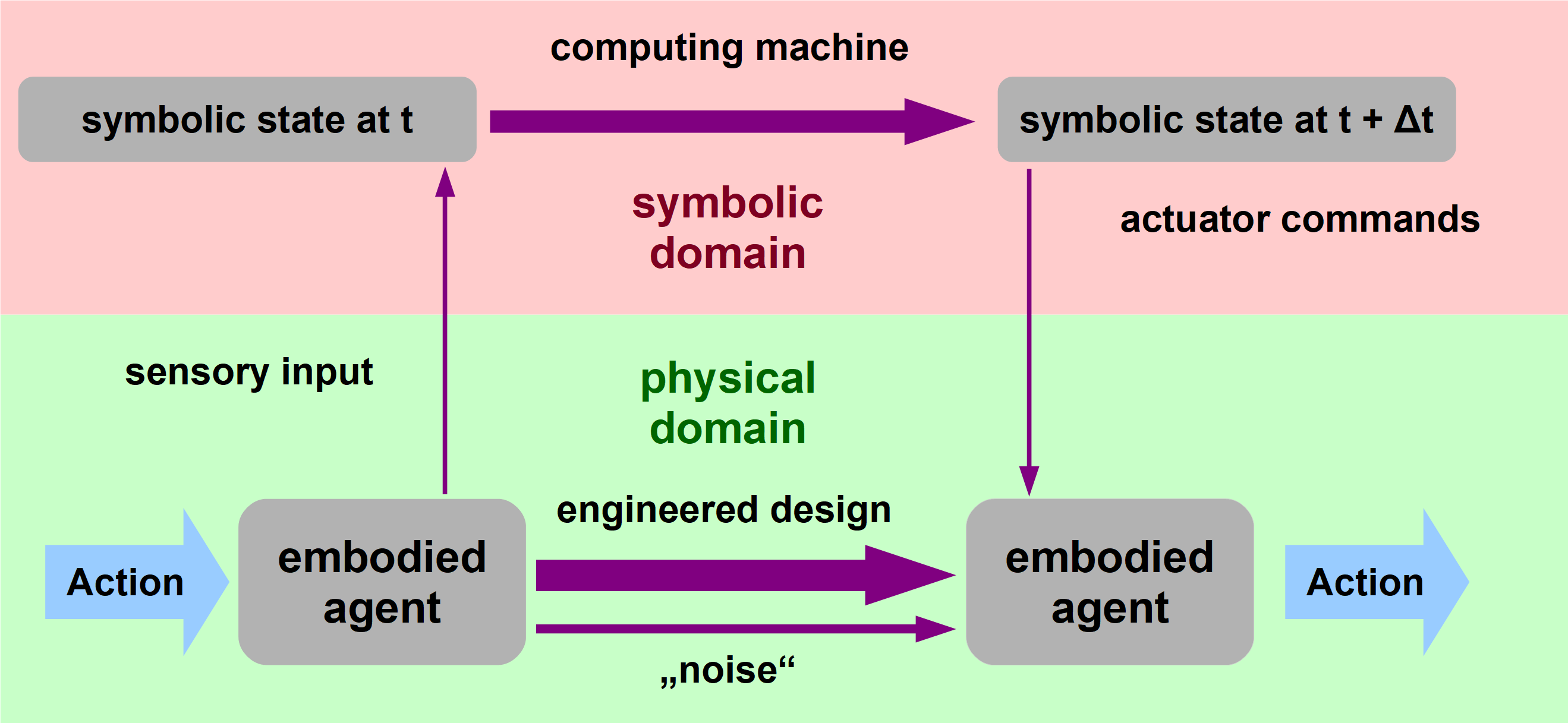} 
 \caption{\textbf{The Conventional Control Architecture and its Separation of Physical and Computational Domains.} Conventional control architecture of an embodied agent, i.e. a controlled system that performs some actions in the physical world. Such an agent requires at least a twofold separation or encapsulation of physical dynamics and control. First, the computing machine including its communication channels with sensors and actuators establishes an interface between abstract and portable algorithms and the computing hardware. The according flow of information is depicted by the vertical and the topmost arrows. The reduced thickness of the vertical arrows visualizes the fact that the sensory bandwidth is still a bottleneck in control. The centred flow of information (decorated with the term "engineered design") represents the part of the control that is facilitated by proper planning and design. Thirdly, there is what we call noise in this context, i.e., the information flow resulting from external influences (which are not necessarily known in all details.) Those parts of the dynamics that are related to the (known) engineered design are modelled, whereas noise stands for the unmodelled parts. Note the importance of a de-physicalization of sensory input and the re-physicalization of the result of computations with control purpose by actuators. This de- and re-physicalization is a characteristic feature of systems with a clear separation between hard- and software and may be obsolete in material-based intelligence (Figure adapted from \cite{fuchslin_morphological_2013}).}
 \label{fig:hardsoft}
\end{figure*}
%%%%%%%%%%%%%%%%%%%%%%%%%%%%%%%%%%%%%%

To be more concrete, we analyze an idealized embodied agent controlled by a conventional digital computer (hardware), as depicted in Fig. \ref{fig:hardsoft}. In this standard architecture, the computation required for control is performed based on an abstract, symbolic representation of reality. This is achieved by converting continuous, real-world sensory data into discrete bit sequences and processing them in discrete time steps using some formal language (software). A key feature of this paradigm is the abstraction from the underlying hardware; the specific physical dynamics of the hardware are rendered irrelevant to the computational outcome. This separation ensures that a program (software) is portable and can run on any compliant machine. Note that this twofold de-physicalization, namely the computational treatment of a symbolic representation of reality with a symbolic or formal language, has to be followed by an according re-physicalization. In more detail, the information flows in the embodied agent can be understood by considering three distinct pathways:

First, the primary control loop (vertical up, top and vertical down arrows) involves a translation between the physical and computational domains. Sensory data from the environment is first \textit{de-physicalized} into a discrete, symbolic representation that can be processed using a symbolic language that runs on a computer. Importantly, the details of the physical dynamics of the computer hardware are irrelevant to the outcome. After some result (e.g. a decision) has been computed, the resulting commands are \textit{re-physicalized} back into physical forces and motions via actuators. These translation steps are technologically demanding and often create a significant bottleneck, limiting the bandwidth of interaction with the world.

Second, a significant part of the agent's success comes from its \textit{engineered design} (center arrow). The goal of good engineering is to facilitate control by equipping the machine with a dynamics that is self-stabilizing and predictable. For example, in aviation, a training plane is designed for "good-natured flight behaviour" to simplify the pilot's task. This deliberate simplification, while ensuring reliability and controller portability, represents an abstention from exploiting the richer, more complex dynamics the physical hardware might naturally possess.

Third, the model includes \textit{noise} (bottom arrow), which represents all unmodeled physical interactions, from external perturbations to incomplete knowledge about the device itself. In conventional robotics, the goal is to minimize the impact of this channel.

A key promise of Material-Based Intelligence is to render the separation between these channels, particularly the primary control loop, obsolete. By embedding computation directly into the material's dynamics, MBI aims to eliminate the costly de- and re-physicalization steps and harness a much broader sensory bandwidth, moving intelligence from the abstract controller into the physical body itself.
%%%%%%%%%%%% FIG 2 %%%%%%%%%%%%%%%%%%%%
%%%%%%%%%%%%%%%%%%%%%%%%%%%%%%%%%%%%%%%%

In conventional devices, control is dominant and mainly exerted by general-purpose computing machines. We claim that the spectrum of machine- and material-based intelligence offers a whole range of systems. Fig. \ref{fig:spectrum} shows our approach to organize them. The two-dimensional figure spans the spectrum of hard- and software separation (horizontal axis) and capability to handle classes of tasks (vertical axis). We identify the term "intelligence" with the size and complexity of these problem classes. We include biological systems, because we are interested in the interplay between algorithmic and physical information processing, thereby interpreting the term "machine" broadly. In addition, we don't discriminate between technological systems that are designed and constructed for a purpose by engineers, and biological systems that use their functionality (reproduction and heredity being major ones) to maintain themselves and are subject to evolution.

To give a better understanding of the intentions motivating Fig. \ref{fig:spectrum}, one can analyze its boundary cases, i.e. the corners of the figure. The lower left corner represents physical objects without any intended or evolved functionality, e.g. a random rock. The upper left corner stands for an object with comprehensive cognitive and computational functionality based entirely on physical embodiment but without any form of programmability. It is hard to conceive such an object. One could think of it as some form of immense implicit hash table that responds to every possible input with an appropriate output. The content of this (implicit) hash table would be immutable and unstructured. The lower right corner stands for abstract statements without free variables, e.g. "1+1=2". Finally, the upper right corner represents a maximally general algorithmic framework, here represented by a Turing machine or any general-purpose programming language that can emulate a Turing machine (e.g. Python, C, etc.). According to the Church - Turing hypothesis, a Turing machine (or an equivalent) encompasses every effectively calculable function; for a thorough, but non-technical discussion, see \cite{sep-church-turing}. 

The figure is structured by a number of "main lines". 
On the right-hand side, we have mathematical reasoning through hierarchies of formal languages. Such languages enable purely algorithmic, hence machine-agnostic solutions. The primary purpose of formalisation is to eliminate dependencies on computational devices and achieve maximum separation between algorithms and the physical processes that realize them. Therefore, the mathematical/formal line is on the right of Fig. \ref{fig:spectrum} and vertical (implying no dependence on material dynamics). It depicts the abstract automata models of the different types of grammars according to the Chomsky~\cite{chomsky_certain_1959} - hierarchy, conventionally used in computer science.

The opposite, left end of the spectrum represents systems with a functionality that is completely embodied and therefore not portable at all. The lower left corner is the starting point of the bio-evolutionary line. This end of the spectrum contains systems, which, at their core, intrinsically master tasks and solve problems through the direct exploitation of their inherent material physics and dynamics, operating without reference to any externalised process- or state information in the form of software and memory (Fig. \ref{fig:spectrum}). Such systems are inherently "arational"; their behavior, while functional, may not be easily described in a formal language or be reducible to a simple algorithm, much like the procedural knowledge involved in riding a bicycle (most individuals know how to ride but lack a language-based description for transmission) or the emergent decision-making of a trained artificial neural network that often defies direct explanation \cite{fuchslin_ai_2022,ropohl_allgemeine_2009}. This lack of a language-based description but physically embodied control can limit portability, but it offers at least two profound advantage: First, in contrast to program code, which must explicitly encode all physics needed, material-based intelligence acts already based on physics, implicitly using all underlying physical laws without the need for explicit instruction \cite{milkowski_morphological_2018, fuchslin_morphological_2013, sitti_physical_2021}. Second, MBI does not require a translation of sensory signals into some form of representation or of results into commands for actuators, see Fig. \ref{fig:hardsoft}). This is not so much a time issue or a question about computational complexity, but, for example, for the control of complex molecular reactions (e.g. in a cell), it is simply not possible to measure and control the position of molecules. But in an embodied system, one can invoke the interplay between reaction, diffusion (in potentially non-trivial geometries) and steric configurations for the orchestration and control of life-like chemistries \cite{soh2010reaction, shaw2010diffusion, seo2022spatiotemporal}. 

Biological evolution has produced systems where form and function are deeply intertwined; from mechanistic single-celled organisms maintaining homeostasis \cite{wadhams_making_2004} or exhibiting simple memory through ionic shifts \cite{boussard_memory_2019}, to complex multi-cellular collectives self-organizing without a central blueprint \cite{theraulaz_spatial_2002, zhang_classical_2025} or displaying robust morphogenetic problem-solving to restore target anatomies \cite{mcmillen_collective_2024, levin_multiscale_2024}. The bio-evolutionary line in Fig. \ref{fig:spectrum} starts on the left (complete dependence on the dynamics of the embedding system), but is gradually bending towards the right. This is because some of the activities of the brain can be regarded as "software". Although brains are not programmed, they can be taught general problem-solving procedures, such as methods for adding numbers, playing chess, or cooking by following a recipe. We placed complex, evolved networks high up on this line. By such networks, we understand all sorts of compositions of interacting nodes, ranging from chemical networks \cite{baltussen2024chemical}, over neuromorphic networks of nanowires \cite{milana_morphological_2022}, random boolean networks as models of cellular/genetic regulation \cite{serra2010dynamics} or networks in social contexts \cite{bessi2015science, lynn2020human} and material science \cite{bonamassa2025logarithmic}. It may be debatable to what extent such networks exhibit a distinction between hard- and software. Note that, e.g. in the case of an organisation, the nodes can be identified with individuals and the edges with their connections/communication lines. Some management approaches favour a setting where nodes and edges are decoupled, meaning that positions are independent of individuals. In the real world, this is rarely the case; the personalities of individuals play a crucial role in the functioning of an organisation and constitute a considerable part of the implicit knowledge stored in a network. 

With regard to the degree of determinism or traceability, the degree of decision-making responsibility, and the susceptibility to new situations, network nodes and subsystems may populate different domains. The interplay of these aspects can be illustrated by coming back to the example of the cyclist. The brain of the cyclist holds the responsibility for deciding where to ride and react on traffic situations. At the other end, the cyclist's legs are complex, integrated subsystems that exhibit foundational aspects of material-based intelligence. For instance, intrinsic mechanical responses are accompanied by finely tuned muscular reflexes in order to stabilize joints - most prominently the knee joint. Intrinsic mechanical response is provided by ligaments, menisci, and bone tissues and deliver an important contribution to stability, by encoding a potential energy landscape in the relatively low-dimensional angular space of the knee. The corresponding subsystem has a high degree of determinism and reliability with respect to the stability target. The next higher processing level is contributed by finely tuned neuron-mediated muscular reflexes triggered by mechanical receptors in the periarticular tissues. Here, the responsibility is widened towards mediating between mechanical state information, the current overarching and purposed muscle action, as well as additional tactile external stimuli that are combined towards a decision of "firing" at the responsible synapses for triggering stabilizing muscle action. In this case, communication is neuron-mediated. The next higher level of coordination mediates between the cyclists decision, where to go, the balance information from the vestibular system, tactile response, and a meaningful muscle action across the whole body. Different cyclists may share action patterns and the cerebellum shows convergence in the solution of the bicycle balance responsibility. However, like in artificial neural networks, the processing state and results are not traceable or within a symbolic language.

The immune system is another example of an information processing biological entity. Since the seminal work of Perelson \cite{perelson1997immunology, hershberg2001immune}, the (adaptive) immune system is also regarded from an information-theoretic perspective. This viewpoint became even more important, since Matzinger \cite{matzinger2002danger} presented a theory in which the immune system is viewed as more than an enumeration of non-self epitopes. Matzinger's work also stimulated discussions in the context of morphological computing \cite{HauserEbookMorphComp2014}, where an engineering perspective was adopted to plausibilize the (still hypothetical) interaction between the molecular immune system and the brain. In short, the immune system collects information about what goes wrong on a chemical level, the brain and nervous system complement this with information about where and the context. Whether or not this information is indeed  linked is unknown, but it would certainly be sensible if evolution had developed a way to do so and engineers certainly would consider it. 

Recent developments investigate the possibility of understanding aspects of the immune system in terms of perceptrons, \cite{scheidegger_modelling_2020}. It is important, however, to realize that the term "perceptron" has to be taken in terms of a data structure and not a material realization of some form of neurons. The perceptron is regarded as a versatile data structure that has found different realisations in biology, probably because, despite its simplicity, it offers a broad range of possibilities for evolutionary fine-tuning. In addition, this data structure is already of value as a single instance, and can be used to set up complex networks. Therefore, a major benefit of perceptrons, independent of their implementation, is that they offer a, in some sense, smooth landscape for evolutionary progress. 

The line of technological development, while driven by the success of computational technology and the clear division of labour between programmable controllers and passive sensors/actuators, exhibits two branches. There are classical computers controlled by algorithms which have a syntax and semantics. From these computers, we distinguished evolutionary or optimisation approaches, in Fig. \ref{fig:spectrum} with artificial neural networks as their endpoint. Classical approaches are on an ascending line, going from left to right. More division between hard- and software coincides with more computational power. 

A note on the now growing field of quantum computing might be in order here. Although in theory, quantum Turing machines are computationally equivalent to classical Tuinng machines (although there are several subtleties to observe, see e.g. \cite{gurevich2022simple}), today's gate-based quantum computers are still far less powerful than their classical counterparts. The technologically already relatively advanced quantum annealers are not Turing complete, but work only on a specific class of optimisation problems, namely quadratically unconstrained binary optimisation (QUBO), and are therefore positioned below more general, e.g., gate-based, quantum computers. The mechanism underlying quantum annealers relies heavily on the details of physics (e.g. with respect to the adiabatic theorem). For a detailed discussion, see \cite{rajak2023quantum}. The left (and lower) part of the technological line accounts for simple tools or devices that, depending on their mechanical or electrical configuration, may serve multiple purposes. The functionality of these systems is basically given by their physical dynamics. There is a considerable conceptual gap between configurable tools and programmable machines. It is this gap that shall be bridged by material intelligence. 

Evolutionary or optimisation approaches are on a line starting from digital computers and going to the left. At first glance, this may look strange. However, optimisation procedures have to be regarded in terms of a "master algorithm" with (at least) two levels of software, \cite{domingos2015master}. To explain this point, we take artificial neural networks as an example. The network (which can be executed on a conventional computer or on an often partly analogue, neuromorphic device, see e.g. \ cite {kimura2021amorphous,gautam2022conductance}) consists of components that are well understood and, in the case of an implementation on conventional hardware, are operated by conventional software. However, the functionality of the neural network is determined by weights (and some other parameters). These weights are the result of a learning process. Although the learning algorithm itself is written in some programming language (and often quite easy to interpret), the weights have no direct interpretation. It is, in fact, on a functional level, hard to say what a neural network does and why it works. The actual software of a neural network sometimes includes billions of parameters. These parameters can be numbers (in which case they take effect through software running the network) or details of the physical setting (in the case of neuromorphic computing). In the latter case, the physical details of the implementation become relevant, and that justifies the direction of the optimisation line. In general, we have to distinguish the process to be optimised and the optimisation algorithm itself. The optimisation of a parameterised process can be achieved with a transparent algorithm. The latter is usually quite independent of the process to be optimised (which may itself be a piece of software, but can also be a technological artefact). 

Note that Fig. \ref{fig:spectrum} does (of course) not capture all controlled processes one observes in nature or technology. As an example, we give one prominent case, which seems to be sporadic with respect to the different lines we have drawn: protein synthesis. Protein synthesis is code-based with respect to the primary structure (the sequence of amino acids). However, the secondary and tertiary structure of proteins is the result of folding (for a discussion of the importance of this fact in the context of the origin of life, see \cite{wills_reflexivity_2019}). Protein folding depends strongly on the details of the involved molecules and their interaction with the environment; this example highlights the complexity of cleanly separating hard- and software in advanced information processing systems.

Based on these considerations and integrating knowledge about various material systems, we propose two complementary viewpoints for MBI: one emphasizing its \textit{architectural foundation} and another describing its \textit{functional manifestations}.

\subsection{Aspect 1: Architectural Foundation of MBI}
\label{ssec:def1}

\begin{formaldef}{Architectural Foundation}{1}
\textbf{Material-Based Intelligence (MBI)} fundamentally describes information processing systems where the principled architectural separation between a passive physical substrate (hardware), discrete data storage units (memory), and abstracted information processing units (software/algorithm) is intentionally minimized or altogether absent \cite{harrison_mind_2022, di_ventra_parallel_2013, mcevoy_materials_2015}. 
\end{formaldef}

In contrast to conventional machine intelligence, which relies on the physical separation of memory and processing and incurs significant energetic and temporal costs from data transfer (the von Neumann bottleneck) \cite{yao_fully_2020, yang_bicoss_2022, mead_neuromorphic_1990}, MBI systems inherently perform computation, control, sensing, and memory operations directly through the intrinsic physical dynamics of the material itself. Information in MBI is not stored in separate memory banks but is encoded within and processed by the material's evolving state variables (\textit{e.g.}, molecular configurations \cite{reiter_memorizing_2020}, structural arrangements \cite{kramar_encoding_2021}, chemical concentrations \cite{kamsma_iontronic_2023}, defect patterns \cite{kos_nematic_2022}, or stress distributions). Processing occurs via the dynamic,in general non-linear, evolution of these states, governed by local interaction rules derived directly from underlying physical laws (\textit{e.g.}, mechanics, thermodynamics, chemical kinetics, electromagnetism). Critically, for sustained activity and complexity, these systems often operate far from thermodynamic equilibrium, constantly exchanging energy with their environment \cite{kuncic_emergent_2018, banda_online_2013, england_dissipative_2015, goettems_physics_2024}. This paradigm uses the material substrate not merely as a passive carrier for a program, but as the active medium in which the 'program' is intrinsically intertwined with the 'processor' \cite{milkowski_morphological_2018}. The computational material is deeply and sensorically embodied within its environment, allowing complex input patterns to arise naturally from this interaction and enabling dynamic morphological transformations to serve as computational primitives \cite{zhao_exploring_2024}.

Key architectural and functional features distinguishing MBI under this viewpoint include:

 \paragraph{Integration vs. Separation}. A direct and intentional minimization of functional distinctions (sensing, memory, computation, actuation) by co-locating these capabilities directly within the material's physical structure and dynamics. This fundamentally contrasts with the spatially separated architectures of traditional machine intelligence \cite{di_ventra_parallel_2013, yao_fully_2020, mead_neuromorphic_1990}.
 \paragraph{Intrinsic Physics as Algorithm}. Computation is defined not as the execution of external instructions, but as the material's dynamic state evolution, which directly and implicitly implements physical laws and relationships \cite{kuncic_emergent_2018, buckley_natural_2024}. The computation \textit{is} the physics in action, making the distinction between output and process ambiguous.
 \paragraph{Embodied State and Data Representation}. Information resides intrinsically in physical state variables of the material system itself (\textit{e.g.}, atomic arrangements, charge distributions, chemical concentrations, defect patterns), rather than being abstract symbols stored in memory units. The material's form embodies its data.
 \paragraph{Distributed and Local Interactions}. Complex global behaviors and problem-solving abilities emerge predominantly from distributed, local physical rules, dynamics, and interactions between proximate material components \cite{rubenstein_programmable_2014, theraulaz_spatial_2002}. Long-distance coordination is achieved through self-propagating phenomena or emergent gradients, not centralized global control signals.
 \paragraph{Non-Equilibrium Operation}. Sustained activity, self-organization, and complex adaptive dynamics typically necessitate a continuous input and throughput of energy, perpetually driving the material system far from thermodynamic equilibrium \cite{england_dissipative_2015, goettems_physics_2024, aprahamian_non-equilibrium_2023}. This energy flow dictates functional existence.

Note that these key architectural features (at least partially) circumvent some essential problems one usually encounters in setting up a representation of reality \cite{fuchslin_morphological_2013}. First, the abstraction or independence from the computational substrate, which underlies the separation of hard- and software, comes at a price: Colloquially expressed, programming languages contain no implicit physics. This requires that all physics has to be encoded again explicitly. Second, the formalisation of complicated boundary conditions, e.g. the shape of geometrically non-trivial objects, is computationally demanding. If this formalisation can be avoided (e.g. by use of soft materials for a gripper), complicated computations can be avoided. Thirdly, nature is inherently parallel, which avoids designing for parallelisation. Finally (and probably of minor importance): Simulations quite often use large amounts of random numbers, which are not entirely trivial to produce. Nature, via temperature, provides randomness for free. 

\subsection{Aspect 2: Functional Manifestation of MBI}
\label{ssec:def2}
As a direct consequence of its distinct architectural foundation (Viewpoint \ref{def:1}),
\begin{formaldef}{Functional Manifestation}{2}
\textbf{Material-Based Intelligence} functionally manifests as the autonomous capacity of a physical system to exhibit dynamic, adaptive, and potentially goal-directed behaviors that emerge from its intrinsic, embodied dynamics, operating over relevant timescales and effectively managing internal and external information.
\end{formaldef}

To make these functional capabilities concrete, we identify a hierarchy of core requirements, derived from a systematic, data-driven analysis of the field \cite{baulin_discovery_2025}, that a material must possess to be considered intelligent. These requirements must operate predominantly through local physical rules, with global coherence emerging collectively.

\paragraph{Local Interaction and Coordinated Distributed Dynamics:} This is the most foundational principle of MBI, stating that intelligent materials must intrinsically sense their local environment and interact primarily with nearby components. Complex global behaviors and functions must emerge from the propagation, integration, and potentially non-local influence of these local physical events, thereby minimizing reliance on centralized control \cite{rubenstein_programmable_2014, pfeifer_self-organization_2007}. Sophisticated \textit{hierarchy} emerges when local dynamics lead to \textit{long-distance coordination} over larger scales, which might occur through physical wave propagation \cite{ziepke_acoustic_2024} or emergent gradients \cite{kramar_encoding_2021}.

\paragraph{Local Active Memory:} Materials must possess intrinsic mechanisms for local information storage that genuinely reflect past states or interactions. This memory is not a passive recording; it must \textit{actively influence} future dynamics and responses \cite{reiter_memorizing_2020}. Information could be encoded via material phase \cite{yu_hydrogels_2020}, charge distribution \cite{kamsma_chemically_2024}, or structural configuration \cite{kramar_encoding_2021, lee_shape_2022}. For behavioral continuity over extended periods, this memory must be robust over relevant timescales and may require mechanisms for persistence or `replication` to ensure information endures beyond transient effects \cite{langton_computation_1990}.

\paragraph{Local Embodied Computation:} Information processing must occur intrinsically through the material's local physical laws and interactions \cite{kuncic_emergent_2018, banda_online_2013, kamsma_brain-inspired_2024}. This necessitates non-linear dynamics capable of transforming signals, integrating information, and making decisions (\textit{e.g.}, via physical thresholding mechanisms like mechanical buckling \cite{xi_emergent_2024-1}), without symbolic processing steps \cite{nakajima_information_2015, kos_nematic_2022}.

\paragraph{Self-Organization for Functional Structure:} The capacity to spontaneously form or reconfigure functional structures from local interactions is fundamental to adaptability \cite{sole_open_2024, osat_non-reciprocal_2023}. This enables systems to build and repair themselves without a complete external blueprint \cite{kriegman_scalable_2020}, as seen in biological development \cite{theraulaz_spatial_2002}.

\paragraph{Self-Correction and Repair:} Autonomous systems must possess mechanisms to detect and recover from errors or damage, restoring functionality and maintaining integrity over time \cite{wool_self-healing_2008, kim_polymeric_2015, terryn_review_2021}. This implies an internal 'sense' of a preferred state, triggering corrective physical feedback \cite{buckley_natural_2024}.

\paragraph{Adaptive Feedback Loops:} Internal feedback cycles are a necessary condition for self-regulation, learning, and persistent goal-seeking. These loops must dynamically link sensing, memory, computation, and action, enabling the system to continuously modulate its behavior \cite{huang_self-regulation_2018}.

\section{Computation in Soft/Active Matter: Intrinsic Transformation of Information}
\label{sec:computation_requirements}
For a material (often derived from soft and active matter) to perform computation intrinsically through its physical dynamics, it must possess specific exploitable characteristics that enable the transformation of input signals or states into distinct output signals or states \cite{milkowski_morphological_2018, kaspar_rise_2021}. This implies a blurring of the lines, where the system's inherent physics themselves perform the computations, without distinction between the 'output' and the 'process'.

\begin{figure*}[htbp]
 \centering
 \includegraphics[width=1\textwidth]{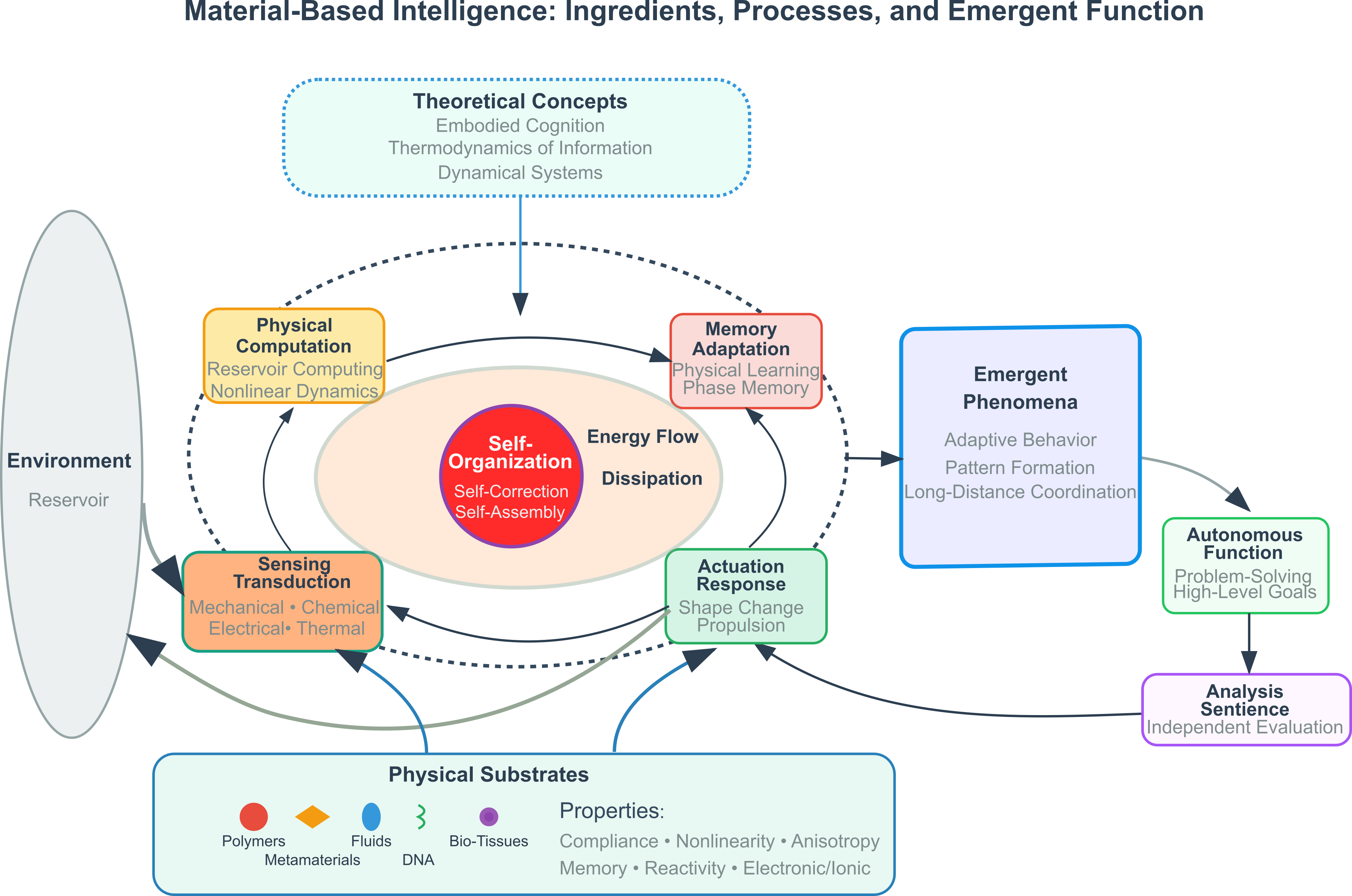} 
 \caption{\textbf{The Material-Based Intelligence (MBI): From Physical Ingredients to Emergent Function.} This diagram provides a view of the MBI paradigm, organised into foundational elements, core processes, and emergent outcomes, illustrating a bottom-up flow from physical substrates to functional intelligence.}
 \label{fig:hierarchical_mbi}
\end{figure*}

The MBI structure, depicted in Figure \ref{fig:hierarchical_mbi}, offers a process-oriented view of the MBI paradigm, from foundational ingredients to emergent function. At its core, the \textbf{Self-Organization} is the central generative mechanism. This is because, unlike engineered systems, the intelligent functionality in MBI is not merely programmed into a static structure; rather, the functional architecture (the intricate web of sensing, computation, memory, and actuation) itself is a dynamic feature that emerges and sustains itself through self-organizing processes. This engine of self-organization, which encompasses self-correction and self-assembly, is what bridges the gap between the basic \textbf{Physical Substrates} and the complex operational loop of an intelligent system. This generative process, however, is not spontaneous and is connected to the environment, such that it is fundamentally a thermodynamic phenomenon. Therefore, \textbf{Energy Flow and Dissipation} are situated alongside self-organization as the essential fuel. MBI systems are open systems operating far from equilibrium, and it is the continuous throughput of energy that drives the self-organization against entropy, allowing for the creation and maintenance of complex, ordered states. From this energy-driven, self-organized core emerges the perception-action loop of MBI, where \textbf{Sensing/Transduction} from the environment feeds into \textbf{Intrinsic Physical Computation}. This computation is deeply intertwined with \textbf{Active Memory and Adaptation}, where the material’s history physically shapes its future responses. This cycle is completed by \textbf{Actuation/Response}, which acts back upon the environment. This entire dynamic interplay, guided by overarching \textbf{Theoretical Concepts}, gives rise to observable \textbf{Emergent Phenomena} such as adaptive behavior and long-distance coordination, which in turn enable the system to perform \textbf{Autonomous Functions} and achieve high-level goals.

Based on the literature and incorporating findings from diverse systems including physical reservoirs \cite{nakajima_information_2015, fernando_pattern_2003}, wave-based computers \cite{marcucci_theory_2020, adamatzky_computing_2011}, and reaction networks \cite{banda_online_2013, lin_intelligent_2021}, several core requirements for realizing information processing mechanisms emerge, as outlined below.

%%%%%%%%%%%%%%%%%%%%%%%%%%%%%%%%%%%%%%%%%
\subsection{Nonlinearity}
\label{subsec:nonlinearity}
As mentioned, a key ingredient for MBI to work stems from nonlinearity. This is required for complex transformations beyond simple superposition or linear filtering, a prerequisite for universal computation and complex function approximation \cite{marcucci_theory_2020, kuncic_emergent_2018}. Nonlinearity can be implemented in a material through a variety of intrinsic responses. For instance the non-Hookean elastic properties of Liquid Crystal Elastomers (LCEs) undergoing large strain deformations \cite{li_computational_2024}, or the viscoelasticity that allows complex temporal integration in biological models \cite{buckley_natural_2024} are simple examples in this framework. Additional examples include non-ideal chemical reaction kinetics found in Reaction-Diffusion Systems \cite{banda_online_2013, lin_intelligent_2021}, inherent threshold phenomena \textit{e.g.}, buckling instabilities in soft metamaterials enabling binary switching \cite{xi_emergent_2024-1}, memristive switching \cite{lee_nanograin_2023}, or polymer phase transitions altering stiffness dramatically \cite{zhao_phase_2021, chen_enormous-stiffness-changing_2022}, or through complex collective interactions leading to non-trivial emergent behaviors \cite{negi_emergent_2022, son_emergent_2024}.

\subsection{History dependence and state memory}
\label{subsec:history}
The material must possess internal state variables that persist over relevant timescales and influence its dynamic evolution. This endows the system with \textit{history dependence}, which is critical for \textit{temporal processing} tasks typical of reservoir computing requiring 'fading memory', where past inputs influence the present but with diminishing strength \cite{nakajima_information_2015, muller_what_2017-1, kaspar_rise_2021}. The physical embodiment of this state can vary from charge distribution in aqueous electrolytes \cite{kamsma_brain-inspired_2024, kamsma_chemically_2024}, to persistent 'structural conformation' in polymers and elastomers \cite{reiter_memorizing_2020}, characteristic liquid crystal defect patterns \cite{kos_nematic_2022, zhang_logic_2022}, dynamically maintained chemical activity patterns \cite{kramar_encoding_2021} or mechanical configurations. This capacity for active memory is crucial for transforming simple stimuli-response into genuine cognitive function. In Sec. \ref{sec:defining_mbi}, based on a dynamical systems approach, we introduced three different types of memory: state in a basin of attraction, change of attractor landscape, perturbations with long transient times. Note that the first possibility constitutes discrete memory states, whereas the second and third variant enable the storage of continuous values which can be discretized by some thresholding process. 

\subsection{Material coupling}
\label{subsec:material}
Components or distinct regions of the MBI must effectively interact to allow for information propagation, complex transformations, and \textit{collective processing} \cite{kuncic_emergent_2018}. This coupling can be realized through diverse physical means: propagation of mechanical forces \cite{louvet_reprogrammable_2024, li_training_2024}, transmission of electrical currents \cite{yao_fully_2020, stern_training_2024}, diffusion of chemical signals \cite{adamatzky_computing_2011, theraulaz_spatial_2002}, or the propagation of acoustic waves \cite{ziepke_acoustic_2024}. The underlying \textit{topology} of these interactions, which defines the network of connectivity, is a fundamental determinant of the system's computational capabilities \cite{kuncic_emergent_2018}.

\subsection{Internal dynamics}
\label{subsec:internal}
The material's dynamic behaviour must not be entirely dictated by the input signal. For effective computation, it needs an "internal life", a sufficiently \textit{rich repertoire of internal states} or \textit{modes of oscillation} that can be excited and modulated by the input, but which are not simply enslaved to it. This provides the \textit{high-dimensional state space} required for complex \textit{non-linear feature mapping}. This characteristic is closely related to 'echo states' in physical reservoir computing, where the material's complex impulse response "echoes" past inputs in a way that reveals hidden computational structures \cite{fernando_pattern_2003}. Without this internal richness (\textit{e.g.}, in a very stiff, over-damped, simple transducer material), the system will lack the complexity needed for higher-order computations.

\subsection{Readout Mechanism}
\label{subsec:readout}

For the computed results to be interpretable, a consistent and reliable mechanism must exist to \textit{measure or interpret specific aspects of the system’s physical state} as the computational output. This can involve measuring bulk properties like electrical conductance, optical transparency, macroscopic mechanical deformation, or the localised concentration of chemicals at specific points or times \cite{fernando_pattern_2003, kim_nanoparticle-based_2020, nakajima_information_2015}. The observability of distinct, input-dependent states is crucial for external users, though internally, this 'output' is inherently part of the 'process' and vice-versa, with no functional separation.

\subsection{Examples}
\label{subsec:examples}
Computational primitives that have been observed or proposed in such systems include physical implementations of basic 'thresholding' operations, \textit{e.g.}, mechanical instability triggers \cite{kim_polymeric_2015}, Boolean 'logic gate' implementations, \textit{e.g.}, via particle collisions in active matter or defects in liquid crystals \cite{adamatzky_computing_2011, kos_nematic_2022, louvet_reprogrammable_2024}, temporal filtering/integration due to material memory effects \cite{nakajima_information_2015, govern_optimal_2014, tanaka_molecular_2018}, 'weighted summation analogies' where physical forces or currents sum at junctions \cite{li_training_2024, banda_online_2013}, and complex spatio-temporal transformations characteristic of physical reservoir computing \cite{marcucci_theory_2020}. Importantly, MBI frequently utilises the inherent \textit{parallelism} and \textit{analogue nature} of physical dynamics, offering potential advantages in 'speed' and 'energy efficiency' over conventional computation for certain tasks \cite{mead_neuromorphic_1990}, in particular, if the computation is performed on a device with a von Neumann architecture.

\section{Strategies for Constructing Emergent Intelligence}
\label{sec:strategies_for_cognition}

Integrating the core characteristics required for Material Intelligence (Section \ref{sec:defining_mbi}) within systems governed by engineered physical dynamics (Section \ref{sec:computation_requirements}) could enable the emergence of sophisticated behaviours far beyond mere responsiveness. The overall MBI paradigm can be viewed as a bottom-up process, flowing from foundational physical properties to high-level autonomous functions, as illustrated in Figure \ref{fig:hierarchical_mbi}.

The strategies to achieve this integration can be understood through the unifying lens of a dynamical systems approach. Within this framework, the vast landscape of MBI research is coalescing around three principal strategies for embodying cognitive function. These distinct pathways are not mutually exclusive but represent different physical philosophies for achieving intelligence. The first leverages \textbf{multistability and ordered states} to create robust logic and memory, analogous to digital computing but realized in physical materials \cite{jiao_mechanical_2023, levin_multiscale_2024}. The second pathway harnesses \textbf{dissipative adaptation}, where intelligent structure emerges as a consequence of a system self-organizing to effectively manage energy flows, a principle central to active matter and life itself \cite{england_dissipative_2015, hanczyc_chemical_2010}. The third and final strategy pursues \textbf{criticality and the "edge of chaos,"} positing that optimal, brain-like information processing occurs when a system is dynamically poised at a phase transition between order and disorder \cite{obyrne_how_2022, zhang_edge_2021}. The following sections detail these three foundational recipes for constructing cognitive matter.

\subsection{Strategy 1: Cognition via Multistability and Ordered States}
This strategy leverages an attractor landscape characterized by multiple deep, well-separated basins of attraction (see Fig. \ref{fig:attractor_landscape_fsm}). Intelligence manifests as the ability of the material to reliably switch between and maintain a set of robust, stable physical configurations. The computation is performed by the state transition itself, driven by a physical input. This approach excels at creating non-volatile memory and deterministic logic, and is the foundation of the \textbf{Self-Assembly Approach}, where the goal is to program local interactions to yield a specific, stable final structure. Evidence for this strategy is found in:

%%%%%%%%%%%%%%%%%% FIG 3 %%%%%%%%%%%%%%%%%%%
\begin{figure*}[htbp]
 \centering
\includegraphics[width=0.8\textwidth]{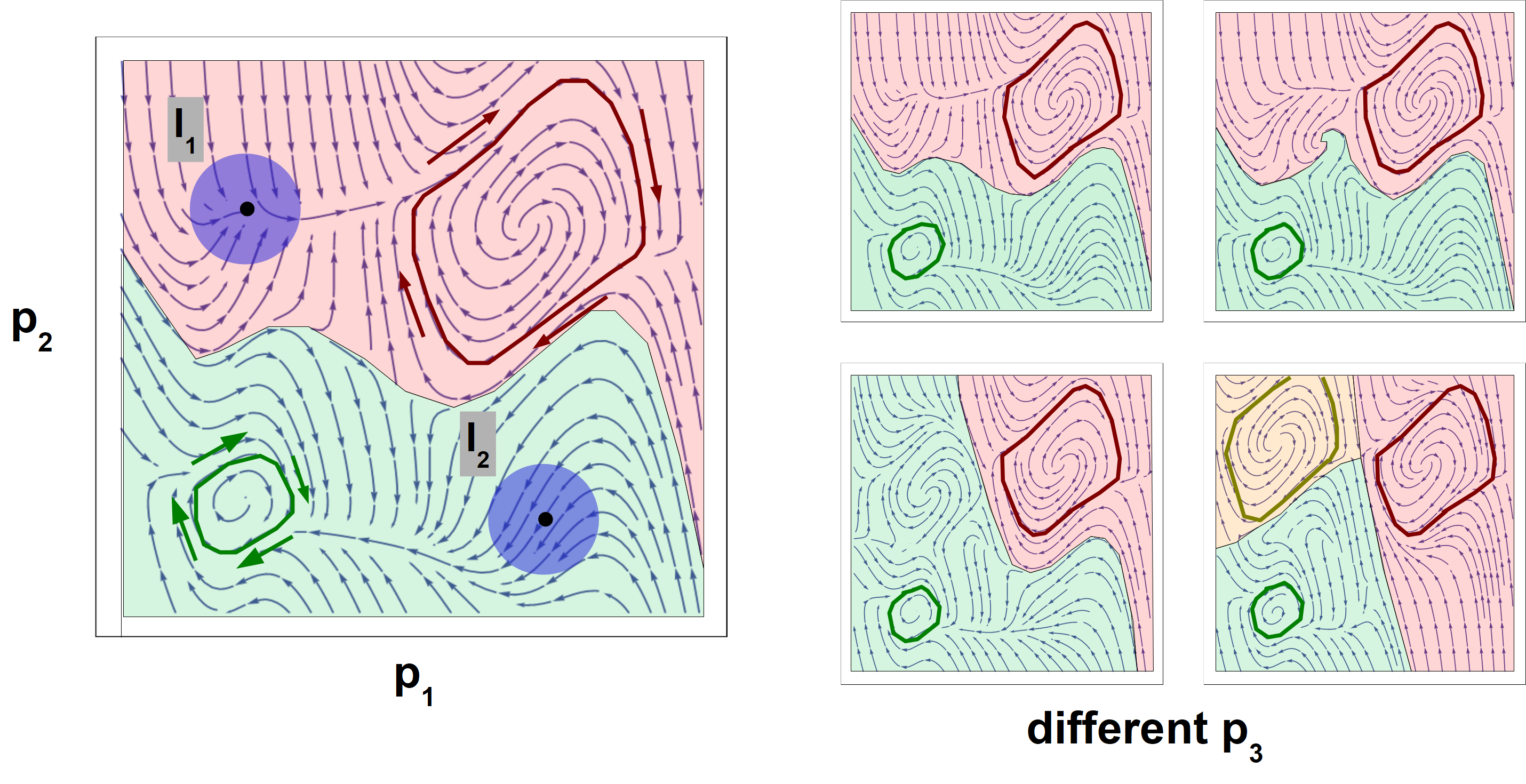} 
 \caption{\textbf{The Dynamical Systems Approach to Material-Based Intelligence.} (Left) Conceptual attractor landscape for a dynamic MBI system with two parameters p$_1$ and p$_2$. The system's state (representing, for example, gaits in a robot or reaction product concentrations) naturally evolves towards an attractor, for devices with a purposeful function a stable state or a limit cycle (\textit{e.g.}, I$_1$, I$_2$). External control input or environmental cues. Assuming a tuning parameter $\alpha$, one can reshape this landscape, altering the basins of attraction (dashed lines) and causing the system to transition between these functional modes (double-headed arrows). This mechanism underpins morphological control, where the detailed physical execution of a pattern is handled autonomously by the system's dynamics. (Right) A Finite State Machine (FSM) can be emulated by chaining multiple such dynamical systems. The final output state (characterised \textit{e.g.} by p$_1^f$ and p$_2^f$) of one dissipative dynamical system (DS$_1$), once settled into an attractor from its initial condition I$_1$, serves as the input (\textit{e.g.}, I$_2$) or as a parameter to tune the attractor landscape of the next dynamical system (DS$_2$). This physical propagation of information can extend to sequences or loops of systems, embodying a discrete computational flow through a continuous physical medium.}
\label{fig:attractor_landscape_fsm}
\end{figure*}
%%%%%%%%%%%%%%%%%%%%%%%%%%%%%%%%%%%%%%%%%%%%%%%

\textbf{Mechanical Metamaterials}, which use bistable unit cells as physical bits and logic gates to perform computation in response to mechanical forces \cite{jiao_mechanical_2023, jiao_mechanical_2023-1, louvet_reprogrammable_2024}.
 
\textbf{Biological Morphogenesis}, where tissues make collective decisions by settling into stable bioelectric patterns representing anatomical goals \cite{levin_multiscale_2024, levin_self-improvising_2024}.

\textbf{DNA Nanotechnology}, where sequence-specific interactions program the self-assembly of complex, stable 3D structures from simpler components \cite{ke_three-dimensional_2012, hadorn_specific_2012}.
 
\textbf{Memristive Systems}, where information is stored in stable high- or low-resistance states, forming the basis of neuromorphic associative memories \cite{kimura2021amorphous, di_ventra_parallel_2013}.

\subsection{Strategy 2: Cognition via Dissipative Adaptation and Self-Organization}
This approach posits that intelligent structure and behavior can emerge as a system self-organizes to more effectively absorb and dissipate energy from an environmental drive. Rooted in non-equilibrium thermodynamics, this paradigm suggests that systems operating far from equilibrium will naturally evolve towards states that are better "adapted" to their energy landscape \cite{england_dissipative_2015, goettems_physics_2024}. Here, the attractor landscape itself is formed and shaped by the dissipative process. This is the core principle behind the \textbf{Developmental Approach}, where living systems use energy to grow, differentiate, and maintain complex forms. Key examples include:

\textbf{Active Matter Systems}, such as self-propelled chemotactic droplets that navigate chemical gradients by continuously dissipating chemical energy to create and maintain Marangoni flows \cite{hanczyc_chemical_2010, boniface_self-propulsion_2019, cejkova_dynamics_2014, horibe_mode_2011}.

\textbf{The Cytoskeleton}, where constant ATP consumption drives the self-organization of functional structures like the mitotic spindle \cite{huber_emergent_2013}.
 
\textbf{Natural Induction}, a theoretical framework where physical networks with viscoelastic (dissipative) elements spontaneously adapt their structure to find exceptionally low-energy solutions to computational problems \cite{buckley_natural_2024}.

\subsection{Strategy 3: Cognition via Criticality and the "Edge of Chaos"}
A third major paradigm suggests that the most potent computational capabilities emerge when a system is poised at a critical point—a phase transition between an ordered state and a chaotic one. From a dynamical systems perspective, this "edge of chaos" is a regime where the attractor landscape becomes flat and complex, with a vast number of shallow, transient states. This is hypothesized to optimize the trade-off between stability (for memory) and flexibility (for computation), leading to maximal information capacity and sensitivity \cite{langton_computation_1990, feng_optimal_2020}. Evidence for this principle is found in:

\textbf{Brain Dynamics}, where cortical activity exhibits statistical signatures of criticality, such as power-law distributions of "neuronal avalanches," a state thought to be optimal for information processing \cite{obyrne_how_2022, srinivasa_criticality_2015}.

\textbf{Artificial Neural Networks}, where optimal performance often correlates with the network's dynamics being tuned to the edge of chaos \cite{zhang_edge_2021, feng_optimal_2020}.

\textbf{Nanowire Networks}, physical neuromorphic systems that exhibit emergent `1/f` noise and other scale-free dynamics characteristic of critical systems \cite{kuncic_emergent_2018}.

\textbf{Collective Behavior}, where swarms like starling flocks exhibit scale-free correlations in their movements, allowing for rapid, system-wide information propagation \cite{cavagna_scale-free_2010, puy_signatures_2024}.

\subsection{Synthesis: Hierarchical and Multi-Paradigm Architectures}
A single principle is unlikely to be sufficient for advanced cognition. The most sophisticated forms of intelligence appear to leverage all these strategies in a hierarchical and context-dependent manner. This aligns with the concept of a "Multiscale Competency Architecture," where different components of a system solve problems at their respective scales of organization, using the most appropriate physical principles \cite{levin_multiscale_2024, seifert_reinforcement_2024}.

A cognitive material might therefore be a hierarchical composite: a substrate of \textbf{multistable metamaterials} could provide robust, long-term memory; a surface layer of \textbf{neuromorphic nanowire networks}, self-tuned to \textbf{criticality}, could perform real-time sensory processing; and the entire system could be powered and maintained by an embedded network driving \textbf{dissipative self-organization}. The theoretical framework of Active Inference provides a mathematical language for describing how such a multi-paradigm system could function, unifying perception, action, and learning under the single imperative of minimizing prediction error (or free energy) \cite{friston_free-energy_2010, parr_active_2022, pezzulo_active_2024}. In this unified vision, the different physical pathways to cognition are not competing alternatives but complementary tools in the toolkit of intelligent matter.

\section{Experimental Approach for Characterizing Material-Based Intelligence}
\label{sec:experimental_approach}

Assessing the degree of MBI and distinguishing it from complex stimulus-responsiveness or computation that relies on an external software scaffold is a profound experimental challenge. Inspired by system identification in engineering and behavioral studies in cognitive science, we propose a systematic experimental methodology, operating within a carefully instrumented "MBI Testing Arena," capable of subjecting candidate MBI systems to diverse, complex, and potentially novel environmental conditions while quantitatively monitoring their inputs, accessible internal states, and physical outputs. This approach demands a focus on measurements that reveal how functionality emerges from correlations between system aspects, going beyond static measurements of independent ingredients.

A crucial step towards a systematic methodology for identifying and characterizing MBI will be the development of large-scale, curated databases specifically for intelligent materials, drawing inspiration from successful paradigms in materials informatics (\textit{\textit{e.g.}}, leveraging property prediction \cite{huang_increasingly_2022} or structure generation \cite{falk_learning_2023, obrien_machine_2024}). These databases must catalog not only constituent materials and macroscopic architectures, but also the intrinsic mechanisms at play, the behavior observed, underlying theoretical concepts, operational parameters, and quantitative performance on benchmark tasks. To facilitate AI-driven design and systematic discovery \cite{chen_metamaterials_2023, paixao_leveraging_2022, bordiga_automated_2024}, such a database needs to encompass "intelligence features" \textit{and their inter-correlations}. This includes quantifying how seemingly \textit{independent ingredients} contribute collectively to an \textit{emergent behavior}. Specific metrics of measurements should include:

\begin{enumerate}
 \item \textit{Information Integration and Complexity:} This involves quantifying the richness of information processing beyond simple one-to-one stimulus-response. Metrics could include integrated information measures \cite{allaire_information-theoretic_2012}, the complexity of sensorimotor loops as quantified via information flow between distributed parameters and actuation patterns \cite{thorsen_microfluidic_2002, rubenstein_programmable_2014}, or the \textit{entropy of accessible states} or behavioral outputs under varied and novel stimuli \cite{langton_computation_1990}. Correlation functions and topological invariants extracted from the material's dynamic response (\textit{e.g.}, in critical systems) can indicate latent computational complexity arising from local interactions \cite{obyrne_how_2022, aguilera_exploring_2018}.

 \item \textit{Memory Fidelity and Utility:} Beyond simply recording states, MBI implies active memory that influences future behavior. Measure retention time, capacity (number of distinguishable states, information-theoretic bits), and, most critically, the degree to which stored information actively and correctly biases future responses and decision-making over extended timescales. This requires observing system behavior under parameter-controlled conditions that specifically trigger memory recall (mechanism for memory readout) or modulate performance, for example by repeatedly driving the system through training patterns, which may necessitate mechanisms like 'replication' of specific information-bearing states \cite{kramar_encoding_2021}. This allows for measurement of the impact of learned \textit{very long} patterns.

 \item \textit{Adaptation and Learning Metrics:} Quantify the learning and adaptive behavior through metrics such as \textit{learning rates} (\textit{e.g.}, trials-to-criterion, speed of error reduction during optimization \cite{li_training_2024, stern_training_2024}), \textit{generalization} to novel stimuli (testing performance on inputs outside the training distribution), and the complexity of problems a material system can learn to solve (\textit{e.g.}, pattern classification with unseen examples \cite{yang_task_2019, beck_dynamic_2025}). The system's ability to self-optimize parameters based on performance correlations without direct human supervision is a key indicator.

 \item \textit{Autonomy and Goal-Directedness:} Characterize the goal directed task execution by assessing how well a system achieves a \textit{high-level goal} (defined via applications, \textit{e.g.}, reaching a target, maintaining homeostasis \cite{man_homeostasis_2019}) under varying and unpredictable conditions \cite{friston_path_2023, seifert_reinforcement_2024}. This includes evaluating the degree to which control is internal versus requiring external intervention to maintain function. Metrics of \textit{autonomous problem solving} (\textit{e.g.}, navigating an unmapped environment or re-solving a perturbed problem state \textit{independent} of prior ingredients used) are key here.

 \item \textit{Robustness and Self-Correction/Self-Maintenance:} Quantify the \textit{robustness} of behavior (\textit{e.g.}, persistence of function) under defined perturbations, damage, or component degradation \cite{wool_self-healing_2008, kriegman_scalable_2020}. Self-healing ability, recovery of functionality after damage, or dynamic recalibration to internal drift demonstrate intrinsic robustness \cite{kim_polymeric_2015}. Metrics here might include time-to-recovery or performance degradation curve upon specific stressors, focusing on how different system ingredients respond.

 \item \textit{Exploring Minimal Models and Mechanisms:} Future efforts must systematically \textit{probe for minimality}. This involves characterizing systems where complex functionalities emerge from a minimal set of underlying physical principles or interacting components \cite{jiao_mechanical_2023}. The 'recipes' for MBI must identify the smallest set of conditions necessary to achieve emergent intelligence, exploring tradeoffs in \textit{performance vs. minimality}. This is a hard-earned measure through systematic variation and careful design (similar to what can be observed in molecular self-assembly processes, for instance when creating new properties \cite{sharma_assembly_2023}).
\end{enumerate}

Machine learning models trained on these comprehensive databases, potentially utilising graph neural network architectures that can directly process the rich graph structure \cite{baulin_discovery_2025}, could then be employed to predict emergent MBI characteristics from material composition and structure, identify promising new material combinations, or suggest novel mechanisms integration based on observed correlations between physical parameters and high-level behaviour. This feedback loop, where theoretical insights guide experimental design and vice versa, facilitates a dynamic cycle of data-driven discovery and hypothesis testing, moving beyond serendipitous findings towards the rational design and synthesis of intelligent matter \cite{howard_evolving_2019, paixao_leveraging_2022, chen_metamaterials_2023, obrien_machine_2024}. The key in evaluating such emerging intelligence is precisely outlining the methods for obtaining and interpreting measurements from experiments or simulations where systems show complex interactions independent of their specific ingredients.

\section{Conclusion}
\label{sec:conclusion}

The development of material-based intelligence demands a profound conceptual and practical shift away from designing materials with isolated, externally controlled functionalities, and towards creating integrated systems where sensing, active memory, embodied computation, self-organization, self-correction (combined with adaptive feedback), and robust adaptive feedback loops emerge synergistically from the materials intrinsic physics and local interactions \cite{kaspar_rise_2021, sitti_physical_2021, mcevoy_materials_2015}. This paradigm inherently embraces the principles of \textit{embodied cognition} and non-equilibrium thermodynamics, viewing the material not as a passive substrate upon which intelligence is imposed, but as an \textit{active, dynamic computational and adaptive medium} where its purpose and \textit{high-level goals} emerge directly from its self-organization. This profound integration dissolves the rigid boundaries of traditional hardware-software divisions \cite{harrison_mind_2022, hughes_embodied_2022}, emphasizing that the 'program' is literally inscribed within and executed by the `physics itself`, making no fundamental distinction between output and process.

The core requirements we have outlined highlight the sheer complexity of this grand challenge: the need for \textit{local interaction rules} that scale into \textit{long-distance coordination}, \textit{dynamically coupled memory} (potentially requiring forms of \textit{replication} for \textit{long-term} persistence) that actively shapes future responses; \textit{computation realized fundamentally through physical dynamics} (ideally with minimal ingredients but emergent complexity); and inherent capacities for self-organization, self-correction, and pervasive adaptive feedback loops. While current technologies have demonstrated individual components—from sophisticated memristive networks capable of rudimentary memory and computation \cite{yao_fully_2020, yang_memristor_2022}, to highly responsive polymers used for actuation \cite{xiao_artificial_2020, xia_dynamic_2022}, and innovative self-healing materials that restore physical integrity \cite{terryn_review_2021}, the deep, autonomous integration of these features within an intelligent material system remains the primary scientific and engineering frontier.

Future progress in material-based intelligence hinges critically on two interconnected pillars. First, the development of new theoretical frameworks rooted in physics, information theory, and complex systems science that can provide a coherent 'recipe' for generating emergent order. Second, the creation of innovative methods for material synthesis, sophisticated automatic experimental platforms, and AI-driven analysis capable of fostering and characterizing these elusive emergent intelligent behaviors \cite{kriegman_scalable_2020, howard_evolving_2019}. This involves devising comprehensive measurements to quantify intrinsic properties beyond mere performance data; developing metrics that explicitly probe correlations between system variables; and assessing the system's \textit{robustness} to external noise, its \textit{adaptability} to novel conditions, and its degree of \textit{autonomy} in problem-solving scenarios \cite{muller_what_2017-1}. Moreover, systematically exploring minimal physical models with core interaction rules will help identify fundamental ingredients from which complexity can spontaneously arise independent of the specific nature of those ingredients, leading to the realization of new kinds of intelligence \cite{jiao_mechanical_2023}. This ambitious research direction promises not only transformative technologies capable of unprecedented \textit{autonomy} and \textit{robustness} for applications in fields like soft robotics \cite{kriegman_scalable_2020, gumuskaya_motile_2024}, medicine, and beyond—but also profound fundamental insights into the physical basis of cognition and the very nature of life itself \cite{levin_multiscale_2024, mcmillen_collective_2024, levin_self-improvising_2024}, propelling us beyond simple biomimicry towards potentially entirely novel forms of non-biological intelligence \cite{rouleau_multiple_2023}. Finally, we point out the role of evolutionary engineering. MBI will not so much rely on abstract planning or coding, but include evolutionary processes. This requires a meta-design in the sense the MBI not only needs to be able to perform complex tasks, but also that the parameter setting enabling these capabilities can be efficiently evolved.  

\begin{acknowledgments}
This publication has been developed thanks to the collaboration/networking within the SoftComp Consortium (\url{https://eu-softcomp.net}). The authors thank the participants to the Workshop " Intelligent Soft Matter" Salou 2024 (\url{https://softmat.net/
intelligent-soft-matter/}). RF gratefully acknowledges support by the EU project HORIZON-EIC Bio-HhOST (Next Generation 3D Tissue Models: Bio-Hybrid Hierarchical Organoid-Synthetic Tissues (Bio-HhOST)
Comprised of Live and Artificial Cells), project number: 101130747.
\end{acknowledgments}

%\bibliography{AI-4,intelliDE,mw_tmp}

\begin{thebibliography}{160}%
\makeatletter
\providecommand \@ifxundefined [1]{%
 \@ifx{#1\undefined}
}%
\providecommand \@ifnum [1]{%
 \ifnum #1\expandafter \@firstoftwo
 \else \expandafter \@secondoftwo
 \fi
}%
\providecommand \@ifx [1]{%
 \ifx #1\expandafter \@firstoftwo
 \else \expandafter \@secondoftwo
 \fi
}%
\providecommand \natexlab [1]{#1}%
\providecommand \enquote  [1]{``#1''}%
\providecommand \bibnamefont  [1]{#1}%
\providecommand \bibfnamefont [1]{#1}%
\providecommand \citenamefont [1]{#1}%
\providecommand \href@noop [0]{\@secondoftwo}%
\providecommand \href [0]{\begingroup \@sanitize@url \@href}%
\providecommand \@href[1]{\@@startlink{#1}\@@href}%
\providecommand \@@href[1]{\endgroup#1\@@endlink}%
\providecommand \@sanitize@url [0]{\catcode `\\12\catcode `\$12\catcode
  `\&12\catcode `\#12\catcode `\^12\catcode `\_12\catcode `\%12\relax}%
\providecommand \@@startlink[1]{}%
\providecommand \@@endlink[0]{}%
\providecommand \url  [0]{\begingroup\@sanitize@url \@url }%
\providecommand \@url [1]{\endgroup\@href {#1}{\urlprefix }}%
\providecommand \urlprefix  [0]{URL }%
\providecommand \Eprint [0]{\href }%
\providecommand \doibase [0]{https://doi.org/}%
\providecommand \selectlanguage [0]{\@gobble}%
\providecommand \bibinfo  [0]{\@secondoftwo}%
\providecommand \bibfield  [0]{\@secondoftwo}%
\providecommand \translation [1]{[#1]}%
\providecommand \BibitemOpen [0]{}%
\providecommand \bibitemStop [0]{}%
\providecommand \bibitemNoStop [0]{.\EOS\space}%
\providecommand \EOS [0]{\spacefactor3000\relax}%
\providecommand \BibitemShut  [1]{\csname bibitem#1\endcsname}%
\let\auto@bib@innerbib\@empty
%</preamble>
\bibitem [{\citenamefont {Bryant}\ and\ \citenamefont
  {Machta}(2023)}]{bryant_physical_2023}%
  \BibitemOpen
  \bibfield  {author} {\bibinfo {author} {\bibfnamefont {S.~J.}\ \bibnamefont
  {Bryant}}\ and\ \bibinfo {author} {\bibfnamefont {B.~B.}\ \bibnamefont
  {Machta}},\ }\bibfield  {title} {\bibinfo {title} {Physical {Constraints} in
  {Intracellular} {Signaling}: {The} {Cost} of {Sending} a {Bit}},\ }\href
  {https://doi.org/10.1103/PhysRevLett.131.068401} {\bibfield  {journal}
  {\bibinfo  {journal} {Phys. Rev. Lett.}\ }\textbf {\bibinfo {volume} {131}},\
  \bibinfo {pages} {068401} (\bibinfo {year} {2023})},\ \bibinfo {note}
  {publisher: American Physical Society}\BibitemShut {NoStop}%
\bibitem [{\citenamefont {Wool}(2008)}]{wool_self-healing_2008}%
  \BibitemOpen
  \bibfield  {author} {\bibinfo {author} {\bibfnamefont {R.~P.}\ \bibnamefont
  {Wool}},\ }\bibfield  {title} {{\selectlanguage {English}\bibinfo {title}
  {Self-healing materials: a review}},\ }\href
  {https://doi.org/10.1039/B711716G} {\bibfield  {journal} {\bibinfo  {journal}
  {Soft Matter}\ }\textbf {\bibinfo {volume} {4}},\ \bibinfo {pages} {400}
  (\bibinfo {year} {2008})},\ \bibinfo {note} {publisher: The Royal Society of
  Chemistry}\BibitemShut {NoStop}%
\bibitem [{\citenamefont {Kaspar}\ \emph {et~al.}(2021)\citenamefont {Kaspar},
  \citenamefont {Ravoo}, \citenamefont {Van Der~Wiel}, \citenamefont {Wegner},\
  and\ \citenamefont {Pernice}}]{kaspar_rise_2021}%
  \BibitemOpen
  \bibfield  {author} {\bibinfo {author} {\bibfnamefont {C.}~\bibnamefont
  {Kaspar}}, \bibinfo {author} {\bibfnamefont {B.~J.}\ \bibnamefont {Ravoo}},
  \bibinfo {author} {\bibfnamefont {W.~G.}\ \bibnamefont {Van Der~Wiel}},
  \bibinfo {author} {\bibfnamefont {S.~V.}\ \bibnamefont {Wegner}},\ and\
  \bibinfo {author} {\bibfnamefont {W.~H.~P.}\ \bibnamefont {Pernice}},\
  }\bibfield  {title} {{\selectlanguage {English}\bibinfo {title} {The rise of
  intelligent matter}},\ }\href {https://doi.org/10.1038/s41586-021-03453-y}
  {\bibfield  {journal} {\bibinfo  {journal} {Nature}\ }\textbf {\bibinfo
  {volume} {594}},\ \bibinfo {pages} {345} (\bibinfo {year}
  {2021})}\BibitemShut {NoStop}%
\bibitem [{\citenamefont {McMillen}\ and\ \citenamefont
  {Levin}(2024)}]{mcmillen_collective_2024}%
  \BibitemOpen
  \bibfield  {author} {\bibinfo {author} {\bibfnamefont {P.}~\bibnamefont
  {McMillen}}\ and\ \bibinfo {author} {\bibfnamefont {M.}~\bibnamefont
  {Levin}},\ }\bibfield  {title} {{\selectlanguage {English}\bibinfo {title}
  {Collective intelligence: {A} unifying concept for integrating biology across
  scales and substrates}},\ }\href {https://doi.org/10.1038/s42003-024-06037-4}
  {\bibfield  {journal} {\bibinfo  {journal} {Commun Biol}\ }\textbf {\bibinfo
  {volume} {7}},\ \bibinfo {pages} {1} (\bibinfo {year} {2024})},\ \bibinfo
  {note} {publisher: Nature Publishing Group}\BibitemShut {NoStop}%
\bibitem [{\citenamefont {Wegst}\ \emph {et~al.}(2015)\citenamefont {Wegst},
  \citenamefont {Bai}, \citenamefont {Saiz}, \citenamefont {Tomsia},\ and\
  \citenamefont {Ritchie}}]{wegst_bioinspired_2015}%
  \BibitemOpen
  \bibfield  {author} {\bibinfo {author} {\bibfnamefont {U.~G.~K.}\
  \bibnamefont {Wegst}}, \bibinfo {author} {\bibfnamefont {H.}~\bibnamefont
  {Bai}}, \bibinfo {author} {\bibfnamefont {E.}~\bibnamefont {Saiz}}, \bibinfo
  {author} {\bibfnamefont {A.~P.}\ \bibnamefont {Tomsia}},\ and\ \bibinfo
  {author} {\bibfnamefont {R.~O.}\ \bibnamefont {Ritchie}},\ }\bibfield
  {title} {{\selectlanguage {English}\bibinfo {title} {Bioinspired structural
  materials}},\ }\href {https://doi.org/10.1038/nmat4089} {\bibfield  {journal}
  {\bibinfo  {journal} {Nature Mater}\ }\textbf {\bibinfo {volume} {14}},\
  \bibinfo {pages} {23} (\bibinfo {year} {2015})},\ \bibinfo {note} {publisher:
  Nature Publishing Group}\BibitemShut {NoStop}%
\bibitem [{\citenamefont {Levin}(2024{\natexlab{a}})}]{levin_multiscale_2024}%
  \BibitemOpen
  \bibfield  {author} {\bibinfo {author} {\bibfnamefont {M.}~\bibnamefont
  {Levin}},\ }\bibfield  {title} {{\selectlanguage {English}\bibinfo {title}
  {The {Multiscale} {Wisdom} of the {Body}: {Collective} {Intelligence} as a
  {Tractable} {Interface} for {Next}-{Generation} {Biomedicine}}},\ }\href
  {https://doi.org/10.1002/bies.202400196} {\bibfield  {journal} {\bibinfo
  {journal} {BioEssays}\ }\textbf {\bibinfo {volume} {47}},\ \bibinfo {pages}
  {e202400196} (\bibinfo {year} {2024}{\natexlab{a}})},\ \bibinfo {note}
  {\_eprint:
  https://onlinelibrary.wiley.com/doi/pdf/10.1002/bies.202400196}\BibitemShut
  {NoStop}%
\bibitem [{\citenamefont {Pfeifer}\ \emph {et~al.}(2007)\citenamefont
  {Pfeifer}, \citenamefont {Lungarella},\ and\ \citenamefont
  {Iida}}]{pfeifer_self-organization_2007}%
  \BibitemOpen
  \bibfield  {author} {\bibinfo {author} {\bibfnamefont {R.}~\bibnamefont
  {Pfeifer}}, \bibinfo {author} {\bibfnamefont {M.}~\bibnamefont
  {Lungarella}},\ and\ \bibinfo {author} {\bibfnamefont {F.}~\bibnamefont
  {Iida}},\ }\bibfield  {title} {\bibinfo {title} {Self-{Organization},
  {Embodiment}, and {Biologically} {Inspired} {Robotics}},\ }\href
  {https://doi.org/10.1126/science.1145803} {\bibfield  {journal} {\bibinfo
  {journal} {Science}\ }\textbf {\bibinfo {volume} {318}},\ \bibinfo {pages}
  {1088} (\bibinfo {year} {2007})},\ \bibinfo {note} {publisher: American
  Association for the Advancement of Science}\BibitemShut {NoStop}%
\bibitem [{\citenamefont {Yao}\ \emph {et~al.}(2020)\citenamefont {Yao},
  \citenamefont {Wu}, \citenamefont {Gao}, \citenamefont {Tang}, \citenamefont
  {Zhang}, \citenamefont {Zhang}, \citenamefont {Yang},\ and\ \citenamefont
  {Qian}}]{yao_fully_2020}%
  \BibitemOpen
  \bibfield  {author} {\bibinfo {author} {\bibfnamefont {P.}~\bibnamefont
  {Yao}}, \bibinfo {author} {\bibfnamefont {H.}~\bibnamefont {Wu}}, \bibinfo
  {author} {\bibfnamefont {B.}~\bibnamefont {Gao}}, \bibinfo {author}
  {\bibfnamefont {J.}~\bibnamefont {Tang}}, \bibinfo {author} {\bibfnamefont
  {Q.}~\bibnamefont {Zhang}}, \bibinfo {author} {\bibfnamefont
  {W.}~\bibnamefont {Zhang}}, \bibinfo {author} {\bibfnamefont {J.~J.}\
  \bibnamefont {Yang}},\ and\ \bibinfo {author} {\bibfnamefont
  {H.}~\bibnamefont {Qian}},\ }\bibfield  {title} {{\selectlanguage
  {English}\bibinfo {title} {Fully hardware-implemented memristor convolutional
  neural network}},\ }\href {https://doi.org/10.1038/s41586-020-1942-4}
  {\bibfield  {journal} {\bibinfo  {journal} {Nature}\ }\textbf {\bibinfo
  {volume} {577}},\ \bibinfo {pages} {641} (\bibinfo {year} {2020})},\ \bibinfo
  {note} {publisher: Nature Publishing Group}\BibitemShut {NoStop}%
\bibitem [{\citenamefont {Yang}\ \emph
  {et~al.}(2022{\natexlab{a}})\citenamefont {Yang}, \citenamefont {Wang},
  \citenamefont {Hao}, \citenamefont {Li}, \citenamefont {Wei}, \citenamefont
  {Deng},\ and\ \citenamefont {Loparo}}]{yang_bicoss_2022}%
  \BibitemOpen
  \bibfield  {author} {\bibinfo {author} {\bibfnamefont {S.}~\bibnamefont
  {Yang}}, \bibinfo {author} {\bibfnamefont {J.}~\bibnamefont {Wang}}, \bibinfo
  {author} {\bibfnamefont {X.}~\bibnamefont {Hao}}, \bibinfo {author}
  {\bibfnamefont {H.}~\bibnamefont {Li}}, \bibinfo {author} {\bibfnamefont
  {X.}~\bibnamefont {Wei}}, \bibinfo {author} {\bibfnamefont {B.}~\bibnamefont
  {Deng}},\ and\ \bibinfo {author} {\bibfnamefont {K.~A.}\ \bibnamefont
  {Loparo}},\ }\bibfield  {title} {\bibinfo {title} {{BiCoSS}: {Toward}
  {Large}-{Scale} {Cognition} {Brain} {With} {Multigranular} {Neuromorphic}
  {Architecture}},\ }\href {https://doi.org/10.1109/TNNLS.2020.3045492}
  {\bibfield  {journal} {\bibinfo  {journal} {IEEE Trans. Neural Netw. Learning
  Syst.}\ }\textbf {\bibinfo {volume} {33}},\ \bibinfo {pages} {2801} (\bibinfo
  {year} {2022}{\natexlab{a}})}\BibitemShut {NoStop}%
\bibitem [{\citenamefont {Mead}(1990)}]{mead_neuromorphic_1990}%
  \BibitemOpen
  \bibfield  {author} {\bibinfo {author} {\bibfnamefont {C.}~\bibnamefont
  {Mead}},\ }\bibfield  {title} {\bibinfo {title} {Neuromorphic electronic
  systems},\ }\href {https://doi.org/10.1109/5.58356} {\bibfield  {journal}
  {\bibinfo  {journal} {Proceedings of the IEEE}\ }\textbf {\bibinfo {volume}
  {78}},\ \bibinfo {pages} {1629} (\bibinfo {year} {1990})},\ \bibinfo {note}
  {conference Name: Proceedings of the IEEE}\BibitemShut {NoStop}%
\bibitem [{\citenamefont {von Neumann}(2012)}]{vonNeumann2012computer}%
  \BibitemOpen
  \bibfield  {author} {\bibinfo {author} {\bibfnamefont {J.}~\bibnamefont {von
  Neumann}},\ }\href@noop {} {\emph {\bibinfo {title} {The Computer and the
  Brain}}},\ \bibinfo {edition} {3rd}\ ed.\ (\bibinfo  {publisher} {Yale
  University Press},\ \bibinfo {year} {2012})\ \bibinfo {note} {foreword by Ray
  Kurzweil}\BibitemShut {NoStop}%
\bibitem [{\citenamefont {Di~Ventra}\ and\ \citenamefont
  {Pershin}(2013)}]{di_ventra_parallel_2013}%
  \BibitemOpen
  \bibfield  {author} {\bibinfo {author} {\bibfnamefont {M.}~\bibnamefont
  {Di~Ventra}}\ and\ \bibinfo {author} {\bibfnamefont {Y.~V.}\ \bibnamefont
  {Pershin}},\ }\bibfield  {title} {{\selectlanguage {English}\bibinfo {title}
  {The parallel approach}},\ }\href {https://doi.org/10.1038/nphys2566}
  {\bibfield  {journal} {\bibinfo  {journal} {Nature Phys}\ }\textbf {\bibinfo
  {volume} {9}},\ \bibinfo {pages} {200} (\bibinfo {year} {2013})}\BibitemShut
  {NoStop}%
\bibitem [{\citenamefont {Kuncic}\ \emph {et~al.}(2018)\citenamefont {Kuncic},
  \citenamefont {Marcus}, \citenamefont {Sanz-Leon}, \citenamefont {Higuchi},
  \citenamefont {Shingaya}, \citenamefont {Li}, \citenamefont {Stieg},
  \citenamefont {Gimzewski}, \citenamefont {Aono},\ and\ \citenamefont
  {Nakayama}}]{kuncic_emergent_2018}%
  \BibitemOpen
  \bibfield  {author} {\bibinfo {author} {\bibfnamefont {Z.}~\bibnamefont
  {Kuncic}}, \bibinfo {author} {\bibfnamefont {I.}~\bibnamefont {Marcus}},
  \bibinfo {author} {\bibfnamefont {P.}~\bibnamefont {Sanz-Leon}}, \bibinfo
  {author} {\bibfnamefont {R.}~\bibnamefont {Higuchi}}, \bibinfo {author}
  {\bibfnamefont {Y.}~\bibnamefont {Shingaya}}, \bibinfo {author}
  {\bibfnamefont {M.}~\bibnamefont {Li}}, \bibinfo {author} {\bibfnamefont
  {A.}~\bibnamefont {Stieg}}, \bibinfo {author} {\bibfnamefont
  {J.}~\bibnamefont {Gimzewski}}, \bibinfo {author} {\bibfnamefont
  {M.}~\bibnamefont {Aono}},\ and\ \bibinfo {author} {\bibfnamefont
  {T.}~\bibnamefont {Nakayama}},\ }\bibfield  {title} {\bibinfo {title}
  {Emergent brain-like complexity from nanowire atomic switch networks:
  {Towards} neuromorphic synthetic intelligence},\ }in\ \href
  {https://doi.org/10.1109/NANO.2018.8626236} {\emph {\bibinfo {booktitle}
  {2018 {IEEE} 18th {International} {Conference} on {Nanotechnology}
  ({IEEE}-{NANO})}}}\ (\bibinfo {year} {2018})\ pp.\ \bibinfo {pages} {1--3},\
  \bibinfo {note} {iSSN: 1944-9380}\BibitemShut {NoStop}%
\bibitem [{\citenamefont {Kindig}(2024)}]{kindig_ai_2024}%
  \BibitemOpen
  \bibfield  {author} {\bibinfo {author} {\bibfnamefont {B.}~\bibnamefont
  {Kindig}},\ }\href
  {https://www.forbes.com/sites/bethkindig/2024/06/20/ai-power-consumption-rapidly-becoming-mission-critical/}
  {{\selectlanguage {English}\bibinfo {title} {{AI} {Power} {Consumption}:
  {Rapidly} {Becoming} {Mission}-{Critical}}}} (\bibinfo {year} {2024}),\
  \bibinfo {note} {section: Consumer Tech}\BibitemShut {NoStop}%
\bibitem [{\citenamefont {Harrison}\ \emph {et~al.}(2022)\citenamefont
  {Harrison}, \citenamefont {Rorot},\ and\ \citenamefont
  {Laukaityte}}]{harrison_mind_2022}%
  \BibitemOpen
  \bibfield  {author} {\bibinfo {author} {\bibfnamefont {D.}~\bibnamefont
  {Harrison}}, \bibinfo {author} {\bibfnamefont {W.}~\bibnamefont {Rorot}},\
  and\ \bibinfo {author} {\bibfnamefont {U.}~\bibnamefont {Laukaityte}},\
  }\bibfield  {title} {{\selectlanguage {English}\bibinfo {title} {Mind the
  matter: {Active} matter, soft robotics, and the making of bio-inspired
  artificial intelligence}},\ }\bibfield  {journal} {\bibinfo  {journal}
  {Front. Neurorobot.}\ }\textbf {\bibinfo {volume} {16}},\ \href
  {https://doi.org/10.3389/fnbot.2022.880724} {10.3389/fnbot.2022.880724}
  (\bibinfo {year} {2022}),\ \bibinfo {note} {publisher: Frontiers}\BibitemShut
  {NoStop}%
\bibitem [{\citenamefont {McEvoy}\ and\ \citenamefont
  {Correll}(2015)}]{mcevoy_materials_2015}%
  \BibitemOpen
  \bibfield  {author} {\bibinfo {author} {\bibfnamefont {M.~A.}\ \bibnamefont
  {McEvoy}}\ and\ \bibinfo {author} {\bibfnamefont {N.}~\bibnamefont
  {Correll}},\ }\bibfield  {title} {\bibinfo {title} {Materials that couple
  sensing, actuation, computation, and communication},\ }\href
  {https://doi.org/10.1126/science.1261689} {\bibfield  {journal} {\bibinfo
  {journal} {Science}\ }\textbf {\bibinfo {volume} {347}},\ \bibinfo {pages}
  {1261689} (\bibinfo {year} {2015})},\ \bibinfo {note} {publisher: American
  Association for the Advancement of Science}\BibitemShut {NoStop}%
\bibitem [{\citenamefont {Miłkowski}(2018)}]{milkowski_morphological_2018}%
  \BibitemOpen
  \bibfield  {author} {\bibinfo {author} {\bibfnamefont {M.}~\bibnamefont
  {Miłkowski}},\ }\bibfield  {title} {{\selectlanguage {English}\bibinfo
  {title} {Morphological {Computation}: {Nothing} but {Physical}
  {Computation}}},\ }\href {https://doi.org/10.3390/e20120942} {\bibfield
  {journal} {\bibinfo  {journal} {Entropy}\ }\textbf {\bibinfo {volume} {20}},\
  \bibinfo {pages} {942} (\bibinfo {year} {2018})},\ \bibinfo {note} {number:
  12 Publisher: Multidisciplinary Digital Publishing Institute}\BibitemShut
  {NoStop}%
\bibitem [{\citenamefont {Pfeifer}\ \emph {et~al.}(2006)\citenamefont
  {Pfeifer}, \citenamefont {Iida},\ and\ \citenamefont
  {Gómez}}]{pfeifer_morphological_2006}%
  \BibitemOpen
  \bibfield  {author} {\bibinfo {author} {\bibfnamefont {R.}~\bibnamefont
  {Pfeifer}}, \bibinfo {author} {\bibfnamefont {F.}~\bibnamefont {Iida}},\ and\
  \bibinfo {author} {\bibfnamefont {G.}~\bibnamefont {Gómez}},\ }\bibfield
  {title} {\bibinfo {title} {Morphological computation for adaptive behavior
  and cognition},\ }\href {https://doi.org/10.1016/j.ics.2005.12.080}
  {\bibfield  {journal} {\bibinfo  {journal} {International Congress Series}\
  }\bibinfo {series} {Brain-{Inspired} {IT} {II}: {Decision} and {Behavioral}
  {Choice} {Organized} by {Natural} and {Artificial} {Brains}. {Invited} and
  selected papers of the 2nd {International} {Conference} on {Brain}-inspired
  {Information} {Technology} held in {Hibikino}, {Kitakyushu}, {Japan} between
  7 and 9 {October} 2005},\ \textbf {\bibinfo {volume} {1291}},\ \bibinfo
  {pages} {22} (\bibinfo {year} {2006})}\BibitemShut {NoStop}%
\bibitem [{\citenamefont {Hauser}\ \emph
  {et~al.}(2011{\natexlab{a}})\citenamefont {Hauser}, \citenamefont {Ijspeert},
  \citenamefont {Füchslin}, \citenamefont {Pfeifer},\ and\ \citenamefont
  {Maass}}]{hauser_towards_2011}%
  \BibitemOpen
  \bibfield  {author} {\bibinfo {author} {\bibfnamefont {H.}~\bibnamefont
  {Hauser}}, \bibinfo {author} {\bibfnamefont {A.~J.}\ \bibnamefont
  {Ijspeert}}, \bibinfo {author} {\bibfnamefont {R.~M.}\ \bibnamefont
  {Füchslin}}, \bibinfo {author} {\bibfnamefont {R.}~\bibnamefont {Pfeifer}},\
  and\ \bibinfo {author} {\bibfnamefont {W.}~\bibnamefont {Maass}},\ }\bibfield
   {title} {{\selectlanguage {English}\bibinfo {title} {Towards a theoretical
  foundation for morphological computation with compliant bodies}},\ }\href
  {https://doi.org/10.1007/s00422-012-0471-0} {\bibfield  {journal} {\bibinfo
  {journal} {Biol Cybern}\ }\textbf {\bibinfo {volume} {105}},\ \bibinfo
  {pages} {355} (\bibinfo {year} {2011}{\natexlab{a}})}\BibitemShut {NoStop}%
\bibitem [{\citenamefont {Hauser}\ \emph
  {et~al.}(2011{\natexlab{b}})\citenamefont {Hauser}, \citenamefont {Ijspeert},
  \citenamefont {Füchslin}, \citenamefont {Pfeifer},\ and\ \citenamefont
  {Maass}}]{hauser_towards_2011-1}%
  \BibitemOpen
  \bibfield  {author} {\bibinfo {author} {\bibfnamefont {H.}~\bibnamefont
  {Hauser}}, \bibinfo {author} {\bibfnamefont {A.~J.}\ \bibnamefont
  {Ijspeert}}, \bibinfo {author} {\bibfnamefont {R.~M.}\ \bibnamefont
  {Füchslin}}, \bibinfo {author} {\bibfnamefont {R.}~\bibnamefont {Pfeifer}},\
  and\ \bibinfo {author} {\bibfnamefont {W.}~\bibnamefont {Maass}},\ }\bibfield
   {title} {{\selectlanguage {English}\bibinfo {title} {Towards a theoretical
  foundation for morphological computation with compliant bodies}},\ }\href
  {https://doi.org/10.1007/s00422-012-0471-0} {\bibfield  {journal} {\bibinfo
  {journal} {Biol Cybern}\ }\textbf {\bibinfo {volume} {105}},\ \bibinfo
  {pages} {355} (\bibinfo {year} {2011}{\natexlab{b}})}\BibitemShut {NoStop}%
\bibitem [{\citenamefont {F{\"u}chslin}\ \emph {et~al.}(2013)\citenamefont
  {F{\"u}chslin}, \citenamefont {Dzyakanchuk}, \citenamefont {Flumini},
  \citenamefont {Hauser}, \citenamefont {Hunt}, \citenamefont {Luchsinger},
  \citenamefont {Reller}, \citenamefont {Scheidegger},\ and\ \citenamefont
  {Walker}}]{fuchslin2013morphological}%
  \BibitemOpen
  \bibfield  {author} {\bibinfo {author} {\bibfnamefont {R.~M.}\ \bibnamefont
  {F{\"u}chslin}}, \bibinfo {author} {\bibfnamefont {A.}~\bibnamefont
  {Dzyakanchuk}}, \bibinfo {author} {\bibfnamefont {D.}~\bibnamefont
  {Flumini}}, \bibinfo {author} {\bibfnamefont {H.}~\bibnamefont {Hauser}},
  \bibinfo {author} {\bibfnamefont {K.~J.}\ \bibnamefont {Hunt}}, \bibinfo
  {author} {\bibfnamefont {R.~H.}\ \bibnamefont {Luchsinger}}, \bibinfo
  {author} {\bibfnamefont {B.}~\bibnamefont {Reller}}, \bibinfo {author}
  {\bibfnamefont {S.}~\bibnamefont {Scheidegger}},\ and\ \bibinfo {author}
  {\bibfnamefont {R.}~\bibnamefont {Walker}},\ }\bibfield  {title} {\bibinfo
  {title} {Morphological computation and morphological control: steps toward a
  formal theory and applications},\ }\href@noop {} {\bibfield  {journal}
  {\bibinfo  {journal} {Artificial life}\ }\textbf {\bibinfo {volume} {19}},\
  \bibinfo {pages} {9} (\bibinfo {year} {2013})}\BibitemShut {NoStop}%
\bibitem [{\citenamefont {Mengaldo}\ \emph {et~al.}(2022)\citenamefont
  {Mengaldo}, \citenamefont {Renda}, \citenamefont {Brunton}, \citenamefont
  {Bächer}, \citenamefont {Calisti}, \citenamefont {Duriez}, \citenamefont
  {Chirikjian},\ and\ \citenamefont {Laschi}}]{mengaldo_concise_2022}%
  \BibitemOpen
  \bibfield  {author} {\bibinfo {author} {\bibfnamefont {G.}~\bibnamefont
  {Mengaldo}}, \bibinfo {author} {\bibfnamefont {F.}~\bibnamefont {Renda}},
  \bibinfo {author} {\bibfnamefont {S.~L.}\ \bibnamefont {Brunton}}, \bibinfo
  {author} {\bibfnamefont {M.}~\bibnamefont {Bächer}}, \bibinfo {author}
  {\bibfnamefont {M.}~\bibnamefont {Calisti}}, \bibinfo {author} {\bibfnamefont
  {C.}~\bibnamefont {Duriez}}, \bibinfo {author} {\bibfnamefont {G.~S.}\
  \bibnamefont {Chirikjian}},\ and\ \bibinfo {author} {\bibfnamefont
  {C.}~\bibnamefont {Laschi}},\ }\bibfield  {title} {{\selectlanguage
  {English}\bibinfo {title} {A concise guide to modelling the physics of
  embodied intelligence in soft robotics}},\ }\href
  {https://doi.org/10.1038/s42254-022-00481-z} {\bibfield  {journal} {\bibinfo
  {journal} {Nat Rev Phys}\ }\textbf {\bibinfo {volume} {4}},\ \bibinfo {pages}
  {595} (\bibinfo {year} {2022})},\ \bibinfo {note} {publisher: Nature
  Publishing Group}\BibitemShut {NoStop}%
\bibitem [{\citenamefont {Zhao}\ \emph {et~al.}(2024)\citenamefont {Zhao},
  \citenamefont {Wu}, \citenamefont {Wang}, \citenamefont {Zhang},
  \citenamefont {Zhong},\ and\ \citenamefont
  {Zhilenkov}}]{zhao_exploring_2024}%
  \BibitemOpen
  \bibfield  {author} {\bibinfo {author} {\bibfnamefont {Z.}~\bibnamefont
  {Zhao}}, \bibinfo {author} {\bibfnamefont {Q.}~\bibnamefont {Wu}}, \bibinfo
  {author} {\bibfnamefont {J.}~\bibnamefont {Wang}}, \bibinfo {author}
  {\bibfnamefont {B.}~\bibnamefont {Zhang}}, \bibinfo {author} {\bibfnamefont
  {C.}~\bibnamefont {Zhong}},\ and\ \bibinfo {author} {\bibfnamefont {A.~A.}\
  \bibnamefont {Zhilenkov}},\ }\bibfield  {title} {{\selectlanguage
  {English}\bibinfo {title} {Exploring {Embodied} {Intelligence} in {Soft}
  {Robotics}: {A} {Review}}},\ }\href
  {https://doi.org/10.3390/biomimetics9040248} {\bibfield  {journal} {\bibinfo
  {journal} {Biomimetics (Basel)}\ }\textbf {\bibinfo {volume} {9}},\ \bibinfo
  {pages} {248} (\bibinfo {year} {2024})}\BibitemShut {NoStop}%
\bibitem [{\citenamefont {Mertan}\ and\ \citenamefont
  {Cheney}(2025)}]{mertan_no-brainer_2025}%
  \BibitemOpen
  \bibfield  {author} {\bibinfo {author} {\bibfnamefont {A.}~\bibnamefont
  {Mertan}}\ and\ \bibinfo {author} {\bibfnamefont {N.}~\bibnamefont
  {Cheney}},\ }\bibfield  {title} {{\selectlanguage {English}\bibinfo {title}
  {No-brainer: {Morphological} {Computation} {Driven} {Adaptive} {Behavior}
  in {Soft} {Robots}}},\ }in\ \href
  {https://doi.org/10.1007/978-3-031-71533-4_6} {{\selectlanguage
  {English}\emph {\bibinfo {booktitle} {From {Animals} to {Animats} 17}}}},\
  \bibinfo {editor} {edited by\ \bibinfo {editor} {\bibfnamefont
  {O.}~\bibnamefont {Brock}}\ and\ \bibinfo {editor} {\bibfnamefont
  {J.}~\bibnamefont {Krichmar}}}\ (\bibinfo  {publisher} {Springer Nature
  Switzerland},\ \bibinfo {address} {Cham},\ \bibinfo {year} {2025})\ pp.\
  \bibinfo {pages} {81--92}\BibitemShut {NoStop}%
\bibitem [{\citenamefont {Li}\ and\ \citenamefont
  {Zhang}(2024)}]{li_computational_2024}%
  \BibitemOpen
  \bibfield  {author} {\bibinfo {author} {\bibfnamefont {W.}~\bibnamefont
  {Li}}\ and\ \bibinfo {author} {\bibfnamefont {X.~S.}\ \bibnamefont {Zhang}},\
  }\bibfield  {title} {{\selectlanguage {English}\bibinfo {title}
  {Computational morphogenesis for liquid crystal elastomer metamaterial}},\
  }\href {https://doi.org/10.1038/s41524-024-01300-y} {\bibfield  {journal}
  {\bibinfo  {journal} {npj Comput Mater}\ }\textbf {\bibinfo {volume} {10}},\
  \bibinfo {pages} {1} (\bibinfo {year} {2024})},\ \bibinfo {note} {publisher:
  Nature Publishing Group}\BibitemShut {NoStop}%
\bibitem [{\citenamefont {Vihmar}\ \emph {et~al.}(2023)\citenamefont {Vihmar},
  \citenamefont {Valdur}, \citenamefont {Banerji},\ and\ \citenamefont
  {Must}}]{vihmar_how_2023}%
  \BibitemOpen
  \bibfield  {author} {\bibinfo {author} {\bibfnamefont {M.}~\bibnamefont
  {Vihmar}}, \bibinfo {author} {\bibfnamefont {K.-A.}\ \bibnamefont {Valdur}},
  \bibinfo {author} {\bibfnamefont {S.}~\bibnamefont {Banerji}},\ and\ \bibinfo
  {author} {\bibfnamefont {I.}~\bibnamefont {Must}},\ }\bibfield  {title}
  {{\selectlanguage {English}\bibinfo {title} {How to measure embodied
  intelligence?}},\ }\href {https://doi.org/10.1088/1757-899X/1292/1/012002}
  {\bibfield  {journal} {\bibinfo  {journal} {IOP Conf. Ser.: Mater. Sci.
  Eng.}\ }\textbf {\bibinfo {volume} {1292}},\ \bibinfo {pages} {012002}
  (\bibinfo {year} {2023})},\ \bibinfo {note} {publisher: IOP
  Publishing}\BibitemShut {NoStop}%
\bibitem [{\citenamefont {Hauser}\ \emph {et~al.}(2023)\citenamefont {Hauser},
  \citenamefont {Nanayakkara},\ and\ \citenamefont
  {Forni}}]{hauser_leveraging_2023}%
  \BibitemOpen
  \bibfield  {author} {\bibinfo {author} {\bibfnamefont {H.}~\bibnamefont
  {Hauser}}, \bibinfo {author} {\bibfnamefont {T.}~\bibnamefont
  {Nanayakkara}},\ and\ \bibinfo {author} {\bibfnamefont {F.}~\bibnamefont
  {Forni}},\ }\bibfield  {title} {\bibinfo {title} {Leveraging {Morphological}
  {Computation} for {Controlling} {Soft} {Robots}: {Learning} from {Nature} to
  {Control} {Soft} {Robots}},\ }\href
  {https://doi.org/10.1109/MCS.2023.3253422} {\bibfield  {journal} {\bibinfo
  {journal} {IEEE Control Systems Magazine}\ }\textbf {\bibinfo {volume}
  {43}},\ \bibinfo {pages} {114} (\bibinfo {year} {2023})}\BibitemShut
  {NoStop}%
\bibitem [{\citenamefont {Hoffmann}\ and\ \citenamefont
  {Müller}(2014)}]{hoffmann_trade-offs_2014}%
  \BibitemOpen
  \bibfield  {author} {\bibinfo {author} {\bibfnamefont {M.}~\bibnamefont
  {Hoffmann}}\ and\ \bibinfo {author} {\bibfnamefont {V.~C.}\ \bibnamefont
  {Müller}},\ }\href {https://doi.org/10.48550/arXiv.1411.2276} {\bibinfo
  {title} {Trade-{Offs} in {Exploiting} {Body} {Morphology} for {Control}: from
  {Simple} {Bodies} and {Model}-{Based} {Control} to {Complex} {Bodies} with
  {Model}-{Free} {Distributed} {Control} {Schemes}}} (\bibinfo {year} {2014}),\
  \bibinfo {note} {arXiv:1411.2276 [cs]}\BibitemShut {NoStop}%
\bibitem [{\citenamefont {Rus}\ and\ \citenamefont
  {Tolley}(2015)}]{rus_design_2015}%
  \BibitemOpen
  \bibfield  {author} {\bibinfo {author} {\bibfnamefont {D.}~\bibnamefont
  {Rus}}\ and\ \bibinfo {author} {\bibfnamefont {M.~T.}\ \bibnamefont
  {Tolley}},\ }\bibfield  {title} {{\selectlanguage {English}\bibinfo {title}
  {Design, fabrication and control of soft robots}},\ }\href
  {https://doi.org/10.1038/nature14543} {\bibfield  {journal} {\bibinfo
  {journal} {Nature}\ }\textbf {\bibinfo {volume} {521}},\ \bibinfo {pages}
  {467} (\bibinfo {year} {2015})},\ \bibinfo {note} {publisher: Nature
  Publishing Group}\BibitemShut {NoStop}%
\bibitem [{\citenamefont {Lu}\ \emph {et~al.}(2018)\citenamefont {Lu},
  \citenamefont {Zhang}, \citenamefont {Yang}, \citenamefont {Huang},
  \citenamefont {Fukuda}, \citenamefont {Wang},\ and\ \citenamefont
  {Shen}}]{lu_bioinspired_2018}%
  \BibitemOpen
  \bibfield  {author} {\bibinfo {author} {\bibfnamefont {H.}~\bibnamefont
  {Lu}}, \bibinfo {author} {\bibfnamefont {M.}~\bibnamefont {Zhang}}, \bibinfo
  {author} {\bibfnamefont {Y.}~\bibnamefont {Yang}}, \bibinfo {author}
  {\bibfnamefont {Q.}~\bibnamefont {Huang}}, \bibinfo {author} {\bibfnamefont
  {T.}~\bibnamefont {Fukuda}}, \bibinfo {author} {\bibfnamefont
  {Z.}~\bibnamefont {Wang}},\ and\ \bibinfo {author} {\bibfnamefont
  {Y.}~\bibnamefont {Shen}},\ }\bibfield  {title} {{\selectlanguage
  {English}\bibinfo {title} {A bioinspired multilegged soft millirobot that
  functions in both dry and wet conditions}},\ }\href
  {https://doi.org/10.1038/s41467-018-06491-9} {\bibfield  {journal} {\bibinfo
  {journal} {Nat Commun}\ }\textbf {\bibinfo {volume} {9}},\ \bibinfo {pages}
  {3944} (\bibinfo {year} {2018})},\ \bibinfo {note} {publisher: Nature
  Publishing Group}\BibitemShut {NoStop}%
\bibitem [{\citenamefont {Hu}\ \emph {et~al.}(2018)\citenamefont {Hu},
  \citenamefont {Lum}, \citenamefont {Mastrangeli},\ and\ \citenamefont
  {Sitti}}]{hu_small-scale_2018}%
  \BibitemOpen
  \bibfield  {author} {\bibinfo {author} {\bibfnamefont {W.}~\bibnamefont
  {Hu}}, \bibinfo {author} {\bibfnamefont {G.~Z.}\ \bibnamefont {Lum}},
  \bibinfo {author} {\bibfnamefont {M.}~\bibnamefont {Mastrangeli}},\ and\
  \bibinfo {author} {\bibfnamefont {M.}~\bibnamefont {Sitti}},\ }\bibfield
  {title} {{\selectlanguage {English}\bibinfo {title} {Small-scale soft-bodied
  robot with multimodal locomotion}},\ }\href
  {https://doi.org/10.1038/nature25443} {\bibfield  {journal} {\bibinfo
  {journal} {Nature}\ }\textbf {\bibinfo {volume} {554}},\ \bibinfo {pages}
  {81} (\bibinfo {year} {2018})},\ \bibinfo {note} {publisher: Nature
  Publishing Group}\BibitemShut {NoStop}%
\bibitem [{\citenamefont {Xiao}\ \emph {et~al.}(2020)\citenamefont {Xiao},
  \citenamefont {Wang}, \citenamefont {Liu}, \citenamefont {Yang},\ and\
  \citenamefont {Zhang}}]{xiao_artificial_2020}%
  \BibitemOpen
  \bibfield  {author} {\bibinfo {author} {\bibfnamefont {M.}~\bibnamefont
  {Xiao}}, \bibinfo {author} {\bibfnamefont {H.}~\bibnamefont {Wang}}, \bibinfo
  {author} {\bibfnamefont {J.}~\bibnamefont {Liu}}, \bibinfo {author}
  {\bibfnamefont {H.}~\bibnamefont {Yang}},\ and\ \bibinfo {author}
  {\bibfnamefont {H.}~\bibnamefont {Zhang}},\ }\bibfield  {title}
  {{\selectlanguage {English}\bibinfo {title} {Artificial visual memory device
  based on a photo-memorizing composite and one-step manufacturing}},\ }\href
  {https://doi.org/10.1039/D0MH00312C} {\bibfield  {journal} {\bibinfo
  {journal} {Mater. Horiz.}\ }\textbf {\bibinfo {volume} {7}},\ \bibinfo
  {pages} {1597} (\bibinfo {year} {2020})},\ \bibinfo {note} {publisher: The
  Royal Society of Chemistry}\BibitemShut {NoStop}%
\bibitem [{\citenamefont {Zhao}\ \emph {et~al.}(2021)\citenamefont {Zhao},
  \citenamefont {Peng}, \citenamefont {Chen}, \citenamefont {Zha},
  \citenamefont {Li}, \citenamefont {Bai}, \citenamefont {Ke}, \citenamefont
  {Bao}, \citenamefont {Yang},\ and\ \citenamefont {Yang}}]{zhao_phase_2021}%
  \BibitemOpen
  \bibfield  {author} {\bibinfo {author} {\bibfnamefont {X.}~\bibnamefont
  {Zhao}}, \bibinfo {author} {\bibfnamefont {L.-M.}\ \bibnamefont {Peng}},
  \bibinfo {author} {\bibfnamefont {Y.}~\bibnamefont {Chen}}, \bibinfo {author}
  {\bibfnamefont {X.-J.}\ \bibnamefont {Zha}}, \bibinfo {author} {\bibfnamefont
  {W.-D.}\ \bibnamefont {Li}}, \bibinfo {author} {\bibfnamefont
  {L.}~\bibnamefont {Bai}}, \bibinfo {author} {\bibfnamefont {K.}~\bibnamefont
  {Ke}}, \bibinfo {author} {\bibfnamefont {R.-Y.}\ \bibnamefont {Bao}},
  \bibinfo {author} {\bibfnamefont {M.-B.}\ \bibnamefont {Yang}},\ and\
  \bibinfo {author} {\bibfnamefont {W.}~\bibnamefont {Yang}},\ }\bibfield
  {title} {{\selectlanguage {English}\bibinfo {title} {Phase change mediated
  mechanically transformative dynamic gel for intelligent control of versatile
  devices}},\ }\href {https://doi.org/10.1039/D0MH02069A} {\bibfield  {journal}
  {\bibinfo  {journal} {Mater. Horiz.}\ }\textbf {\bibinfo {volume} {8}},\
  \bibinfo {pages} {1230} (\bibinfo {year} {2021})},\ \bibinfo {note}
  {publisher: The Royal Society of Chemistry}\BibitemShut {NoStop}%
\bibitem [{\citenamefont {Xia}\ \emph {et~al.}(2022)\citenamefont {Xia},
  \citenamefont {Jin}, \citenamefont {Pan}, \citenamefont {Zhang},
  \citenamefont {Yang}, \citenamefont {Su}, \citenamefont {Zhao}, \citenamefont
  {Wang},\ and\ \citenamefont {Zhang}}]{xia_dynamic_2022}%
  \BibitemOpen
  \bibfield  {author} {\bibinfo {author} {\bibfnamefont {N.}~\bibnamefont
  {Xia}}, \bibinfo {author} {\bibfnamefont {D.}~\bibnamefont {Jin}}, \bibinfo
  {author} {\bibfnamefont {C.}~\bibnamefont {Pan}}, \bibinfo {author}
  {\bibfnamefont {J.}~\bibnamefont {Zhang}}, \bibinfo {author} {\bibfnamefont
  {Z.}~\bibnamefont {Yang}}, \bibinfo {author} {\bibfnamefont {L.}~\bibnamefont
  {Su}}, \bibinfo {author} {\bibfnamefont {J.}~\bibnamefont {Zhao}}, \bibinfo
  {author} {\bibfnamefont {L.}~\bibnamefont {Wang}},\ and\ \bibinfo {author}
  {\bibfnamefont {L.}~\bibnamefont {Zhang}},\ }\bibfield  {title}
  {{\selectlanguage {English}\bibinfo {title} {Dynamic morphological
  transformations in soft architected materials via buckling instability
  encoded heterogeneous magnetization}},\ }\href
  {https://doi.org/10.1038/s41467-022-35212-6} {\bibfield  {journal} {\bibinfo
  {journal} {Nat Commun}\ }\textbf {\bibinfo {volume} {13}},\ \bibinfo {pages}
  {7514} (\bibinfo {year} {2022})}\BibitemShut {NoStop}%
\bibitem [{\citenamefont {Reiter}(2020)}]{reiter_memorizing_2020}%
  \BibitemOpen
  \bibfield  {author} {\bibinfo {author} {\bibfnamefont {G.}~\bibnamefont
  {Reiter}},\ }\bibfield  {title} {\bibinfo {title} {The memorizing capacity of
  polymers},\ }\href {https://doi.org/10.1063/1.5139621} {\bibfield  {journal}
  {\bibinfo  {journal} {The Journal of Chemical Physics}\ }\textbf {\bibinfo
  {volume} {152}},\ \bibinfo {pages} {150901} (\bibinfo {year}
  {2020})}\BibitemShut {NoStop}%
\bibitem [{\citenamefont {Loeffler}\ \emph {et~al.}(2023)\citenamefont
  {Loeffler}, \citenamefont {Diaz-Alvarez}, \citenamefont {Zhu}, \citenamefont
  {Ganesh}, \citenamefont {Shine}, \citenamefont {Nakayama},\ and\
  \citenamefont {Kuncic}}]{loeffler_neuromorphic_2023}%
  \BibitemOpen
  \bibfield  {author} {\bibinfo {author} {\bibfnamefont {A.}~\bibnamefont
  {Loeffler}}, \bibinfo {author} {\bibfnamefont {A.}~\bibnamefont
  {Diaz-Alvarez}}, \bibinfo {author} {\bibfnamefont {R.}~\bibnamefont {Zhu}},
  \bibinfo {author} {\bibfnamefont {N.}~\bibnamefont {Ganesh}}, \bibinfo
  {author} {\bibfnamefont {J.~M.}\ \bibnamefont {Shine}}, \bibinfo {author}
  {\bibfnamefont {T.}~\bibnamefont {Nakayama}},\ and\ \bibinfo {author}
  {\bibfnamefont {Z.}~\bibnamefont {Kuncic}},\ }\bibfield  {title} {\bibinfo
  {title} {Neuromorphic learning, working memory, and metaplasticity in
  nanowire networks},\ }\href {https://doi.org/10.1126/sciadv.adg3289}
  {\bibfield  {journal} {\bibinfo  {journal} {Science Advances}\ }\textbf
  {\bibinfo {volume} {9}},\ \bibinfo {pages} {eadg3289} (\bibinfo {year}
  {2023})},\ \bibinfo {note} {publisher: American Association for the
  Advancement of Science}\BibitemShut {NoStop}%
\bibitem [{\citenamefont {Kamsma}\ \emph
  {et~al.}(2024{\natexlab{a}})\citenamefont {Kamsma}, \citenamefont {Kim},
  \citenamefont {Kim}, \citenamefont {Boon}, \citenamefont {Spitoni},
  \citenamefont {Park},\ and\ \citenamefont
  {Van~Roij}}]{kamsma_brain-inspired_2024}%
  \BibitemOpen
  \bibfield  {author} {\bibinfo {author} {\bibfnamefont {T.~M.}\ \bibnamefont
  {Kamsma}}, \bibinfo {author} {\bibfnamefont {J.}~\bibnamefont {Kim}},
  \bibinfo {author} {\bibfnamefont {K.}~\bibnamefont {Kim}}, \bibinfo {author}
  {\bibfnamefont {W.~Q.}\ \bibnamefont {Boon}}, \bibinfo {author}
  {\bibfnamefont {C.}~\bibnamefont {Spitoni}}, \bibinfo {author} {\bibfnamefont
  {J.}~\bibnamefont {Park}},\ and\ \bibinfo {author} {\bibfnamefont
  {R.}~\bibnamefont {Van~Roij}},\ }\bibfield  {title} {{\selectlanguage
  {English}\bibinfo {title} {Brain-inspired computing with fluidic iontronic
  nanochannels}},\ }\href {https://doi.org/10.1073/pnas.2320242121} {\bibfield
  {journal} {\bibinfo  {journal} {Proc. Natl. Acad. Sci. U.S.A.}\ }\textbf
  {\bibinfo {volume} {121}},\ \bibinfo {pages} {e2320242121} (\bibinfo {year}
  {2024}{\natexlab{a}})}\BibitemShut {NoStop}%
\bibitem [{\citenamefont {Soto}\ \emph {et~al.}(2023)\citenamefont {Soto},
  \citenamefont {Tsui}, \citenamefont {Surappa}, \citenamefont {Ahmed},
  \citenamefont {Wang}, \citenamefont {Kılınç}, \citenamefont {Akin},\ and\
  \citenamefont {Demirci}}]{soto_programmable_2023}%
  \BibitemOpen
  \bibfield  {author} {\bibinfo {author} {\bibfnamefont {F.}~\bibnamefont
  {Soto}}, \bibinfo {author} {\bibfnamefont {A.}~\bibnamefont {Tsui}}, \bibinfo
  {author} {\bibfnamefont {S.}~\bibnamefont {Surappa}}, \bibinfo {author}
  {\bibfnamefont {R.}~\bibnamefont {Ahmed}}, \bibinfo {author} {\bibfnamefont
  {J.}~\bibnamefont {Wang}}, \bibinfo {author} {\bibfnamefont {U.}~\bibnamefont
  {Kılınç}}, \bibinfo {author} {\bibfnamefont {D.}~\bibnamefont {Akin}},\
  and\ \bibinfo {author} {\bibfnamefont {U.}~\bibnamefont {Demirci}},\
  }\bibfield  {title} {{\selectlanguage {English}\bibinfo {title} {Programmable
  {Shape} {Morphing} {Metasponge}}},\ }\href
  {https://doi.org/10.1002/aisy.202300043} {\bibfield  {journal} {\bibinfo
  {journal} {Advanced Intelligent Systems}\ }\textbf {\bibinfo {volume} {5}},\
  \bibinfo {pages} {2300043} (\bibinfo {year} {2023})},\ \bibinfo {note}
  {\_eprint:
  https://onlinelibrary.wiley.com/doi/pdf/10.1002/aisy.202300043}\BibitemShut
  {NoStop}%
\bibitem [{\citenamefont {Tanaka}\ \emph {et~al.}(2018)\citenamefont {Tanaka},
  \citenamefont {Akai-Kasaya}, \citenamefont {TermehYousefi}, \citenamefont
  {Hong}, \citenamefont {Fu}, \citenamefont {Tamukoh}, \citenamefont {Tanaka},
  \citenamefont {Asai},\ and\ \citenamefont {Ogawa}}]{tanaka_molecular_2018}%
  \BibitemOpen
  \bibfield  {author} {\bibinfo {author} {\bibfnamefont {H.}~\bibnamefont
  {Tanaka}}, \bibinfo {author} {\bibfnamefont {M.}~\bibnamefont {Akai-Kasaya}},
  \bibinfo {author} {\bibfnamefont {A.}~\bibnamefont {TermehYousefi}}, \bibinfo
  {author} {\bibfnamefont {L.}~\bibnamefont {Hong}}, \bibinfo {author}
  {\bibfnamefont {L.}~\bibnamefont {Fu}}, \bibinfo {author} {\bibfnamefont
  {H.}~\bibnamefont {Tamukoh}}, \bibinfo {author} {\bibfnamefont
  {D.}~\bibnamefont {Tanaka}}, \bibinfo {author} {\bibfnamefont
  {T.}~\bibnamefont {Asai}},\ and\ \bibinfo {author} {\bibfnamefont
  {T.}~\bibnamefont {Ogawa}},\ }\bibfield  {title} {{\selectlanguage
  {English}\bibinfo {title} {A molecular neuromorphic network device consisting
  of single-walled carbon nanotubes complexed with polyoxometalate}},\ }\href
  {https://doi.org/10.1038/s41467-018-04886-2} {\bibfield  {journal} {\bibinfo
  {journal} {Nat Commun}\ }\textbf {\bibinfo {volume} {9}},\ \bibinfo {pages}
  {2693} (\bibinfo {year} {2018})},\ \bibinfo {note} {publisher: Nature
  Publishing Group}\BibitemShut {NoStop}%
\bibitem [{\citenamefont {Feng}\ \emph {et~al.}(2025)\citenamefont {Feng},
  \citenamefont {Wu}, \citenamefont {Zou}, \citenamefont {Cheng}, \citenamefont
  {Zhao}, \citenamefont {Zhang}, \citenamefont {Lu}, \citenamefont {Wang},
  \citenamefont {Wang}, \citenamefont {Wang}, \citenamefont {Guo},
  \citenamefont {Qian}, \citenamefont {Zhu}, \citenamefont {Xu}, \citenamefont
  {Dai},\ and\ \citenamefont {Liu}}]{feng_memristive_2025}%
  \BibitemOpen
  \bibfield  {author} {\bibinfo {author} {\bibfnamefont {Z.}~\bibnamefont
  {Feng}}, \bibinfo {author} {\bibfnamefont {Z.}~\bibnamefont {Wu}}, \bibinfo
  {author} {\bibfnamefont {J.}~\bibnamefont {Zou}}, \bibinfo {author}
  {\bibfnamefont {L.}~\bibnamefont {Cheng}}, \bibinfo {author} {\bibfnamefont
  {X.}~\bibnamefont {Zhao}}, \bibinfo {author} {\bibfnamefont {X.}~\bibnamefont
  {Zhang}}, \bibinfo {author} {\bibfnamefont {J.}~\bibnamefont {Lu}}, \bibinfo
  {author} {\bibfnamefont {C.}~\bibnamefont {Wang}}, \bibinfo {author}
  {\bibfnamefont {Y.}~\bibnamefont {Wang}}, \bibinfo {author} {\bibfnamefont
  {H.}~\bibnamefont {Wang}}, \bibinfo {author} {\bibfnamefont {W.}~\bibnamefont
  {Guo}}, \bibinfo {author} {\bibfnamefont {Z.}~\bibnamefont {Qian}}, \bibinfo
  {author} {\bibfnamefont {Y.}~\bibnamefont {Zhu}}, \bibinfo {author}
  {\bibfnamefont {Z.}~\bibnamefont {Xu}}, \bibinfo {author} {\bibfnamefont
  {Y.}~\bibnamefont {Dai}},\ and\ \bibinfo {author} {\bibfnamefont
  {Q.}~\bibnamefont {Liu}},\ }\bibfield  {title} {{\selectlanguage
  {English}\bibinfo {title} {Memristive {Bellman} solver for
  decision-making}},\ }\href {https://doi.org/10.1038/s41467-025-60085-w}
  {\bibfield  {journal} {\bibinfo  {journal} {Nat Commun}\ }\textbf {\bibinfo
  {volume} {16}},\ \bibinfo {pages} {4925} (\bibinfo {year} {2025})},\ \bibinfo
  {note} {publisher: Nature Publishing Group}\BibitemShut {NoStop}%
\bibitem [{\citenamefont {Ionov}(2015)}]{ionov:l:2015}%
  \BibitemOpen
  \bibfield  {author} {\bibinfo {author} {\bibfnamefont {L.}~\bibnamefont
  {Ionov}},\ }\bibfield  {title} {\bibinfo {title} {Polymeric {{Actuators}}},\
  }\href {https://doi.org/10.1021/la503407z} {\bibfield  {journal} {\bibinfo
  {journal} {Langmuir}\ }\textbf {\bibinfo {volume} {31}},\ \bibinfo {pages}
  {5015} (\bibinfo {year} {2015})}\BibitemShut {NoStop}%
\bibitem [{\citenamefont {Chen}\ \emph {et~al.}(2021)\citenamefont {Chen},
  \citenamefont {Pauly},\ and\ \citenamefont {Reis}}]{chen:n:2021}%
  \BibitemOpen
  \bibfield  {author} {\bibinfo {author} {\bibfnamefont {T.}~\bibnamefont
  {Chen}}, \bibinfo {author} {\bibfnamefont {M.}~\bibnamefont {Pauly}},\ and\
  \bibinfo {author} {\bibfnamefont {P.~M.}\ \bibnamefont {Reis}},\ }\bibfield
  {title} {\bibinfo {title} {A reprogrammable mechanical metamaterial with
  stable memory},\ }\href {https://doi.org/10.1038/s41586-020-03123-5}
  {\bibfield  {journal} {\bibinfo  {journal} {Nature}\ }\textbf {\bibinfo
  {volume} {589}},\ \bibinfo {pages} {386} (\bibinfo {year}
  {2021})}\BibitemShut {NoStop}%
\bibitem [{\citenamefont {Xia}\ \emph {et~al.}(2021)\citenamefont {Xia},
  \citenamefont {He}, \citenamefont {Zhang}, \citenamefont {Liu},\ and\
  \citenamefont {Leng}}]{xia:am:2021}%
  \BibitemOpen
  \bibfield  {author} {\bibinfo {author} {\bibfnamefont {Y.}~\bibnamefont
  {Xia}}, \bibinfo {author} {\bibfnamefont {Y.}~\bibnamefont {He}}, \bibinfo
  {author} {\bibfnamefont {F.}~\bibnamefont {Zhang}}, \bibinfo {author}
  {\bibfnamefont {Y.}~\bibnamefont {Liu}},\ and\ \bibinfo {author}
  {\bibfnamefont {J.}~\bibnamefont {Leng}},\ }\bibfield  {title} {\bibinfo
  {title} {A {{Review}} of {{Shape Memory Polymers}} and {{Composites}}:
  {{Mechanisms}}, {{Materials}}, and {{Applications}}},\ }\href
  {https://doi.org/10.1002/adma.202000713} {\bibfield  {journal} {\bibinfo
  {journal} {Advanced Materials}\ }\textbf {\bibinfo {volume} {33}},\ \bibinfo
  {pages} {2000713} (\bibinfo {year} {2021})}\BibitemShut {NoStop}%
\bibitem [{\citenamefont {Zhang}\ \emph
  {et~al.}(2025{\natexlab{a}})\citenamefont {Zhang}, \citenamefont {Li},
  \citenamefont {Guo}, \citenamefont {Li}, \citenamefont {Li}, \citenamefont
  {Li}, \citenamefont {Yi},\ and\ \citenamefont {Xie}}]{zhang:aaem:2025}%
  \BibitemOpen
  \bibfield  {author} {\bibinfo {author} {\bibfnamefont {H.}~\bibnamefont
  {Zhang}}, \bibinfo {author} {\bibfnamefont {L.}~\bibnamefont {Li}}, \bibinfo
  {author} {\bibfnamefont {A.}~\bibnamefont {Guo}}, \bibinfo {author}
  {\bibfnamefont {J.}~\bibnamefont {Li}}, \bibinfo {author} {\bibfnamefont
  {Y.-T.}\ \bibnamefont {Li}}, \bibinfo {author} {\bibfnamefont
  {W.}~\bibnamefont {Li}}, \bibinfo {author} {\bibfnamefont {M.}~\bibnamefont
  {Yi}},\ and\ \bibinfo {author} {\bibfnamefont {L.}~\bibnamefont {Xie}},\
  }\bibfield  {title} {\bibinfo {title} {{{P3OT-Based Organic Polymer
  Memristors}} for {{Artificial Synaptic Behavior}} and {{Neuromorphic
  Computing Applications}}},\ }\href {https://doi.org/10.1021/acsaelm.4c02278}
  {\bibfield  {journal} {\bibinfo  {journal} {ACS Appl. Electron. Mater.}\
  }\textbf {\bibinfo {volume} {7}},\ \bibinfo {pages} {2001} (\bibinfo {year}
  {2025}{\natexlab{a}})}\BibitemShut {NoStop}%
\bibitem [{\citenamefont {Wang}\ \emph {et~al.}(2024)\citenamefont {Wang},
  \citenamefont {Wang}, \citenamefont {Chen}, \citenamefont {Liu},
  \citenamefont {Wang}, \citenamefont {Yuan}, \citenamefont {Ma}, \citenamefont
  {Xu}, \citenamefont {Cheng}, \citenamefont {Ji}, \citenamefont {Yang},
  \citenamefont {Shuai}, \citenamefont {Ye}, \citenamefont {Wang},
  \citenamefont {Jiao},\ and\ \citenamefont {Liu}}]{wang_robo-matter_2024}%
  \BibitemOpen
  \bibfield  {author} {\bibinfo {author} {\bibfnamefont {J.}~\bibnamefont
  {Wang}}, \bibinfo {author} {\bibfnamefont {G.}~\bibnamefont {Wang}}, \bibinfo
  {author} {\bibfnamefont {H.}~\bibnamefont {Chen}}, \bibinfo {author}
  {\bibfnamefont {Y.}~\bibnamefont {Liu}}, \bibinfo {author} {\bibfnamefont
  {P.}~\bibnamefont {Wang}}, \bibinfo {author} {\bibfnamefont {D.}~\bibnamefont
  {Yuan}}, \bibinfo {author} {\bibfnamefont {X.}~\bibnamefont {Ma}}, \bibinfo
  {author} {\bibfnamefont {X.}~\bibnamefont {Xu}}, \bibinfo {author}
  {\bibfnamefont {Z.}~\bibnamefont {Cheng}}, \bibinfo {author} {\bibfnamefont
  {B.}~\bibnamefont {Ji}}, \bibinfo {author} {\bibfnamefont {M.}~\bibnamefont
  {Yang}}, \bibinfo {author} {\bibfnamefont {J.}~\bibnamefont {Shuai}},
  \bibinfo {author} {\bibfnamefont {F.}~\bibnamefont {Ye}}, \bibinfo {author}
  {\bibfnamefont {J.}~\bibnamefont {Wang}}, \bibinfo {author} {\bibfnamefont
  {Y.}~\bibnamefont {Jiao}},\ and\ \bibinfo {author} {\bibfnamefont
  {L.}~\bibnamefont {Liu}},\ }\bibfield  {title} {{\selectlanguage
  {English}\bibinfo {title} {Robo-{Matter} towards reconfigurable
  multifunctional smart materials}},\ }\href
  {https://doi.org/10.1038/s41467-024-53123-6} {\bibfield  {journal} {\bibinfo
  {journal} {Nat Commun}\ }\textbf {\bibinfo {volume} {15}},\ \bibinfo {pages}
  {8853} (\bibinfo {year} {2024})},\ \bibinfo {note} {publisher: Nature
  Publishing Group}\BibitemShut {NoStop}%
\bibitem [{\citenamefont {Alapan}\ \emph {et~al.}(2019)\citenamefont {Alapan},
  \citenamefont {Yigit}, \citenamefont {Beker}, \citenamefont {Demirörs},\
  and\ \citenamefont {Sitti}}]{alapan_shape-encoded_2019}%
  \BibitemOpen
  \bibfield  {author} {\bibinfo {author} {\bibfnamefont {Y.}~\bibnamefont
  {Alapan}}, \bibinfo {author} {\bibfnamefont {B.}~\bibnamefont {Yigit}},
  \bibinfo {author} {\bibfnamefont {O.}~\bibnamefont {Beker}}, \bibinfo
  {author} {\bibfnamefont {A.~F.}\ \bibnamefont {Demirörs}},\ and\ \bibinfo
  {author} {\bibfnamefont {M.}~\bibnamefont {Sitti}},\ }\bibfield  {title}
  {{\selectlanguage {English}\bibinfo {title} {Shape-encoded dynamic assembly
  of mobile micromachines}},\ }\href
  {https://doi.org/10.1038/s41563-019-0407-3} {\bibfield  {journal} {\bibinfo
  {journal} {Nat. Mater.}\ }\textbf {\bibinfo {volume} {18}},\ \bibinfo {pages}
  {1244} (\bibinfo {year} {2019})},\ \bibinfo {note} {publisher: Nature
  Publishing Group}\BibitemShut {NoStop}%
\bibitem [{\citenamefont {Calvino}\ and\ \citenamefont
  {Weder}(2018)}]{calvino_microcapsule-containing_2018}%
  \BibitemOpen
  \bibfield  {author} {\bibinfo {author} {\bibfnamefont {C.}~\bibnamefont
  {Calvino}}\ and\ \bibinfo {author} {\bibfnamefont {C.}~\bibnamefont
  {Weder}},\ }\bibfield  {title} {{\selectlanguage {English}\bibinfo {title}
  {Microcapsule-{Containing} {Self}-{Reporting} {Polymers}}},\ }\href
  {https://doi.org/10.1002/smll.201802489} {\bibfield  {journal} {\bibinfo
  {journal} {Small}\ }\textbf {\bibinfo {volume} {14}},\ \bibinfo {pages}
  {1802489} (\bibinfo {year} {2018})},\ \bibinfo {note} {\_eprint:
  https://onlinelibrary.wiley.com/doi/pdf/10.1002/smll.201802489}\BibitemShut
  {NoStop}%
\bibitem [{\citenamefont {Bayat}\ \emph {et~al.}(2024)\citenamefont {Bayat},
  \citenamefont {Mardani}, \citenamefont {Roghani-Mamaqani},\ and\
  \citenamefont {Hoogenboom}}]{bayat_self-indicating_2024}%
  \BibitemOpen
  \bibfield  {author} {\bibinfo {author} {\bibfnamefont {M.}~\bibnamefont
  {Bayat}}, \bibinfo {author} {\bibfnamefont {H.}~\bibnamefont {Mardani}},
  \bibinfo {author} {\bibfnamefont {H.}~\bibnamefont {Roghani-Mamaqani}},\ and\
  \bibinfo {author} {\bibfnamefont {R.}~\bibnamefont {Hoogenboom}},\ }\bibfield
   {title} {{\selectlanguage {English}\bibinfo {title} {Self-indicating
  polymers: a pathway to intelligent materials}},\ }\href
  {https://doi.org/10.1039/D3CS00431G} {\bibfield  {journal} {\bibinfo
  {journal} {Chem. Soc. Rev.}\ }\textbf {\bibinfo {volume} {53}},\ \bibinfo
  {pages} {4045} (\bibinfo {year} {2024})},\ \bibinfo {note} {publisher: The
  Royal Society of Chemistry}\BibitemShut {NoStop}%
\bibitem [{\citenamefont {Jiao}\ \emph
  {et~al.}(2023{\natexlab{a}})\citenamefont {Jiao}, \citenamefont {Mueller},
  \citenamefont {Raney}, \citenamefont {Zheng},\ and\ \citenamefont
  {Alavi}}]{jiao_mechanical_2023}%
  \BibitemOpen
  \bibfield  {author} {\bibinfo {author} {\bibfnamefont {P.}~\bibnamefont
  {Jiao}}, \bibinfo {author} {\bibfnamefont {J.}~\bibnamefont {Mueller}},
  \bibinfo {author} {\bibfnamefont {J.~R.}\ \bibnamefont {Raney}}, \bibinfo
  {author} {\bibfnamefont {X.~R.}\ \bibnamefont {Zheng}},\ and\ \bibinfo
  {author} {\bibfnamefont {A.~H.}\ \bibnamefont {Alavi}},\ }\bibfield  {title}
  {{\selectlanguage {English}\bibinfo {title} {Mechanical metamaterials and
  beyond}},\ }\href {https://doi.org/10.1038/s41467-023-41679-8} {\bibfield
  {journal} {\bibinfo  {journal} {Nat Commun}\ }\textbf {\bibinfo {volume}
  {14}},\ \bibinfo {pages} {6004} (\bibinfo {year} {2023}{\natexlab{a}})},\
  \bibinfo {note} {publisher: Nature Publishing Group}\BibitemShut {NoStop}%
\bibitem [{\citenamefont {Luo}\ \emph {et~al.}(2023)\citenamefont {Luo},
  \citenamefont {Shao}, \citenamefont {Ji}, \citenamefont {Chen}, \citenamefont
  {Wang}, \citenamefont {Wu}, \citenamefont {Kong}, \citenamefont {Guo},
  \citenamefont {Wei}, \citenamefont {Zhao}, \citenamefont {Liu},\ and\
  \citenamefont {Wei}}]{luo_highly_2023}%
  \BibitemOpen
  \bibfield  {author} {\bibinfo {author} {\bibfnamefont {S.}~\bibnamefont
  {Luo}}, \bibinfo {author} {\bibfnamefont {L.}~\bibnamefont {Shao}}, \bibinfo
  {author} {\bibfnamefont {D.}~\bibnamefont {Ji}}, \bibinfo {author}
  {\bibfnamefont {Y.}~\bibnamefont {Chen}}, \bibinfo {author} {\bibfnamefont
  {X.}~\bibnamefont {Wang}}, \bibinfo {author} {\bibfnamefont {Y.}~\bibnamefont
  {Wu}}, \bibinfo {author} {\bibfnamefont {D.}~\bibnamefont {Kong}}, \bibinfo
  {author} {\bibfnamefont {M.}~\bibnamefont {Guo}}, \bibinfo {author}
  {\bibfnamefont {D.}~\bibnamefont {Wei}}, \bibinfo {author} {\bibfnamefont
  {Y.}~\bibnamefont {Zhao}}, \bibinfo {author} {\bibfnamefont {Y.}~\bibnamefont
  {Liu}},\ and\ \bibinfo {author} {\bibfnamefont {D.}~\bibnamefont {Wei}},\
  }\bibfield  {title} {\bibinfo {title} {Highly {Bionic}
  {Neurotransmitter}-{Communicated} {Neurons} {Following}
  {Integrate}-and-{Fire} {Dynamics}},\ }\href
  {https://doi.org/10.1021/acs.nanolett.3c00799} {\bibfield  {journal}
  {\bibinfo  {journal} {Nano Lett.}\ }\textbf {\bibinfo {volume} {23}},\
  \bibinfo {pages} {4974} (\bibinfo {year} {2023})},\ \bibinfo {note}
  {publisher: American Chemical Society}\BibitemShut {NoStop}%
\bibitem [{\citenamefont {Bordiga}\ \emph {et~al.}(2024)\citenamefont
  {Bordiga}, \citenamefont {Medina}, \citenamefont {Jafarzadeh}, \citenamefont
  {Bösch}, \citenamefont {Adams}, \citenamefont {Tournat},\ and\ \citenamefont
  {Bertoldi}}]{bordiga_automated_2024}%
  \BibitemOpen
  \bibfield  {author} {\bibinfo {author} {\bibfnamefont {G.}~\bibnamefont
  {Bordiga}}, \bibinfo {author} {\bibfnamefont {E.}~\bibnamefont {Medina}},
  \bibinfo {author} {\bibfnamefont {S.}~\bibnamefont {Jafarzadeh}}, \bibinfo
  {author} {\bibfnamefont {C.}~\bibnamefont {Bösch}}, \bibinfo {author}
  {\bibfnamefont {R.~P.}\ \bibnamefont {Adams}}, \bibinfo {author}
  {\bibfnamefont {V.}~\bibnamefont {Tournat}},\ and\ \bibinfo {author}
  {\bibfnamefont {K.}~\bibnamefont {Bertoldi}},\ }\bibfield  {title}
  {{\selectlanguage {English}\bibinfo {title} {Automated discovery of
  reprogrammable nonlinear dynamic metamaterials}},\ }\href
  {https://doi.org/10.1038/s41563-024-02008-6} {\bibfield  {journal} {\bibinfo
  {journal} {Nat. Mater.}\ }\textbf {\bibinfo {volume} {23}},\ \bibinfo {pages}
  {1486} (\bibinfo {year} {2024})},\ \bibinfo {note} {publisher: Nature
  Publishing Group}\BibitemShut {NoStop}%
\bibitem [{\citenamefont {Sabelhaus}\ \emph {et~al.}(2022)\citenamefont
  {Sabelhaus}, \citenamefont {Mehta}, \citenamefont {Wertz},\ and\
  \citenamefont {Majidi}}]{sabelhaus_-situ_2022}%
  \BibitemOpen
  \bibfield  {author} {\bibinfo {author} {\bibfnamefont {A.~P.}\ \bibnamefont
  {Sabelhaus}}, \bibinfo {author} {\bibfnamefont {R.~K.}\ \bibnamefont
  {Mehta}}, \bibinfo {author} {\bibfnamefont {A.~T.}\ \bibnamefont {Wertz}},\
  and\ \bibinfo {author} {\bibfnamefont {C.}~\bibnamefont {Majidi}},\
  }\bibfield  {title} {\bibinfo {title} {In-{Situ} {Sensing} and {Dynamics}
  {Predictions} for {Electrothermally}-{Actuated} {Soft} {Robot} {Limbs}},\
  }\href {https://doi.org/10.3389/frobt.2022.888261} {\bibfield  {journal}
  {\bibinfo  {journal} {Front. Robot. AI}\ }\textbf {\bibinfo {volume} {9}},\
  \bibinfo {pages} {888261} (\bibinfo {year} {2022})},\ \bibinfo {note}
  {arXiv:2111.04851 [cs]}\BibitemShut {NoStop}%
\bibitem [{\citenamefont {Buckley}\ \emph {et~al.}(2024)\citenamefont
  {Buckley}, \citenamefont {Lewens}, \citenamefont {Levin}, \citenamefont
  {Millidge}, \citenamefont {Tschantz},\ and\ \citenamefont
  {Watson}}]{buckley_natural_2024}%
  \BibitemOpen
  \bibfield  {author} {\bibinfo {author} {\bibfnamefont {C.~L.}\ \bibnamefont
  {Buckley}}, \bibinfo {author} {\bibfnamefont {T.}~\bibnamefont {Lewens}},
  \bibinfo {author} {\bibfnamefont {M.}~\bibnamefont {Levin}}, \bibinfo
  {author} {\bibfnamefont {B.}~\bibnamefont {Millidge}}, \bibinfo {author}
  {\bibfnamefont {A.}~\bibnamefont {Tschantz}},\ and\ \bibinfo {author}
  {\bibfnamefont {R.~A.}\ \bibnamefont {Watson}},\ }\bibfield  {title}
  {{\selectlanguage {English}\bibinfo {title} {Natural {Induction}:
  {Spontaneous} {Adaptive} {Organisation} without {Natural} {Selection}}},\
  }\href {https://doi.org/10.3390/e26090765} {\bibfield  {journal} {\bibinfo
  {journal} {Entropy}\ }\textbf {\bibinfo {volume} {26}},\ \bibinfo {pages}
  {765} (\bibinfo {year} {2024})},\ \bibinfo {note} {number: 9 Publisher:
  Multidisciplinary Digital Publishing Institute}\BibitemShut {NoStop}%
\bibitem [{\citenamefont {Kim}\ \emph {et~al.}(2015)\citenamefont {Kim},
  \citenamefont {Baker},\ and\ \citenamefont {Phillips}}]{kim_polymeric_2015}%
  \BibitemOpen
  \bibfield  {author} {\bibinfo {author} {\bibfnamefont {H.}~\bibnamefont
  {Kim}}, \bibinfo {author} {\bibfnamefont {M.~S.}\ \bibnamefont {Baker}},\
  and\ \bibinfo {author} {\bibfnamefont {S.~T.}\ \bibnamefont {Phillips}},\
  }\bibfield  {title} {{\selectlanguage {English}\bibinfo {title} {Polymeric
  materials that convert local fleeting signals into global macroscopic
  responses}},\ }\href {https://doi.org/10.1039/C5SC00701A} {\bibfield
  {journal} {\bibinfo  {journal} {Chem. Sci.}\ }\textbf {\bibinfo {volume}
  {6}},\ \bibinfo {pages} {3388} (\bibinfo {year} {2015})},\ \bibinfo {note}
  {publisher: The Royal Society of Chemistry}\BibitemShut {NoStop}%
\bibitem [{\citenamefont {Kramar}\ and\ \citenamefont
  {Alim}(2021)}]{kramar_encoding_2021}%
  \BibitemOpen
  \bibfield  {author} {\bibinfo {author} {\bibfnamefont {M.}~\bibnamefont
  {Kramar}}\ and\ \bibinfo {author} {\bibfnamefont {K.}~\bibnamefont {Alim}},\
  }\bibfield  {title} {\bibinfo {title} {Encoding memory in tube diameter
  hierarchy of living flow network},\ }\href
  {https://doi.org/10.1073/pnas.2007815118} {\bibfield  {journal} {\bibinfo
  {journal} {Proceedings of the National Academy of Sciences}\ }\textbf
  {\bibinfo {volume} {118}},\ \bibinfo {pages} {e2007815118} (\bibinfo {year}
  {2021})},\ \bibinfo {note} {publisher: Proceedings of the National Academy of
  Sciences}\BibitemShut {NoStop}%
\bibitem [{\citenamefont {Yao}\ \emph {et~al.}(2022)\citenamefont {Yao},
  \citenamefont {Kos}, \citenamefont {Zhang}, \citenamefont {Luo},
  \citenamefont {Serra}, \citenamefont {Steager}, \citenamefont {Ravnik},\ and\
  \citenamefont {Stebe}}]{yao_nematic_2022}%
  \BibitemOpen
  \bibfield  {author} {\bibinfo {author} {\bibfnamefont {T.}~\bibnamefont
  {Yao}}, \bibinfo {author} {\bibfnamefont {Z.}~\bibnamefont {Kos}}, \bibinfo
  {author} {\bibfnamefont {Q.~X.}\ \bibnamefont {Zhang}}, \bibinfo {author}
  {\bibfnamefont {Y.}~\bibnamefont {Luo}}, \bibinfo {author} {\bibfnamefont
  {F.}~\bibnamefont {Serra}}, \bibinfo {author} {\bibfnamefont {E.~B.}\
  \bibnamefont {Steager}}, \bibinfo {author} {\bibfnamefont {M.}~\bibnamefont
  {Ravnik}},\ and\ \bibinfo {author} {\bibfnamefont {K.~J.}\ \bibnamefont
  {Stebe}},\ }\bibfield  {title} {{\selectlanguage {English}\bibinfo {title}
  {Nematic {Colloidal} {Micro}-{Robots} as {Physically} {Intelligent}
  {Systems}}},\ }\href {https://doi.org/10.1002/adfm.202205546} {\bibfield
  {journal} {\bibinfo  {journal} {Advanced Functional Materials}\ }\textbf
  {\bibinfo {volume} {32}},\ \bibinfo {pages} {2205546} (\bibinfo {year}
  {2022})},\ \bibinfo {note} {\_eprint:
  https://onlinelibrary.wiley.com/doi/pdf/10.1002/adfm.202205546}\BibitemShut
  {NoStop}%
\bibitem [{\citenamefont {Kos}\ and\ \citenamefont
  {Dunkel}(2022)}]{kos_nematic_2022}%
  \BibitemOpen
  \bibfield  {author} {\bibinfo {author} {\bibfnamefont {Z.}~\bibnamefont
  {Kos}}\ and\ \bibinfo {author} {\bibfnamefont {J.}~\bibnamefont {Dunkel}},\
  }\bibfield  {title} {\bibinfo {title} {Nematic bits and universal logic
  gates},\ }\href {https://doi.org/10.1126/sciadv.abp8371} {\bibfield
  {journal} {\bibinfo  {journal} {Science Advances}\ }\textbf {\bibinfo
  {volume} {8}},\ \bibinfo {pages} {eabp8371} (\bibinfo {year} {2022})},\
  \bibinfo {note} {publisher: American Association for the Advancement of
  Science}\BibitemShut {NoStop}%
\bibitem [{\citenamefont {Chen}\ \emph {et~al.}(2022)\citenamefont {Chen},
  \citenamefont {Zhao}, \citenamefont {Huang}, \citenamefont {Zhou},\ and\
  \citenamefont {Liu}}]{chen_enormous-stiffness-changing_2022}%
  \BibitemOpen
  \bibfield  {author} {\bibinfo {author} {\bibfnamefont {L.}~\bibnamefont
  {Chen}}, \bibinfo {author} {\bibfnamefont {C.}~\bibnamefont {Zhao}}, \bibinfo
  {author} {\bibfnamefont {J.}~\bibnamefont {Huang}}, \bibinfo {author}
  {\bibfnamefont {J.}~\bibnamefont {Zhou}},\ and\ \bibinfo {author}
  {\bibfnamefont {M.}~\bibnamefont {Liu}},\ }\bibfield  {title}
  {{\selectlanguage {English}\bibinfo {title} {Enormous-stiffness-changing
  polymer networks by glass transition mediated microphase separation}},\
  }\href {https://doi.org/10.1038/s41467-022-34677-9} {\bibfield  {journal}
  {\bibinfo  {journal} {Nat Commun}\ }\textbf {\bibinfo {volume} {13}},\
  \bibinfo {pages} {6821} (\bibinfo {year} {2022})},\ \bibinfo {note}
  {publisher: Nature Publishing Group}\BibitemShut {NoStop}%
\bibitem [{\citenamefont {Hu}\ \emph {et~al.}(2025)\citenamefont {Hu},
  \citenamefont {Ma}, \citenamefont {Kim}, \citenamefont {Kim}, \citenamefont
  {Ye}, \citenamefont {Pané}, \citenamefont {Bao}, \citenamefont {Style},\
  and\ \citenamefont {Isa}}]{hu_self-reporting_2025}%
  \BibitemOpen
  \bibfield  {author} {\bibinfo {author} {\bibfnamefont {M.}~\bibnamefont
  {Hu}}, \bibinfo {author} {\bibfnamefont {Z.}~\bibnamefont {Ma}}, \bibinfo
  {author} {\bibfnamefont {M.}~\bibnamefont {Kim}}, \bibinfo {author}
  {\bibfnamefont {D.}~\bibnamefont {Kim}}, \bibinfo {author} {\bibfnamefont
  {S.}~\bibnamefont {Ye}}, \bibinfo {author} {\bibfnamefont {S.}~\bibnamefont
  {Pané}}, \bibinfo {author} {\bibfnamefont {Y.}~\bibnamefont {Bao}}, \bibinfo
  {author} {\bibfnamefont {R.~W.}\ \bibnamefont {Style}},\ and\ \bibinfo
  {author} {\bibfnamefont {L.}~\bibnamefont {Isa}},\ }\bibfield  {title}
  {{\selectlanguage {English}\bibinfo {title} {Self-{Reporting} {Multiple}
  {Microscopic} {Stresses} {Through} {Tunable} {Microcapsule} {Arrays}}},\
  }\href {https://doi.org/10.1002/adma.202410945} {\bibfield  {journal}
  {\bibinfo  {journal} {Advanced Materials}\ }\textbf {\bibinfo {volume}
  {37}},\ \bibinfo {pages} {2410945} (\bibinfo {year} {2025})},\ \bibinfo
  {note} {\_eprint:
  https://onlinelibrary.wiley.com/doi/pdf/10.1002/adma.202410945}\BibitemShut
  {NoStop}%
\bibitem [{\citenamefont {Milana}\ \emph {et~al.}(2022)\citenamefont {Milana},
  \citenamefont {Van~Raemdonck}, \citenamefont {Casla}, \citenamefont
  {De~Volder}, \citenamefont {Reynaerts},\ and\ \citenamefont
  {Gorissen}}]{milana_morphological_2022}%
  \BibitemOpen
  \bibfield  {author} {\bibinfo {author} {\bibfnamefont {E.}~\bibnamefont
  {Milana}}, \bibinfo {author} {\bibfnamefont {B.}~\bibnamefont
  {Van~Raemdonck}}, \bibinfo {author} {\bibfnamefont {A.~S.}\ \bibnamefont
  {Casla}}, \bibinfo {author} {\bibfnamefont {M.}~\bibnamefont {De~Volder}},
  \bibinfo {author} {\bibfnamefont {D.}~\bibnamefont {Reynaerts}},\ and\
  \bibinfo {author} {\bibfnamefont {B.}~\bibnamefont {Gorissen}},\ }\bibfield
  {title} {{\selectlanguage {English}\bibinfo {title} {Morphological {Control}
  of {Cilia}-{Inspired} {Asymmetric} {Movements} {Using} {Nonlinear} {Soft}
  {Inflatable} {Actuators}}},\ }\bibfield  {journal} {\bibinfo  {journal}
  {Front. Robot. AI}\ }\textbf {\bibinfo {volume} {8}},\ \href
  {https://doi.org/10.3389/frobt.2021.788067} {10.3389/frobt.2021.788067}
  (\bibinfo {year} {2022}),\ \bibinfo {note} {publisher: Frontiers}\BibitemShut
  {NoStop}%
\bibitem [{\citenamefont {Ke}\ \emph {et~al.}(2012)\citenamefont {Ke},
  \citenamefont {Ong}, \citenamefont {Shih},\ and\ \citenamefont
  {Yin}}]{ke_three-dimensional_2012}%
  \BibitemOpen
  \bibfield  {author} {\bibinfo {author} {\bibfnamefont {Y.}~\bibnamefont
  {Ke}}, \bibinfo {author} {\bibfnamefont {L.~L.}\ \bibnamefont {Ong}},
  \bibinfo {author} {\bibfnamefont {W.~M.}\ \bibnamefont {Shih}},\ and\
  \bibinfo {author} {\bibfnamefont {P.}~\bibnamefont {Yin}},\ }\bibfield
  {title} {\bibinfo {title} {Three-{Dimensional} {Structures}
  {Self}-{Assembled} from {DNA} {Bricks}},\ }\href
  {https://doi.org/10.1126/science.1227268} {\bibfield  {journal} {\bibinfo
  {journal} {Science}\ }\textbf {\bibinfo {volume} {338}},\ \bibinfo {pages}
  {1177} (\bibinfo {year} {2012})},\ \bibinfo {note} {publisher: American
  Association for the Advancement of Science}\BibitemShut {NoStop}%
\bibitem [{\citenamefont {Scheidegger}\ \emph {et~al.}(2020)\citenamefont
  {Scheidegger}, \citenamefont {Mikos},\ and\ \citenamefont
  {Fellermann}}]{scheidegger_modelling_2020}%
  \BibitemOpen
  \bibfield  {author} {\bibinfo {author} {\bibfnamefont {S.}~\bibnamefont
  {Scheidegger}}, \bibinfo {author} {\bibfnamefont {A.}~\bibnamefont {Mikos}},\
  and\ \bibinfo {author} {\bibfnamefont {H.}~\bibnamefont {Fellermann}},\
  }\bibfield  {title} {{\selectlanguage {English}\bibinfo {title} {Modelling
  {Artificial} {Immune} – {Tumor} {Ecosystem} {Interaction} {During}
  {Radiation} {Therapy} {Using} a {Perceptron} – {Based} {Antigen} {Pattern}
  {Recognition}}},\ }in\ \href {https://doi.org/10.1162/isal_a_00271}
  {{\selectlanguage {English}\emph {\bibinfo {booktitle} {The 2020 {Conference}
  on {Artificial} {Life}}}}}\ (\bibinfo  {publisher} {MIT Press},\ \bibinfo
  {address} {Online},\ \bibinfo {year} {2020})\ pp.\ \bibinfo {pages}
  {541--548}\BibitemShut {NoStop}%
\bibitem [{\citenamefont {Rauba}\ \emph {et~al.}(2024)\citenamefont {Rauba},
  \citenamefont {Seedat}, \citenamefont {Kacprzyk},\ and\ \citenamefont
  {Schaar}}]{rauba_self-healing_2024}%
  \BibitemOpen
  \bibfield  {author} {\bibinfo {author} {\bibfnamefont {P.}~\bibnamefont
  {Rauba}}, \bibinfo {author} {\bibfnamefont {N.}~\bibnamefont {Seedat}},
  \bibinfo {author} {\bibfnamefont {K.}~\bibnamefont {Kacprzyk}},\ and\
  \bibinfo {author} {\bibfnamefont {M.~v.~d.}\ \bibnamefont {Schaar}},\ }\href
  {https://doi.org/10.48550/arXiv.2411.00186} {\bibinfo {title} {Self-{Healing}
  {Machine} {Learning}: {A} {Framework} for {Autonomous} {Adaptation} in
  {Real}-{World} {Environments}}} (\bibinfo {year} {2024}),\ \bibinfo {note}
  {arXiv:2411.00186 [cs]}\BibitemShut {NoStop}%
\bibitem [{\citenamefont {Lee}\ \emph {et~al.}(2022)\citenamefont {Lee},
  \citenamefont {Calcaterra}, \citenamefont {Lee}, \citenamefont {Hadibrata},
  \citenamefont {Lee}, \citenamefont {Oh}, \citenamefont {Aydin}, \citenamefont
  {Glotzer},\ and\ \citenamefont {Mirkin}}]{lee_shape_2022}%
  \BibitemOpen
  \bibfield  {author} {\bibinfo {author} {\bibfnamefont {S.}~\bibnamefont
  {Lee}}, \bibinfo {author} {\bibfnamefont {H.~A.}\ \bibnamefont {Calcaterra}},
  \bibinfo {author} {\bibfnamefont {S.}~\bibnamefont {Lee}}, \bibinfo {author}
  {\bibfnamefont {W.}~\bibnamefont {Hadibrata}}, \bibinfo {author}
  {\bibfnamefont {B.}~\bibnamefont {Lee}}, \bibinfo {author} {\bibfnamefont
  {E.}~\bibnamefont {Oh}}, \bibinfo {author} {\bibfnamefont {K.}~\bibnamefont
  {Aydin}}, \bibinfo {author} {\bibfnamefont {S.~C.}\ \bibnamefont {Glotzer}},\
  and\ \bibinfo {author} {\bibfnamefont {C.~A.}\ \bibnamefont {Mirkin}},\
  }\bibfield  {title} {{\selectlanguage {English}\bibinfo {title} {Shape memory
  in self-adapting colloidal crystals}},\ }\href
  {https://doi.org/10.1038/s41586-022-05232-9} {\bibfield  {journal} {\bibinfo
  {journal} {Nature}\ }\textbf {\bibinfo {volume} {610}},\ \bibinfo {pages}
  {674} (\bibinfo {year} {2022})},\ \bibinfo {note} {publisher: Nature
  Publishing Group}\BibitemShut {NoStop}%
\bibitem [{\citenamefont {Jiao}\ \emph
  {et~al.}(2023{\natexlab{b}})\citenamefont {Jiao}, \citenamefont {Mueller},
  \citenamefont {Raney}, \citenamefont {Zheng},\ and\ \citenamefont
  {Alavi}}]{jiao_mechanical_2023-1}%
  \BibitemOpen
  \bibfield  {author} {\bibinfo {author} {\bibfnamefont {P.}~\bibnamefont
  {Jiao}}, \bibinfo {author} {\bibfnamefont {J.}~\bibnamefont {Mueller}},
  \bibinfo {author} {\bibfnamefont {J.~R.}\ \bibnamefont {Raney}}, \bibinfo
  {author} {\bibfnamefont {X.~R.}\ \bibnamefont {Zheng}},\ and\ \bibinfo
  {author} {\bibfnamefont {A.~H.}\ \bibnamefont {Alavi}},\ }\bibfield  {title}
  {{\selectlanguage {English}\bibinfo {title} {Mechanical metamaterials and
  beyond}},\ }\href {https://doi.org/10.1038/s41467-023-41679-8} {\bibfield
  {journal} {\bibinfo  {journal} {Nat Commun}\ }\textbf {\bibinfo {volume}
  {14}},\ \bibinfo {pages} {6004} (\bibinfo {year} {2023}{\natexlab{b}})},\
  \bibinfo {note} {publisher: Nature Publishing Group}\BibitemShut {NoStop}%
\bibitem [{\citenamefont {Adamatzky}\ \emph {et~al.}(2011)\citenamefont
  {Adamatzky}, \citenamefont {Kitson}, \citenamefont {Costello}, \citenamefont
  {Matranga},\ and\ \citenamefont {Younger}}]{adamatzky_computing_2011}%
  \BibitemOpen
  \bibfield  {author} {\bibinfo {author} {\bibfnamefont {A.}~\bibnamefont
  {Adamatzky}}, \bibinfo {author} {\bibfnamefont {S.}~\bibnamefont {Kitson}},
  \bibinfo {author} {\bibfnamefont {B.~D.~L.}\ \bibnamefont {Costello}},
  \bibinfo {author} {\bibfnamefont {M.~A.}\ \bibnamefont {Matranga}},\ and\
  \bibinfo {author} {\bibfnamefont {D.}~\bibnamefont {Younger}},\ }\bibfield
  {title} {{\selectlanguage {English}\bibinfo {title} {Computing with liquid
  crystal fingers: {Models} of geometric and logical computation}},\ }\href
  {https://doi.org/10.1103/PhysRevE.84.061702} {\bibfield  {journal} {\bibinfo
  {journal} {Phys. Rev. E}\ }\textbf {\bibinfo {volume} {84}},\ \bibinfo
  {pages} {061702} (\bibinfo {year} {2011})}\BibitemShut {NoStop}%
\bibitem [{\citenamefont {Wright}\ \emph {et~al.}(2022)\citenamefont {Wright},
  \citenamefont {Onodera}, \citenamefont {Stein}, \citenamefont {Wang},
  \citenamefont {Schachter}, \citenamefont {Hu},\ and\ \citenamefont
  {McMahon}}]{wright_deep_2022}%
  \BibitemOpen
  \bibfield  {author} {\bibinfo {author} {\bibfnamefont {L.~G.}\ \bibnamefont
  {Wright}}, \bibinfo {author} {\bibfnamefont {T.}~\bibnamefont {Onodera}},
  \bibinfo {author} {\bibfnamefont {M.~M.}\ \bibnamefont {Stein}}, \bibinfo
  {author} {\bibfnamefont {T.}~\bibnamefont {Wang}}, \bibinfo {author}
  {\bibfnamefont {D.~T.}\ \bibnamefont {Schachter}}, \bibinfo {author}
  {\bibfnamefont {Z.}~\bibnamefont {Hu}},\ and\ \bibinfo {author}
  {\bibfnamefont {P.~L.}\ \bibnamefont {McMahon}},\ }\bibfield  {title}
  {{\selectlanguage {English}\bibinfo {title} {Deep physical neural networks
  trained with backpropagation}},\ }\href
  {https://doi.org/10.1038/s41586-021-04223-6} {\bibfield  {journal} {\bibinfo
  {journal} {Nature}\ }\textbf {\bibinfo {volume} {601}},\ \bibinfo {pages}
  {549} (\bibinfo {year} {2022})},\ \bibinfo {note} {publisher: Nature
  Publishing Group}\BibitemShut {NoStop}%
\bibitem [{\citenamefont {Li}\ and\ \citenamefont
  {Mao}(2024)}]{li_training_2024}%
  \BibitemOpen
  \bibfield  {author} {\bibinfo {author} {\bibfnamefont {S.}~\bibnamefont
  {Li}}\ and\ \bibinfo {author} {\bibfnamefont {X.}~\bibnamefont {Mao}},\
  }\bibfield  {title} {{\selectlanguage {English}\bibinfo {title} {Training
  all-mechanical neural networks for task learning through in situ
  backpropagation}},\ }\href {https://doi.org/10.1038/s41467-024-54849-z}
  {\bibfield  {journal} {\bibinfo  {journal} {Nat Commun}\ }\textbf {\bibinfo
  {volume} {15}},\ \bibinfo {pages} {10528} (\bibinfo {year} {2024})},\
  \bibinfo {note} {publisher: Nature Publishing Group}\BibitemShut {NoStop}%
\bibitem [{\citenamefont {Fernando}\ and\ \citenamefont
  {Sojakka}(2003)}]{fernando_pattern_2003}%
  \BibitemOpen
  \bibfield  {author} {\bibinfo {author} {\bibfnamefont {C.}~\bibnamefont
  {Fernando}}\ and\ \bibinfo {author} {\bibfnamefont {S.}~\bibnamefont
  {Sojakka}},\ }\bibfield  {title} {{\selectlanguage {English}\bibinfo {title}
  {Pattern {Recognition} in a {Bucket}}},\ }in\ \href
  {https://doi.org/10.1007/978-3-540-39432-7_63} {{\selectlanguage
  {English}\emph {\bibinfo {booktitle} {Advances in {Artificial} {Life}}}}},\
  \bibinfo {editor} {edited by\ \bibinfo {editor} {\bibfnamefont
  {W.}~\bibnamefont {Banzhaf}}, \bibinfo {editor} {\bibfnamefont
  {J.}~\bibnamefont {Ziegler}}, \bibinfo {editor} {\bibfnamefont
  {T.}~\bibnamefont {Christaller}}, \bibinfo {editor} {\bibfnamefont
  {P.}~\bibnamefont {Dittrich}},\ and\ \bibinfo {editor} {\bibfnamefont
  {J.~T.}\ \bibnamefont {Kim}}}\ (\bibinfo  {publisher} {Springer},\ \bibinfo
  {address} {Berlin, Heidelberg},\ \bibinfo {year} {2003})\ pp.\ \bibinfo
  {pages} {588--597}\BibitemShut {NoStop}%
\bibitem [{\citenamefont {Nakajima}\ \emph {et~al.}(2015)\citenamefont
  {Nakajima}, \citenamefont {Hauser}, \citenamefont {Li},\ and\ \citenamefont
  {Pfeifer}}]{nakajima_information_2015}%
  \BibitemOpen
  \bibfield  {author} {\bibinfo {author} {\bibfnamefont {K.}~\bibnamefont
  {Nakajima}}, \bibinfo {author} {\bibfnamefont {H.}~\bibnamefont {Hauser}},
  \bibinfo {author} {\bibfnamefont {T.}~\bibnamefont {Li}},\ and\ \bibinfo
  {author} {\bibfnamefont {R.}~\bibnamefont {Pfeifer}},\ }\bibfield  {title}
  {{\selectlanguage {English}\bibinfo {title} {Information processing via
  physical soft body}},\ }\href {https://doi.org/10.1038/srep10487} {\bibfield
  {journal} {\bibinfo  {journal} {Sci Rep}\ }\textbf {\bibinfo {volume} {5}},\
  \bibinfo {pages} {10487} (\bibinfo {year} {2015})},\ \bibinfo {note}
  {publisher: Nature Publishing Group}\BibitemShut {NoStop}%
\bibitem [{\citenamefont {Hanczyc}\ and\ \citenamefont
  {Ikegami}(2010)}]{hanczyc_chemical_2010}%
  \BibitemOpen
  \bibfield  {author} {\bibinfo {author} {\bibfnamefont {M.~M.}\ \bibnamefont
  {Hanczyc}}\ and\ \bibinfo {author} {\bibfnamefont {T.}~\bibnamefont
  {Ikegami}},\ }\bibfield  {title} {\bibinfo {title} {Chemical {Basis} for
  {Minimal} {Cognition}},\ }\href {https://doi.org/10.1162/artl_a_00002}
  {\bibfield  {journal} {\bibinfo  {journal} {Artificial Life}\ }\textbf
  {\bibinfo {volume} {16}},\ \bibinfo {pages} {233} (\bibinfo {year}
  {2010})}\BibitemShut {NoStop}%
\bibitem [{\citenamefont {Čejková}\ \emph {et~al.}(2014)\citenamefont
  {Čejková}, \citenamefont {Novák}, \citenamefont {Štěpánek},\ and\
  \citenamefont {Hanczyc}}]{cejkova_dynamics_2014}%
  \BibitemOpen
  \bibfield  {author} {\bibinfo {author} {\bibfnamefont {J.}~\bibnamefont
  {Čejková}}, \bibinfo {author} {\bibfnamefont {M.}~\bibnamefont {Novák}},
  \bibinfo {author} {\bibfnamefont {F.}~\bibnamefont {Štěpánek}},\ and\
  \bibinfo {author} {\bibfnamefont {M.~M.}\ \bibnamefont {Hanczyc}},\
  }\bibfield  {title} {\bibinfo {title} {Dynamics of {Chemotactic} {Droplets}
  in {Salt} {Concentration} {Gradients}},\ }\href
  {https://doi.org/10.1021/la502624f} {\bibfield  {journal} {\bibinfo
  {journal} {Langmuir}\ }\textbf {\bibinfo {volume} {30}},\ \bibinfo {pages}
  {11937} (\bibinfo {year} {2014})},\ \bibinfo {note} {publisher: American
  Chemical Society}\BibitemShut {NoStop}%
\bibitem [{\citenamefont {Horibe}\ \emph {et~al.}(2011)\citenamefont {Horibe},
  \citenamefont {Hanczyc},\ and\ \citenamefont {Ikegami}}]{horibe_mode_2011}%
  \BibitemOpen
  \bibfield  {author} {\bibinfo {author} {\bibfnamefont {N.}~\bibnamefont
  {Horibe}}, \bibinfo {author} {\bibfnamefont {M.~M.}\ \bibnamefont
  {Hanczyc}},\ and\ \bibinfo {author} {\bibfnamefont {T.}~\bibnamefont
  {Ikegami}},\ }\bibfield  {title} {{\selectlanguage {English}\bibinfo {title}
  {Mode {Switching} and {Collective} {Behavior} in {Chemical} {Oil}
  {Droplets}}},\ }\href {https://doi.org/10.3390/e13030709} {\bibfield
  {journal} {\bibinfo  {journal} {Entropy}\ }\textbf {\bibinfo {volume} {13}},\
  \bibinfo {pages} {709} (\bibinfo {year} {2011})}\BibitemShut {NoStop}%
\bibitem [{\citenamefont {Füchslin}\ and\ \citenamefont
  {Flumini}(2022)}]{fuchslin_ai_2022}%
  \BibitemOpen
  \bibfield  {author} {\bibinfo {author} {\bibfnamefont {D.}~\bibnamefont
  {Füchslin}}\ and\ \bibinfo {author} {\bibfnamefont {R.~M.}\ \bibnamefont
  {Flumini}},\ }\bibfield  {title} {{\selectlanguage {English}\bibinfo {title}
  {{AI} as {Arational} {Intelligence}?}},\ }\bibfield  {journal} {\bibinfo
  {journal} {Archives of Data Science}\ }\bibinfo {series} {A},\ \textbf
  {\bibinfo {volume} {8}},\ \href {https://doi.org/10.5445/IR/1000150236}
  {10.5445/IR/1000150236} (\bibinfo {year} {2022}),\ \bibinfo {note}
  {publisher: Karlsruher Institut für Technologie (KIT)}\BibitemShut {NoStop}%
\bibitem [{\citenamefont {Huang}\ \emph {et~al.}(2018)\citenamefont {Huang},
  \citenamefont {Lei}, \citenamefont {He}, \citenamefont {Xu}, \citenamefont
  {Williams}, \citenamefont {Hu}, \citenamefont {McNeil}, \citenamefont {Ruso},
  \citenamefont {Liu}, \citenamefont {Guo},\ and\ \citenamefont
  {Wang}}]{huang_self-regulation_2018}%
  \BibitemOpen
  \bibfield  {author} {\bibinfo {author} {\bibfnamefont {Z.}~\bibnamefont
  {Huang}}, \bibinfo {author} {\bibfnamefont {K.}~\bibnamefont {Lei}}, \bibinfo
  {author} {\bibfnamefont {D.}~\bibnamefont {He}}, \bibinfo {author}
  {\bibfnamefont {Y.}~\bibnamefont {Xu}}, \bibinfo {author} {\bibfnamefont
  {J.}~\bibnamefont {Williams}}, \bibinfo {author} {\bibfnamefont
  {L.}~\bibnamefont {Hu}}, \bibinfo {author} {\bibfnamefont {M.}~\bibnamefont
  {McNeil}}, \bibinfo {author} {\bibfnamefont {J.~M.}\ \bibnamefont {Ruso}},
  \bibinfo {author} {\bibfnamefont {Z.}~\bibnamefont {Liu}}, \bibinfo {author}
  {\bibfnamefont {Z.}~\bibnamefont {Guo}},\ and\ \bibinfo {author}
  {\bibfnamefont {Z.}~\bibnamefont {Wang}},\ }\bibfield  {title} {\bibinfo
  {title} {Self-regulation in chemical and bio-engineering materials for
  intelligent systems},\ }\href {https://doi.org/10.1049/trit.2018.0004}
  {\bibfield  {journal} {\bibinfo  {journal} {CAAI Trans Intell Technol}\
  }\textbf {\bibinfo {volume} {3}},\ \bibinfo {pages} {40} (\bibinfo {year}
  {2018})}\BibitemShut {NoStop}%
\bibitem [{\citenamefont {Srinivasa}\ \emph {et~al.}(2015)\citenamefont
  {Srinivasa}, \citenamefont {Stepp},\ and\ \citenamefont
  {Cruz-Albrecht}}]{srinivasa_criticality_2015}%
  \BibitemOpen
  \bibfield  {author} {\bibinfo {author} {\bibfnamefont {N.}~\bibnamefont
  {Srinivasa}}, \bibinfo {author} {\bibfnamefont {N.~D.}\ \bibnamefont
  {Stepp}},\ and\ \bibinfo {author} {\bibfnamefont {J.}~\bibnamefont
  {Cruz-Albrecht}},\ }\bibfield  {title} {{\selectlanguage {English}\bibinfo
  {title} {Criticality as a {Set}-{Point} for {Adaptive} {Behavior} in
  {Neuromorphic} {Hardware}}},\ }\bibfield  {journal} {\bibinfo  {journal}
  {Front. Neurosci.}\ }\textbf {\bibinfo {volume} {9}},\ \href
  {https://doi.org/10.3389/fnins.2015.00449} {10.3389/fnins.2015.00449}
  (\bibinfo {year} {2015}),\ \bibinfo {note} {publisher: Frontiers}\BibitemShut
  {NoStop}%
\bibitem [{\citenamefont {Wadhams}\ and\ \citenamefont
  {Armitage}(2004)}]{wadhams_making_2004}%
  \BibitemOpen
  \bibfield  {author} {\bibinfo {author} {\bibfnamefont {G.~H.}\ \bibnamefont
  {Wadhams}}\ and\ \bibinfo {author} {\bibfnamefont {J.~P.}\ \bibnamefont
  {Armitage}},\ }\bibfield  {title} {{\selectlanguage {English}\bibinfo {title}
  {Making sense of it all: bacterial chemotaxis}},\ }\href
  {https://doi.org/10.1038/nrm1524} {\bibfield  {journal} {\bibinfo  {journal}
  {Nat Rev Mol Cell Biol}\ }\textbf {\bibinfo {volume} {5}},\ \bibinfo {pages}
  {1024} (\bibinfo {year} {2004})}\BibitemShut {NoStop}%
\bibitem [{\citenamefont {Füchslin}\ \emph
  {et~al.}(2013{\natexlab{a}})\citenamefont {Füchslin}, \citenamefont
  {Dzyakanchuk}, \citenamefont {Flumini}, \citenamefont {Hauser}, \citenamefont
  {Hunt}, \citenamefont {Luchsinger}, \citenamefont {Reller}, \citenamefont
  {Scheidegger},\ and\ \citenamefont {Walker}}]{fuchslin_morphological_2013}%
  \BibitemOpen
  \bibfield  {author} {\bibinfo {author} {\bibfnamefont {R.~M.}\ \bibnamefont
  {Füchslin}}, \bibinfo {author} {\bibfnamefont {A.}~\bibnamefont
  {Dzyakanchuk}}, \bibinfo {author} {\bibfnamefont {D.}~\bibnamefont
  {Flumini}}, \bibinfo {author} {\bibfnamefont {H.}~\bibnamefont {Hauser}},
  \bibinfo {author} {\bibfnamefont {K.~J.}\ \bibnamefont {Hunt}}, \bibinfo
  {author} {\bibfnamefont {R.~H.}\ \bibnamefont {Luchsinger}}, \bibinfo
  {author} {\bibfnamefont {B.}~\bibnamefont {Reller}}, \bibinfo {author}
  {\bibfnamefont {S.}~\bibnamefont {Scheidegger}},\ and\ \bibinfo {author}
  {\bibfnamefont {R.}~\bibnamefont {Walker}},\ }\bibfield  {title} {\bibinfo
  {title} {Morphological {Computation} and {Morphological} {Control}: {Steps}
  {Toward} a {Formal} {Theory} and {Applications}},\ }\href
  {https://doi.org/10.1162/ARTL_a_00079} {\bibfield  {journal} {\bibinfo
  {journal} {Artificial Life}\ }\textbf {\bibinfo {volume} {19}},\ \bibinfo
  {pages} {9} (\bibinfo {year} {2013}{\natexlab{a}})}\BibitemShut {NoStop}%
\bibitem [{\citenamefont {Copeland}(2024)}]{sep-church-turing}%
  \BibitemOpen
  \bibfield  {author} {\bibinfo {author} {\bibfnamefont {B.~J.}\ \bibnamefont
  {Copeland}},\ }\bibfield  {title} {\bibinfo {title} {{The Church-Turing
  Thesis}},\ }in\ \href@noop {} {\emph {\bibinfo {booktitle} {The {Stanford}
  Encyclopedia of Philosophy}}},\ \bibinfo {editor} {edited by\ \bibinfo
  {editor} {\bibfnamefont {E.~N.}\ \bibnamefont {Zalta}}\ and\ \bibinfo
  {editor} {\bibfnamefont {U.}~\bibnamefont {Nodelman}}}\ (\bibinfo
  {publisher} {Metaphysics Research Lab, Stanford University},\ \bibinfo {year}
  {2024})\ \bibinfo {edition} {{W}inter 2024}\ ed.\BibitemShut {Stop}%
\bibitem [{\citenamefont {Chomsky}(1959)}]{chomsky_certain_1959}%
  \BibitemOpen
  \bibfield  {author} {\bibinfo {author} {\bibfnamefont {N.}~\bibnamefont
  {Chomsky}},\ }\bibfield  {title} {{\selectlanguage {English}\bibinfo {title}
  {On certain formal properties of grammars}},\ }\href
  {https://doi.org/10.1016/s0019-9958(59)90362-6} {\bibfield  {journal}
  {\bibinfo  {journal} {Information and Control}\ }\textbf {\bibinfo {volume}
  {2}},\ \bibinfo {pages} {137} (\bibinfo {year} {1959})},\ \bibinfo {note}
  {publisher: Elsevier BV}\BibitemShut {NoStop}%
\bibitem [{\citenamefont {Ropohl}(2009)}]{ropohl_allgemeine_2009}%
  \BibitemOpen
  \bibfield  {author} {\bibinfo {author} {\bibfnamefont {G.}~\bibnamefont
  {Ropohl}},\ }\href@noop {} {{\selectlanguage {German}\emph {\bibinfo {title}
  {Allgemeine {Technologie}: eine {Systemtheorie} der {Technik}}}}}\ (\bibinfo
  {publisher} {KIT Scientific Publishing},\ \bibinfo {year} {2009})\ \bibinfo
  {note} {google-Books-ID: bZjToyzgRzIC}\BibitemShut {NoStop}%
\bibitem [{\citenamefont {Sitti}(2021)}]{sitti_physical_2021}%
  \BibitemOpen
  \bibfield  {author} {\bibinfo {author} {\bibfnamefont {M.}~\bibnamefont
  {Sitti}},\ }\bibfield  {title} {\bibinfo {title} {Physical intelligence as a
  new paradigm},\ }\href {https://doi.org/10.1016/j.eml.2021.101340} {\bibfield
   {journal} {\bibinfo  {journal} {Extreme Mechanics Letters}\ }\textbf
  {\bibinfo {volume} {46}},\ \bibinfo {pages} {101340} (\bibinfo {year}
  {2021})}\BibitemShut {NoStop}%
\bibitem [{\citenamefont {Soh}\ \emph {et~al.}(2010)\citenamefont {Soh},
  \citenamefont {Byrska}, \citenamefont {Kandere-Grzybowska},\ and\
  \citenamefont {Grzybowski}}]{soh2010reaction}%
  \BibitemOpen
  \bibfield  {author} {\bibinfo {author} {\bibfnamefont {S.}~\bibnamefont
  {Soh}}, \bibinfo {author} {\bibfnamefont {M.}~\bibnamefont {Byrska}},
  \bibinfo {author} {\bibfnamefont {K.}~\bibnamefont {Kandere-Grzybowska}},\
  and\ \bibinfo {author} {\bibfnamefont {B.~A.}\ \bibnamefont {Grzybowski}},\
  }\bibfield  {title} {\bibinfo {title} {Reaction-diffusion systems in
  intracellular molecular transport and control},\ }\href@noop {} {\bibfield
  {journal} {\bibinfo  {journal} {Angewandte Chemie International Edition}\
  }\textbf {\bibinfo {volume} {49}},\ \bibinfo {pages} {4170} (\bibinfo {year}
  {2010})}\BibitemShut {NoStop}%
\bibitem [{\citenamefont {Shaw}\ and\ \citenamefont
  {Packard}(2010)}]{shaw2010diffusion}%
  \BibitemOpen
  \bibfield  {author} {\bibinfo {author} {\bibfnamefont {R.}~\bibnamefont
  {Shaw}}\ and\ \bibinfo {author} {\bibfnamefont {N.}~\bibnamefont {Packard}},\
  }\bibfield  {title} {\bibinfo {title} {Diffusion of shapes},\ }\href@noop {}
  {\bibfield  {journal} {\bibinfo  {journal} {Artificial Life and Evolutionary
  Computation}\ ,\ \bibinfo {pages} {33}} (\bibinfo {year} {2010})}\BibitemShut
  {NoStop}%
\bibitem [{\citenamefont {Seo}\ and\ \citenamefont
  {Lee}(2022)}]{seo2022spatiotemporal}%
  \BibitemOpen
  \bibfield  {author} {\bibinfo {author} {\bibfnamefont {H.}~\bibnamefont
  {Seo}}\ and\ \bibinfo {author} {\bibfnamefont {H.}~\bibnamefont {Lee}},\
  }\bibfield  {title} {\bibinfo {title} {Spatiotemporal control of
  signal-driven enzymatic reaction in artificial cell-like polymersomes},\
  }\href@noop {} {\bibfield  {journal} {\bibinfo  {journal} {Nature
  communications}\ }\textbf {\bibinfo {volume} {13}},\ \bibinfo {pages} {5179}
  (\bibinfo {year} {2022})}\BibitemShut {NoStop}%
\bibitem [{\citenamefont {Boussard}\ \emph {et~al.}(2019)\citenamefont
  {Boussard}, \citenamefont {Delescluse}, \citenamefont {Pérez-Escudero},\
  and\ \citenamefont {Dussutour}}]{boussard_memory_2019}%
  \BibitemOpen
  \bibfield  {author} {\bibinfo {author} {\bibfnamefont {A.}~\bibnamefont
  {Boussard}}, \bibinfo {author} {\bibfnamefont {J.}~\bibnamefont
  {Delescluse}}, \bibinfo {author} {\bibfnamefont {A.}~\bibnamefont
  {Pérez-Escudero}},\ and\ \bibinfo {author} {\bibfnamefont {A.}~\bibnamefont
  {Dussutour}},\ }\bibfield  {title} {\bibinfo {title} {Memory inception and
  preservation in slime moulds: the quest for a common mechanism},\ }\href
  {https://doi.org/10.1098/rstb.2018.0368} {\bibfield  {journal} {\bibinfo
  {journal} {Philosophical Transactions of the Royal Society B: Biological
  Sciences}\ }\textbf {\bibinfo {volume} {374}},\ \bibinfo {pages} {20180368}
  (\bibinfo {year} {2019})},\ \bibinfo {note} {publisher: Royal
  Society}\BibitemShut {NoStop}%
\bibitem [{\citenamefont {Theraulaz}\ \emph {et~al.}(2002)\citenamefont
  {Theraulaz}, \citenamefont {Bonabeau}, \citenamefont {Nicolis}, \citenamefont
  {Solé}, \citenamefont {Fourcassié}, \citenamefont {Blanco}, \citenamefont
  {Fournier}, \citenamefont {Joly}, \citenamefont {Fernández}, \citenamefont
  {Grimal}, \citenamefont {Dalle},\ and\ \citenamefont
  {Deneubourg}}]{theraulaz_spatial_2002}%
  \BibitemOpen
  \bibfield  {author} {\bibinfo {author} {\bibfnamefont {G.}~\bibnamefont
  {Theraulaz}}, \bibinfo {author} {\bibfnamefont {E.}~\bibnamefont {Bonabeau}},
  \bibinfo {author} {\bibfnamefont {S.~C.}\ \bibnamefont {Nicolis}}, \bibinfo
  {author} {\bibfnamefont {R.~V.}\ \bibnamefont {Solé}}, \bibinfo {author}
  {\bibfnamefont {V.}~\bibnamefont {Fourcassié}}, \bibinfo {author}
  {\bibfnamefont {S.}~\bibnamefont {Blanco}}, \bibinfo {author} {\bibfnamefont
  {R.}~\bibnamefont {Fournier}}, \bibinfo {author} {\bibfnamefont {J.-L.}\
  \bibnamefont {Joly}}, \bibinfo {author} {\bibfnamefont {P.}~\bibnamefont
  {Fernández}}, \bibinfo {author} {\bibfnamefont {A.}~\bibnamefont {Grimal}},
  \bibinfo {author} {\bibfnamefont {P.}~\bibnamefont {Dalle}},\ and\ \bibinfo
  {author} {\bibfnamefont {J.-L.}\ \bibnamefont {Deneubourg}},\ }\bibfield
  {title} {\bibinfo {title} {Spatial patterns in ant colonies},\ }\href
  {https://doi.org/10.1073/pnas.152302199} {\bibfield  {journal} {\bibinfo
  {journal} {Proceedings of the National Academy of Sciences}\ }\textbf
  {\bibinfo {volume} {99}},\ \bibinfo {pages} {9645} (\bibinfo {year}
  {2002})},\ \bibinfo {note} {publisher: Proceedings of the National Academy of
  Sciences}\BibitemShut {NoStop}%
\bibitem [{\citenamefont {Zhang}\ \emph
  {et~al.}(2025{\natexlab{b}})\citenamefont {Zhang}, \citenamefont
  {Goldstein},\ and\ \citenamefont {Levin}}]{zhang_classical_2025}%
  \BibitemOpen
  \bibfield  {author} {\bibinfo {author} {\bibfnamefont {T.}~\bibnamefont
  {Zhang}}, \bibinfo {author} {\bibfnamefont {A.}~\bibnamefont {Goldstein}},\
  and\ \bibinfo {author} {\bibfnamefont {M.}~\bibnamefont {Levin}},\ }\bibfield
   {title} {{\selectlanguage {English}\bibinfo {title} {Classical sorting
  algorithms as a model of morphogenesis: {Self}-sorting arrays reveal
  unexpected competencies in a minimal model of basal intelligence}},\ }\href
  {https://doi.org/10.1177/10597123241269740} {\bibfield  {journal} {\bibinfo
  {journal} {Adaptive Behavior}\ }\textbf {\bibinfo {volume} {33}},\ \bibinfo
  {pages} {25} (\bibinfo {year} {2025}{\natexlab{b}})},\ \bibinfo {note}
  {publisher: SAGE Publications Ltd STM}\BibitemShut {NoStop}%
\bibitem [{\citenamefont {Baltussen}\ \emph {et~al.}(2024)\citenamefont
  {Baltussen}, \citenamefont {de~Jong}, \citenamefont {Duez}, \citenamefont
  {Robinson},\ and\ \citenamefont {Huck}}]{baltussen2024chemical}%
  \BibitemOpen
  \bibfield  {author} {\bibinfo {author} {\bibfnamefont {M.~G.}\ \bibnamefont
  {Baltussen}}, \bibinfo {author} {\bibfnamefont {T.~J.}\ \bibnamefont
  {de~Jong}}, \bibinfo {author} {\bibfnamefont {Q.}~\bibnamefont {Duez}},
  \bibinfo {author} {\bibfnamefont {W.~E.}\ \bibnamefont {Robinson}},\ and\
  \bibinfo {author} {\bibfnamefont {W.~T.}\ \bibnamefont {Huck}},\ }\bibfield
  {title} {\bibinfo {title} {Chemical reservoir computation in a
  self-organizing reaction network},\ }\href@noop {} {\bibfield  {journal}
  {\bibinfo  {journal} {Nature}\ }\textbf {\bibinfo {volume} {631}},\ \bibinfo
  {pages} {549} (\bibinfo {year} {2024})}\BibitemShut {NoStop}%
\bibitem [{\citenamefont {Serra}\ \emph {et~al.}(2010)\citenamefont {Serra},
  \citenamefont {Villani}, \citenamefont {Barbieri}, \citenamefont {Kauffman},\
  and\ \citenamefont {Colacci}}]{serra2010dynamics}%
  \BibitemOpen
  \bibfield  {author} {\bibinfo {author} {\bibfnamefont {R.}~\bibnamefont
  {Serra}}, \bibinfo {author} {\bibfnamefont {M.}~\bibnamefont {Villani}},
  \bibinfo {author} {\bibfnamefont {A.}~\bibnamefont {Barbieri}}, \bibinfo
  {author} {\bibfnamefont {S.~A.}\ \bibnamefont {Kauffman}},\ and\ \bibinfo
  {author} {\bibfnamefont {A.}~\bibnamefont {Colacci}},\ }\bibfield  {title}
  {\bibinfo {title} {On the dynamics of random boolean networks subject to
  noise: attractors, ergodic sets and cell types},\ }\href@noop {} {\bibfield
  {journal} {\bibinfo  {journal} {Journal of theoretical biology}\ }\textbf
  {\bibinfo {volume} {265}},\ \bibinfo {pages} {185} (\bibinfo {year}
  {2010})}\BibitemShut {NoStop}%
\bibitem [{\citenamefont {Bessi}\ \emph {et~al.}(2015)\citenamefont {Bessi},
  \citenamefont {Coletto}, \citenamefont {Davidescu}, \citenamefont {Scala},
  \citenamefont {Caldarelli},\ and\ \citenamefont
  {Quattrociocchi}}]{bessi2015science}%
  \BibitemOpen
  \bibfield  {author} {\bibinfo {author} {\bibfnamefont {A.}~\bibnamefont
  {Bessi}}, \bibinfo {author} {\bibfnamefont {M.}~\bibnamefont {Coletto}},
  \bibinfo {author} {\bibfnamefont {G.~A.}\ \bibnamefont {Davidescu}}, \bibinfo
  {author} {\bibfnamefont {A.}~\bibnamefont {Scala}}, \bibinfo {author}
  {\bibfnamefont {G.}~\bibnamefont {Caldarelli}},\ and\ \bibinfo {author}
  {\bibfnamefont {W.}~\bibnamefont {Quattrociocchi}},\ }\bibfield  {title}
  {\bibinfo {title} {Science vs conspiracy: Collective narratives in the age of
  misinformation},\ }\href@noop {} {\bibfield  {journal} {\bibinfo  {journal}
  {PloS one}\ }\textbf {\bibinfo {volume} {10}},\ \bibinfo {pages} {e0118093}
  (\bibinfo {year} {2015})}\BibitemShut {NoStop}%
\bibitem [{\citenamefont {Lynn}\ \emph {et~al.}(2020)\citenamefont {Lynn},
  \citenamefont {Papadopoulos}, \citenamefont {Kahn},\ and\ \citenamefont
  {Bassett}}]{lynn2020human}%
  \BibitemOpen
  \bibfield  {author} {\bibinfo {author} {\bibfnamefont {C.~W.}\ \bibnamefont
  {Lynn}}, \bibinfo {author} {\bibfnamefont {L.}~\bibnamefont {Papadopoulos}},
  \bibinfo {author} {\bibfnamefont {A.~E.}\ \bibnamefont {Kahn}},\ and\
  \bibinfo {author} {\bibfnamefont {D.~S.}\ \bibnamefont {Bassett}},\
  }\bibfield  {title} {\bibinfo {title} {Human information processing in
  complex networks},\ }\href@noop {} {\bibfield  {journal} {\bibinfo  {journal}
  {Nature Physics}\ }\textbf {\bibinfo {volume} {16}},\ \bibinfo {pages} {965}
  (\bibinfo {year} {2020})}\BibitemShut {NoStop}%
\bibitem [{\citenamefont {Bonamassa}\ \emph {et~al.}(2025)\citenamefont
  {Bonamassa}, \citenamefont {R{\'a}th}, \citenamefont {P{\'o}sfai},
  \citenamefont {Ab{\'e}rt}, \citenamefont {Keliger}, \citenamefont {Szegedy},
  \citenamefont {Kert{\'e}sz}, \citenamefont {Lov{\'a}sz},\ and\ \citenamefont
  {Barab{\'a}si}}]{bonamassa2025logarithmic}%
  \BibitemOpen
  \bibfield  {author} {\bibinfo {author} {\bibfnamefont {I.}~\bibnamefont
  {Bonamassa}}, \bibinfo {author} {\bibfnamefont {B.}~\bibnamefont {R{\'a}th}},
  \bibinfo {author} {\bibfnamefont {M.}~\bibnamefont {P{\'o}sfai}}, \bibinfo
  {author} {\bibfnamefont {M.}~\bibnamefont {Ab{\'e}rt}}, \bibinfo {author}
  {\bibfnamefont {D.}~\bibnamefont {Keliger}}, \bibinfo {author} {\bibfnamefont
  {B.}~\bibnamefont {Szegedy}}, \bibinfo {author} {\bibfnamefont
  {J.}~\bibnamefont {Kert{\'e}sz}}, \bibinfo {author} {\bibfnamefont
  {L.}~\bibnamefont {Lov{\'a}sz}},\ and\ \bibinfo {author} {\bibfnamefont
  {A.-L.}\ \bibnamefont {Barab{\'a}si}},\ }\bibfield  {title} {\bibinfo {title}
  {Logarithmic kinetics and bundling in random packings of elongated 3d
  physical links},\ }\href@noop {} {\bibfield  {journal} {\bibinfo  {journal}
  {Proceedings of the National Academy of Sciences}\ }\textbf {\bibinfo
  {volume} {122}},\ \bibinfo {pages} {e2427145122} (\bibinfo {year}
  {2025})}\BibitemShut {NoStop}%
\bibitem [{\citenamefont {Perelson}\ and\ \citenamefont
  {Weisbuch}(1997)}]{perelson1997immunology}%
  \BibitemOpen
  \bibfield  {author} {\bibinfo {author} {\bibfnamefont {A.~S.}\ \bibnamefont
  {Perelson}}\ and\ \bibinfo {author} {\bibfnamefont {G.}~\bibnamefont
  {Weisbuch}},\ }\bibfield  {title} {\bibinfo {title} {Immunology for
  physicists},\ }\href@noop {} {\bibfield  {journal} {\bibinfo  {journal}
  {Reviews of modern physics}\ }\textbf {\bibinfo {volume} {69}},\ \bibinfo
  {pages} {1219} (\bibinfo {year} {1997})}\BibitemShut {NoStop}%
\bibitem [{\citenamefont {Hershberg}\ and\ \citenamefont
  {Efroni}(2001)}]{hershberg2001immune}%
  \BibitemOpen
  \bibfield  {author} {\bibinfo {author} {\bibfnamefont {U.}~\bibnamefont
  {Hershberg}}\ and\ \bibinfo {author} {\bibfnamefont {S.}~\bibnamefont
  {Efroni}},\ }\bibfield  {title} {\bibinfo {title} {The immune system and
  other cognitive systems},\ }\href@noop {} {\bibfield  {journal} {\bibinfo
  {journal} {Complexity}\ }\textbf {\bibinfo {volume} {6}},\ \bibinfo {pages}
  {14} (\bibinfo {year} {2001})}\BibitemShut {NoStop}%
\bibitem [{\citenamefont {Matzinger}(2002)}]{matzinger2002danger}%
  \BibitemOpen
  \bibfield  {author} {\bibinfo {author} {\bibfnamefont {P.}~\bibnamefont
  {Matzinger}},\ }\bibfield  {title} {\bibinfo {title} {The danger model: a
  renewed sense of self},\ }\href@noop {} {\bibfield  {journal} {\bibinfo
  {journal} {science}\ }\textbf {\bibinfo {volume} {296}},\ \bibinfo {pages}
  {301} (\bibinfo {year} {2002})}\BibitemShut {NoStop}%
\bibitem [{\citenamefont {Füchslin}\ \emph
  {et~al.}(2013{\natexlab{b}})\citenamefont {Füchslin}, \citenamefont
  {Flumini}, \citenamefont {Hauser}, \citenamefont {Hunt}, \citenamefont
  {Luchsinger},\ and\ \citenamefont {Scheidegger}}]{HauserEbookMorphComp2014}%
  \BibitemOpen
  \bibfield  {author} {\bibinfo {author} {\bibfnamefont {R.~M.}\ \bibnamefont
  {Füchslin}}, \bibinfo {author} {\bibfnamefont {D.}~\bibnamefont {Flumini}},
  \bibinfo {author} {\bibfnamefont {H.}~\bibnamefont {Hauser}}, \bibinfo
  {author} {\bibfnamefont {K.~J.}\ \bibnamefont {Hunt}}, \bibinfo {author}
  {\bibfnamefont {R.~H.}\ \bibnamefont {Luchsinger}},\ and\ \bibinfo {author}
  {\bibfnamefont {S.}~\bibnamefont {Scheidegger}},\ }\bibfield  {title}
  {{\selectlanguage {English}\bibinfo {title} {Morphological control as guiding
  principle in physiology and medical applications}}\ }(\bibinfo {year}
  {2013})\BibitemShut {NoStop}%
\bibitem [{\citenamefont {Gurevich}\ and\ \citenamefont
  {Blass}(2022)}]{gurevich2022simple}%
  \BibitemOpen
  \bibfield  {author} {\bibinfo {author} {\bibfnamefont {Y.}~\bibnamefont
  {Gurevich}}\ and\ \bibinfo {author} {\bibfnamefont {A.}~\bibnamefont
  {Blass}},\ }\bibfield  {title} {\bibinfo {title} {Simple circuit simulations
  of classical and quantum turing machines},\ }\href@noop {} {\bibfield
  {journal} {\bibinfo  {journal} {Proceedings of the Royal Society A}\ }\textbf
  {\bibinfo {volume} {478}},\ \bibinfo {pages} {20210891} (\bibinfo {year}
  {2022})}\BibitemShut {NoStop}%
\bibitem [{\citenamefont {Rajak}\ \emph {et~al.}(2023)\citenamefont {Rajak},
  \citenamefont {Suzuki}, \citenamefont {Dutta},\ and\ \citenamefont
  {Chakrabarti}}]{rajak2023quantum}%
  \BibitemOpen
  \bibfield  {author} {\bibinfo {author} {\bibfnamefont {A.}~\bibnamefont
  {Rajak}}, \bibinfo {author} {\bibfnamefont {S.}~\bibnamefont {Suzuki}},
  \bibinfo {author} {\bibfnamefont {A.}~\bibnamefont {Dutta}},\ and\ \bibinfo
  {author} {\bibfnamefont {B.~K.}\ \bibnamefont {Chakrabarti}},\ }\bibfield
  {title} {\bibinfo {title} {Quantum annealing: An overview},\ }\href@noop {}
  {\bibfield  {journal} {\bibinfo  {journal} {Philosophical Transactions of the
  Royal Society A}\ }\textbf {\bibinfo {volume} {381}},\ \bibinfo {pages}
  {20210417} (\bibinfo {year} {2023})}\BibitemShut {NoStop}%
\bibitem [{\citenamefont {Domingos}(2015)}]{domingos2015master}%
  \BibitemOpen
  \bibfield  {author} {\bibinfo {author} {\bibfnamefont {P.}~\bibnamefont
  {Domingos}},\ }\href@noop {} {\emph {\bibinfo {title} {The master algorithm:
  How the quest for the ultimate learning machine will remake our world}}}\
  (\bibinfo  {publisher} {Basic Books},\ \bibinfo {year} {2015})\BibitemShut
  {NoStop}%
\bibitem [{\citenamefont {Wills}(2019)}]{wills_reflexivity_2019}%
  \BibitemOpen
  \bibfield  {author} {\bibinfo {author} {\bibfnamefont {P.~R.}\ \bibnamefont
  {Wills}},\ }\bibfield  {title} {\bibinfo {title} {Reflexivity, coding and
  quantum biology},\ }\href {https://doi.org/10.1016/j.biosystems.2019.104027}
  {\bibfield  {journal} {\bibinfo  {journal} {Biosystems}\ }\textbf {\bibinfo
  {volume} {185}},\ \bibinfo {pages} {104027} (\bibinfo {year}
  {2019})}\BibitemShut {NoStop}%
\bibitem [{\citenamefont {Kamsma}\ \emph {et~al.}(2023)\citenamefont {Kamsma},
  \citenamefont {Boon}, \citenamefont {ter Rele}, \citenamefont {Spitoni},\
  and\ \citenamefont {van Roij}}]{kamsma_iontronic_2023}%
  \BibitemOpen
  \bibfield  {author} {\bibinfo {author} {\bibfnamefont {T.}~\bibnamefont
  {Kamsma}}, \bibinfo {author} {\bibfnamefont {W.}~\bibnamefont {Boon}},
  \bibinfo {author} {\bibfnamefont {T.}~\bibnamefont {ter Rele}}, \bibinfo
  {author} {\bibfnamefont {C.}~\bibnamefont {Spitoni}},\ and\ \bibinfo {author}
  {\bibfnamefont {R.}~\bibnamefont {van Roij}},\ }\bibfield  {title} {\bibinfo
  {title} {Iontronic {Neuromorphic} {Signaling} with {Conical} {Microfluidic}
  {Memristors}},\ }\href {https://doi.org/10.1103/PhysRevLett.130.268401}
  {\bibfield  {journal} {\bibinfo  {journal} {Phys. Rev. Lett.}\ }\textbf
  {\bibinfo {volume} {130}},\ \bibinfo {pages} {268401} (\bibinfo {year}
  {2023})},\ \bibinfo {note} {publisher: American Physical Society}\BibitemShut
  {NoStop}%
\bibitem [{\citenamefont {Banda}\ \emph {et~al.}(2013)\citenamefont {Banda},
  \citenamefont {Teuscher},\ and\ \citenamefont {Lakin}}]{banda_online_2013}%
  \BibitemOpen
  \bibfield  {author} {\bibinfo {author} {\bibfnamefont {P.}~\bibnamefont
  {Banda}}, \bibinfo {author} {\bibfnamefont {C.}~\bibnamefont {Teuscher}},\
  and\ \bibinfo {author} {\bibfnamefont {M.~R.}\ \bibnamefont {Lakin}},\
  }\bibfield  {title} {\bibinfo {title} {Online {Learning} in a {Chemical}
  {Perceptron}},\ }\href {https://doi.org/10.1162/ARTL_a_00105} {\bibfield
  {journal} {\bibinfo  {journal} {Artificial Life}\ }\textbf {\bibinfo {volume}
  {19}},\ \bibinfo {pages} {195} (\bibinfo {year} {2013})}\BibitemShut
  {NoStop}%
\bibitem [{\citenamefont {England}(2015)}]{england_dissipative_2015}%
  \BibitemOpen
  \bibfield  {author} {\bibinfo {author} {\bibfnamefont {J.~L.}\ \bibnamefont
  {England}},\ }\bibfield  {title} {{\selectlanguage {English}\bibinfo {title}
  {Dissipative adaptation in driven self-assembly}},\ }\href
  {https://doi.org/10.1038/nnano.2015.250} {\bibfield  {journal} {\bibinfo
  {journal} {Nature Nanotech}\ }\textbf {\bibinfo {volume} {10}},\ \bibinfo
  {pages} {919} (\bibinfo {year} {2015})},\ \bibinfo {note} {publisher: Nature
  Publishing Group}\BibitemShut {NoStop}%
\bibitem [{\citenamefont {Goettems}(2024)}]{goettems_physics_2024}%
  \BibitemOpen
  \bibfield  {author} {\bibinfo {author} {\bibfnamefont {E.~I.}\ \bibnamefont
  {Goettems}},\ }{\selectlanguage {English}\emph {\bibinfo {title} {On the
  physics of dissipative systems: classical dynamics and quantum dissipative
  adaptation}}},\ \href
  {https://doi.org/10.11606/T.76.2024.tde-23042024-111403} {\bibinfo {type}
  {text}},\ \bibinfo  {school} {Universidade de São Paulo} (\bibinfo {year}
  {2024})\BibitemShut {NoStop}%
\bibitem [{\citenamefont {Rubenstein}\ \emph {et~al.}(2014)\citenamefont
  {Rubenstein}, \citenamefont {Cornejo},\ and\ \citenamefont
  {Nagpal}}]{rubenstein_programmable_2014}%
  \BibitemOpen
  \bibfield  {author} {\bibinfo {author} {\bibfnamefont {M.}~\bibnamefont
  {Rubenstein}}, \bibinfo {author} {\bibfnamefont {A.}~\bibnamefont
  {Cornejo}},\ and\ \bibinfo {author} {\bibfnamefont {R.}~\bibnamefont
  {Nagpal}},\ }\bibfield  {title} {\bibinfo {title} {Programmable self-assembly
  in a thousand-robot swarm},\ }\href {https://doi.org/10.1126/science.1254295}
  {\bibfield  {journal} {\bibinfo  {journal} {Science}\ }\textbf {\bibinfo
  {volume} {345}},\ \bibinfo {pages} {795} (\bibinfo {year} {2014})},\ \bibinfo
  {note} {publisher: American Association for the Advancement of
  Science}\BibitemShut {NoStop}%
\bibitem [{\citenamefont {Aprahamian}\ and\ \citenamefont
  {Goldup}(2023)}]{aprahamian_non-equilibrium_2023}%
  \BibitemOpen
  \bibfield  {author} {\bibinfo {author} {\bibfnamefont {I.}~\bibnamefont
  {Aprahamian}}\ and\ \bibinfo {author} {\bibfnamefont {S.~M.}\ \bibnamefont
  {Goldup}},\ }\bibfield  {title} {\bibinfo {title} {Non-equilibrium {Steady}
  {States} in {Catalysis}, {Molecular} {Motors}, and {Supramolecular}
  {Materials}: {Why} {Networks} and {Language} {Matter}},\ }\href
  {https://doi.org/10.1021/jacs.2c12665} {\bibfield  {journal} {\bibinfo
  {journal} {J. Am. Chem. Soc.}\ }\textbf {\bibinfo {volume} {145}},\ \bibinfo
  {pages} {14169} (\bibinfo {year} {2023})},\ \bibinfo {note} {publisher:
  American Chemical Society}\BibitemShut {NoStop}%
\bibitem [{\citenamefont {Baulin}\ \emph {et~al.}(2025)\citenamefont {Baulin},
  \citenamefont {Cook}, \citenamefont {Friedman}, \citenamefont {Lumiruusu},
  \citenamefont {Pashea}, \citenamefont {Rahman},\ and\ \citenamefont
  {Waldeck}}]{baulin_discovery_2025}%
  \BibitemOpen
  \bibfield  {author} {\bibinfo {author} {\bibfnamefont {V.}~\bibnamefont
  {Baulin}}, \bibinfo {author} {\bibfnamefont {A.}~\bibnamefont {Cook}},
  \bibinfo {author} {\bibfnamefont {D.}~\bibnamefont {Friedman}}, \bibinfo
  {author} {\bibfnamefont {J.}~\bibnamefont {Lumiruusu}}, \bibinfo {author}
  {\bibfnamefont {A.}~\bibnamefont {Pashea}}, \bibinfo {author} {\bibfnamefont
  {S.}~\bibnamefont {Rahman}},\ and\ \bibinfo {author} {\bibfnamefont
  {B.}~\bibnamefont {Waldeck}},\ }\href
  {https://doi.org/10.48550/arXiv.2505.17500} {\bibinfo {title} {The
  {Discovery} {Engine}: {A} {Framework} for {AI}-{Driven} {Synthesis} and
  {Navigation} of {Scientific} {Knowledge} {Landscapes}}} (\bibinfo {year}
  {2025}),\ \bibinfo {note} {arXiv:2505.17500 [cond-mat]}\BibitemShut {NoStop}%
\bibitem [{\citenamefont {Ziepke}\ \emph {et~al.}(2024)\citenamefont {Ziepke},
  \citenamefont {Maryshev}, \citenamefont {Aranson},\ and\ \citenamefont
  {Frey}}]{ziepke_acoustic_2024}%
  \BibitemOpen
  \bibfield  {author} {\bibinfo {author} {\bibfnamefont {A.}~\bibnamefont
  {Ziepke}}, \bibinfo {author} {\bibfnamefont {I.}~\bibnamefont {Maryshev}},
  \bibinfo {author} {\bibfnamefont {I.~S.}\ \bibnamefont {Aranson}},\ and\
  \bibinfo {author} {\bibfnamefont {E.}~\bibnamefont {Frey}},\ }\href
  {https://doi.org/10.48550/arXiv.2410.02940} {\bibinfo {title} {Acoustic
  signaling enables collective perception and control in active matter
  systems}} (\bibinfo {year} {2024}),\ \bibinfo {note} {arXiv:2410.02940
  [cond-mat]}\BibitemShut {NoStop}%
\bibitem [{\citenamefont {Yu}\ \emph {et~al.}(2020)\citenamefont {Yu},
  \citenamefont {Guo}, \citenamefont {Cui}, \citenamefont {Li}, \citenamefont
  {Ye}, \citenamefont {Kurokawa},\ and\ \citenamefont
  {Gong}}]{yu_hydrogels_2020}%
  \BibitemOpen
  \bibfield  {author} {\bibinfo {author} {\bibfnamefont {C.}~\bibnamefont
  {Yu}}, \bibinfo {author} {\bibfnamefont {H.}~\bibnamefont {Guo}}, \bibinfo
  {author} {\bibfnamefont {K.}~\bibnamefont {Cui}}, \bibinfo {author}
  {\bibfnamefont {X.}~\bibnamefont {Li}}, \bibinfo {author} {\bibfnamefont
  {Y.~N.}\ \bibnamefont {Ye}}, \bibinfo {author} {\bibfnamefont
  {T.}~\bibnamefont {Kurokawa}},\ and\ \bibinfo {author} {\bibfnamefont
  {J.~P.}\ \bibnamefont {Gong}},\ }\bibfield  {title} {{\selectlanguage
  {English}\bibinfo {title} {Hydrogels as dynamic memory with forgetting
  ability}},\ }\href {https://doi.org/10.1073/pnas.2006842117} {\bibfield
  {journal} {\bibinfo  {journal} {Proc. Natl. Acad. Sci. U.S.A.}\ }\textbf
  {\bibinfo {volume} {117}},\ \bibinfo {pages} {18962} (\bibinfo {year}
  {2020})}\BibitemShut {NoStop}%
\bibitem [{\citenamefont {Kamsma}\ \emph
  {et~al.}(2024{\natexlab{b}})\citenamefont {Kamsma}, \citenamefont {Klop},
  \citenamefont {Boon}, \citenamefont {Spitoni}, \citenamefont {Rueckauer},\
  and\ \citenamefont {Roij}}]{kamsma_chemically_2024}%
  \BibitemOpen
  \bibfield  {author} {\bibinfo {author} {\bibfnamefont {T.~M.}\ \bibnamefont
  {Kamsma}}, \bibinfo {author} {\bibfnamefont {M.~S.}\ \bibnamefont {Klop}},
  \bibinfo {author} {\bibfnamefont {W.~Q.}\ \bibnamefont {Boon}}, \bibinfo
  {author} {\bibfnamefont {C.}~\bibnamefont {Spitoni}}, \bibinfo {author}
  {\bibfnamefont {B.}~\bibnamefont {Rueckauer}},\ and\ \bibinfo {author}
  {\bibfnamefont {R.~v.}\ \bibnamefont {Roij}},\ }\href
  {https://doi.org/10.48550/arXiv.2406.03195} {\bibinfo {title} {Chemically
  {Regulated} {Conical} {Channel} {Synapse} for {Neuromorphic} and {Sensing}
  {Applications}}} (\bibinfo {year} {2024}{\natexlab{b}}),\ \bibinfo {note}
  {arXiv:2406.03195 [cond-mat]}\BibitemShut {NoStop}%
\bibitem [{\citenamefont {Langton}(1990)}]{langton_computation_1990}%
  \BibitemOpen
  \bibfield  {author} {\bibinfo {author} {\bibfnamefont {C.~G.}\ \bibnamefont
  {Langton}},\ }\bibfield  {title} {{\selectlanguage {English}\bibinfo {title}
  {Computation at the edge of chaos: {Phase} transitions and emergent
  computation}},\ }\href {https://doi.org/10.1016/0167-2789(90)90064-V}
  {\bibfield  {journal} {\bibinfo  {journal} {Physica D: Nonlinear Phenomena}\
  }\textbf {\bibinfo {volume} {42}},\ \bibinfo {pages} {12} (\bibinfo {year}
  {1990})}\BibitemShut {NoStop}%
\bibitem [{\citenamefont {Xi}\ \emph {et~al.}(2024)\citenamefont {Xi},
  \citenamefont {Jones}, \citenamefont {Huang}, \citenamefont {Marzin},\ and\
  \citenamefont {Brun}}]{xi_emergent_2024-1}%
  \BibitemOpen
  \bibfield  {author} {\bibinfo {author} {\bibfnamefont {Y.}~\bibnamefont
  {Xi}}, \bibinfo {author} {\bibfnamefont {T.~J.}\ \bibnamefont {Jones}},
  \bibinfo {author} {\bibfnamefont {R.}~\bibnamefont {Huang}}, \bibinfo
  {author} {\bibfnamefont {T.}~\bibnamefont {Marzin}},\ and\ \bibinfo {author}
  {\bibfnamefont {P.-T.}\ \bibnamefont {Brun}},\ }\href
  {https://doi.org/10.48550/arXiv.2404.10614} {\bibinfo {title} {Emergent
  intelligence of buckling-driven elasto-active structures}} (\bibinfo {year}
  {2024}),\ \bibinfo {note} {arXiv:2404.10614 [cond-mat]}\BibitemShut {NoStop}%
\bibitem [{\citenamefont {Solé}\ \emph {et~al.}(2024)\citenamefont {Solé},
  \citenamefont {Conde–Pueyo}, \citenamefont {Pla–Mauri}, \citenamefont
  {Garcia–Ojalvo}, \citenamefont {Montserrat},\ and\ \citenamefont
  {Levin}}]{sole_open_2024}%
  \BibitemOpen
  \bibfield  {author} {\bibinfo {author} {\bibfnamefont {R.}~\bibnamefont
  {Solé}}, \bibinfo {author} {\bibfnamefont {N.}~\bibnamefont
  {Conde–Pueyo}}, \bibinfo {author} {\bibfnamefont {J.}~\bibnamefont
  {Pla–Mauri}}, \bibinfo {author} {\bibfnamefont {J.}~\bibnamefont
  {Garcia–Ojalvo}}, \bibinfo {author} {\bibfnamefont {N.}~\bibnamefont
  {Montserrat}},\ and\ \bibinfo {author} {\bibfnamefont {M.}~\bibnamefont
  {Levin}},\ }\bibfield  {title} {{\selectlanguage {English}\bibinfo {title}
  {Open problems in synthetic multicellularity}},\ }\href
  {https://doi.org/10.1038/s41540-024-00477-8} {\bibfield  {journal} {\bibinfo
  {journal} {npj Syst Biol Appl}\ }\textbf {\bibinfo {volume} {10}},\ \bibinfo
  {pages} {1} (\bibinfo {year} {2024})},\ \bibinfo {note} {publisher: Nature
  Publishing Group}\BibitemShut {NoStop}%
\bibitem [{\citenamefont {Osat}\ and\ \citenamefont
  {Golestanian}(2023)}]{osat_non-reciprocal_2023}%
  \BibitemOpen
  \bibfield  {author} {\bibinfo {author} {\bibfnamefont {S.}~\bibnamefont
  {Osat}}\ and\ \bibinfo {author} {\bibfnamefont {R.}~\bibnamefont
  {Golestanian}},\ }\bibfield  {title} {{\selectlanguage {English}\bibinfo
  {title} {Non-reciprocal multifarious self-organization}},\ }\href
  {https://doi.org/10.1038/s41565-022-01258-2} {\bibfield  {journal} {\bibinfo
  {journal} {Nat. Nanotechnol.}\ }\textbf {\bibinfo {volume} {18}},\ \bibinfo
  {pages} {79} (\bibinfo {year} {2023})},\ \bibinfo {note} {publisher: Nature
  Publishing Group}\BibitemShut {NoStop}%
\bibitem [{\citenamefont {Kriegman}\ \emph {et~al.}(2020)\citenamefont
  {Kriegman}, \citenamefont {Blackiston}, \citenamefont {Levin},\ and\
  \citenamefont {Bongard}}]{kriegman_scalable_2020}%
  \BibitemOpen
  \bibfield  {author} {\bibinfo {author} {\bibfnamefont {S.}~\bibnamefont
  {Kriegman}}, \bibinfo {author} {\bibfnamefont {D.}~\bibnamefont
  {Blackiston}}, \bibinfo {author} {\bibfnamefont {M.}~\bibnamefont {Levin}},\
  and\ \bibinfo {author} {\bibfnamefont {J.}~\bibnamefont {Bongard}},\
  }\bibfield  {title} {\bibinfo {title} {A scalable pipeline for designing
  reconfigurable organisms},\ }\href {https://doi.org/10.1073/pnas.1910837117}
  {\bibfield  {journal} {\bibinfo  {journal} {Proceedings of the National
  Academy of Sciences}\ }\textbf {\bibinfo {volume} {117}},\ \bibinfo {pages}
  {1853} (\bibinfo {year} {2020})},\ \bibinfo {note} {publisher: Proceedings of
  the National Academy of Sciences}\BibitemShut {NoStop}%
\bibitem [{\citenamefont {Terryn}\ \emph {et~al.}(2021)\citenamefont {Terryn},
  \citenamefont {Langenbach}, \citenamefont {Roels}, \citenamefont {Brancart},
  \citenamefont {Bakkali-Hassani}, \citenamefont {Poutrel}, \citenamefont
  {Georgopoulou}, \citenamefont {George~Thuruthel}, \citenamefont {Safaei},
  \citenamefont {Ferrentino}, \citenamefont {Sebastian}, \citenamefont
  {Norvez}, \citenamefont {Iida}, \citenamefont {Bosman}, \citenamefont
  {Tournilhac}, \citenamefont {Clemens}, \citenamefont {Van~Assche},\ and\
  \citenamefont {Vanderborght}}]{terryn_review_2021}%
  \BibitemOpen
  \bibfield  {author} {\bibinfo {author} {\bibfnamefont {S.}~\bibnamefont
  {Terryn}}, \bibinfo {author} {\bibfnamefont {J.}~\bibnamefont {Langenbach}},
  \bibinfo {author} {\bibfnamefont {E.}~\bibnamefont {Roels}}, \bibinfo
  {author} {\bibfnamefont {J.}~\bibnamefont {Brancart}}, \bibinfo {author}
  {\bibfnamefont {C.}~\bibnamefont {Bakkali-Hassani}}, \bibinfo {author}
  {\bibfnamefont {Q.-A.}\ \bibnamefont {Poutrel}}, \bibinfo {author}
  {\bibfnamefont {A.}~\bibnamefont {Georgopoulou}}, \bibinfo {author}
  {\bibfnamefont {T.}~\bibnamefont {George~Thuruthel}}, \bibinfo {author}
  {\bibfnamefont {A.}~\bibnamefont {Safaei}}, \bibinfo {author} {\bibfnamefont
  {P.}~\bibnamefont {Ferrentino}}, \bibinfo {author} {\bibfnamefont
  {T.}~\bibnamefont {Sebastian}}, \bibinfo {author} {\bibfnamefont
  {S.}~\bibnamefont {Norvez}}, \bibinfo {author} {\bibfnamefont
  {F.}~\bibnamefont {Iida}}, \bibinfo {author} {\bibfnamefont {A.~W.}\
  \bibnamefont {Bosman}}, \bibinfo {author} {\bibfnamefont {F.}~\bibnamefont
  {Tournilhac}}, \bibinfo {author} {\bibfnamefont {F.}~\bibnamefont {Clemens}},
  \bibinfo {author} {\bibfnamefont {G.}~\bibnamefont {Van~Assche}},\ and\
  \bibinfo {author} {\bibfnamefont {B.}~\bibnamefont {Vanderborght}},\
  }\bibfield  {title} {\bibinfo {title} {A review on self-healing polymers for
  soft robotics},\ }\href {https://doi.org/10.1016/j.mattod.2021.01.009}
  {\bibfield  {journal} {\bibinfo  {journal} {Materials Today}\ }\textbf
  {\bibinfo {volume} {47}},\ \bibinfo {pages} {187} (\bibinfo {year}
  {2021})}\BibitemShut {NoStop}%
\bibitem [{\citenamefont {Marcucci}\ \emph {et~al.}(2020)\citenamefont
  {Marcucci}, \citenamefont {Pierangeli},\ and\ \citenamefont
  {Conti}}]{marcucci_theory_2020}%
  \BibitemOpen
  \bibfield  {author} {\bibinfo {author} {\bibfnamefont {G.}~\bibnamefont
  {Marcucci}}, \bibinfo {author} {\bibfnamefont {D.}~\bibnamefont
  {Pierangeli}},\ and\ \bibinfo {author} {\bibfnamefont {C.}~\bibnamefont
  {Conti}},\ }\bibfield  {title} {\bibinfo {title} {Theory of {Neuromorphic}
  {Computing} by {Waves}: {Machine} {Learning} by {Rogue} {Waves}, {Dispersive}
  {Shocks}, and {Solitons}},\ }\href
  {https://doi.org/10.1103/PhysRevLett.125.093901} {\bibfield  {journal}
  {\bibinfo  {journal} {Phys. Rev. Lett.}\ }\textbf {\bibinfo {volume} {125}},\
  \bibinfo {pages} {093901} (\bibinfo {year} {2020})},\ \bibinfo {note}
  {publisher: American Physical Society}\BibitemShut {NoStop}%
\bibitem [{\citenamefont {Lin}\ and\ \citenamefont
  {Keidar}(2021)}]{lin_intelligent_2021}%
  \BibitemOpen
  \bibfield  {author} {\bibinfo {author} {\bibfnamefont {L.}~\bibnamefont
  {Lin}}\ and\ \bibinfo {author} {\bibfnamefont {M.}~\bibnamefont {Keidar}},\
  }\href {https://doi.org/10.48550/arXiv.2109.02735} {\bibinfo {title} {An
  {Intelligent} {Material} with {Chemical} {Pathway} {Networks}}} (\bibinfo
  {year} {2021}),\ \bibinfo {note} {arXiv:2109.02735 [cs]}\BibitemShut
  {NoStop}%
\bibitem [{\citenamefont {Lee}\ \emph {et~al.}(2023)\citenamefont {Lee},
  \citenamefont {Kim}, \citenamefont {Kim}, \citenamefont {Kim}, \citenamefont
  {Park}, \citenamefont {So}, \citenamefont {Lee}, \citenamefont {Hwang},\ and\
  \citenamefont {Park}}]{lee_nanograin_2023}%
  \BibitemOpen
  \bibfield  {author} {\bibinfo {author} {\bibfnamefont {H.-C.}\ \bibnamefont
  {Lee}}, \bibinfo {author} {\bibfnamefont {J.}~\bibnamefont {Kim}}, \bibinfo
  {author} {\bibfnamefont {H.-R.}\ \bibnamefont {Kim}}, \bibinfo {author}
  {\bibfnamefont {K.-H.}\ \bibnamefont {Kim}}, \bibinfo {author} {\bibfnamefont
  {K.-J.}\ \bibnamefont {Park}}, \bibinfo {author} {\bibfnamefont {J.-P.}\
  \bibnamefont {So}}, \bibinfo {author} {\bibfnamefont {J.~M.}\ \bibnamefont
  {Lee}}, \bibinfo {author} {\bibfnamefont {M.-S.}\ \bibnamefont {Hwang}},\
  and\ \bibinfo {author} {\bibfnamefont {H.-G.}\ \bibnamefont {Park}},\
  }\bibfield  {title} {{\selectlanguage {English}\bibinfo {title} {Nanograin
  network memory with reconfigurable percolation paths for synaptic
  interactions}},\ }\href {https://doi.org/10.1038/s41377-023-01168-5}
  {\bibfield  {journal} {\bibinfo  {journal} {Light Sci Appl}\ }\textbf
  {\bibinfo {volume} {12}},\ \bibinfo {pages} {118} (\bibinfo {year}
  {2023})}\BibitemShut {NoStop}%
\bibitem [{\citenamefont {Negi}\ \emph {et~al.}(2022)\citenamefont {Negi},
  \citenamefont {Winkler},\ and\ \citenamefont {Gompper}}]{negi_emergent_2022}%
  \BibitemOpen
  \bibfield  {author} {\bibinfo {author} {\bibfnamefont {R.~S.}\ \bibnamefont
  {Negi}}, \bibinfo {author} {\bibfnamefont {R.~G.}\ \bibnamefont {Winkler}},\
  and\ \bibinfo {author} {\bibfnamefont {G.}~\bibnamefont {Gompper}},\
  }\bibfield  {title} {{\selectlanguage {English}\bibinfo {title} {Emergent
  collective behavior of active {Brownian} particles with visual perception}},\
  }\href {https://doi.org/10.1039/D2SM00736C} {\bibfield  {journal} {\bibinfo
  {journal} {Soft Matter}\ }\textbf {\bibinfo {volume} {18}},\ \bibinfo {pages}
  {6167} (\bibinfo {year} {2022})}\BibitemShut {NoStop}%
\bibitem [{\citenamefont {Son}\ \emph {et~al.}(2024)\citenamefont {Son},
  \citenamefont {Bowal}, \citenamefont {Mahadevan},\ and\ \citenamefont
  {Kim}}]{son_emergent_2024}%
  \BibitemOpen
  \bibfield  {author} {\bibinfo {author} {\bibfnamefont {K.}~\bibnamefont
  {Son}}, \bibinfo {author} {\bibfnamefont {K.}~\bibnamefont {Bowal}}, \bibinfo
  {author} {\bibfnamefont {L.}~\bibnamefont {Mahadevan}},\ and\ \bibinfo
  {author} {\bibfnamefont {H.-Y.}\ \bibnamefont {Kim}},\ }\href
  {https://doi.org/10.48550/arXiv.2411.08163} {\bibinfo {title} {Emergent
  functional dynamics of link-bots}} (\bibinfo {year} {2024}),\ \bibinfo {note}
  {arXiv:2411.08163 [cond-mat]}\BibitemShut {NoStop}%
\bibitem [{\citenamefont {Müller}\ and\ \citenamefont
  {Hoffmann}(2017)}]{muller_what_2017-1}%
  \BibitemOpen
  \bibfield  {author} {\bibinfo {author} {\bibfnamefont {V.~C.}\ \bibnamefont
  {Müller}}\ and\ \bibinfo {author} {\bibfnamefont {M.}~\bibnamefont
  {Hoffmann}},\ }\bibfield  {title} {\bibinfo {title} {What {Is}
  {Morphological} {Computation}? {On} {How} the {Body} {Contributes} to
  {Cognition} and {Control}},\ }\href {https://doi.org/10.1162/ARTL_a_00219}
  {\bibfield  {journal} {\bibinfo  {journal} {Artificial Life}\ }\textbf
  {\bibinfo {volume} {23}},\ \bibinfo {pages} {1} (\bibinfo {year}
  {2017})}\BibitemShut {NoStop}%
\bibitem [{\citenamefont {Zhang}\ \emph {et~al.}(2022)\citenamefont {Zhang},
  \citenamefont {Mozaffari},\ and\ \citenamefont
  {De~Pablo}}]{zhang_logic_2022}%
  \BibitemOpen
  \bibfield  {author} {\bibinfo {author} {\bibfnamefont {R.}~\bibnamefont
  {Zhang}}, \bibinfo {author} {\bibfnamefont {A.}~\bibnamefont {Mozaffari}},\
  and\ \bibinfo {author} {\bibfnamefont {J.~J.}\ \bibnamefont {De~Pablo}},\
  }\bibfield  {title} {{\selectlanguage {English}\bibinfo {title} {Logic
  operations with active topological defects}},\ }\href
  {https://doi.org/10.1126/sciadv.abg9060} {\bibfield  {journal} {\bibinfo
  {journal} {Sci. Adv.}\ }\textbf {\bibinfo {volume} {8}},\ \bibinfo {pages}
  {eabg9060} (\bibinfo {year} {2022})}\BibitemShut {NoStop}%
\bibitem [{\citenamefont {Louvet}\ \emph {et~al.}(2024)\citenamefont {Louvet},
  \citenamefont {Omidvar},\ and\ \citenamefont
  {Serra-Garcia}}]{louvet_reprogrammable_2024}%
  \BibitemOpen
  \bibfield  {author} {\bibinfo {author} {\bibfnamefont {T.}~\bibnamefont
  {Louvet}}, \bibinfo {author} {\bibfnamefont {P.}~\bibnamefont {Omidvar}},\
  and\ \bibinfo {author} {\bibfnamefont {M.}~\bibnamefont {Serra-Garcia}},\
  }\href {https://doi.org/10.48550/arXiv.2409.20425} {\bibinfo {title}
  {Reprogrammable, in-materia matrix-vector multiplication with floppy modes}}
  (\bibinfo {year} {2024}),\ \bibinfo {note} {arXiv:2409.20425
  [cond-mat]}\BibitemShut {NoStop}%
\bibitem [{\citenamefont {Stern}\ \emph {et~al.}(2024)\citenamefont {Stern},
  \citenamefont {Dillavou}, \citenamefont {Jayaraman}, \citenamefont {Durian},\
  and\ \citenamefont {Liu}}]{stern_training_2024}%
  \BibitemOpen
  \bibfield  {author} {\bibinfo {author} {\bibfnamefont {M.}~\bibnamefont
  {Stern}}, \bibinfo {author} {\bibfnamefont {S.}~\bibnamefont {Dillavou}},
  \bibinfo {author} {\bibfnamefont {D.}~\bibnamefont {Jayaraman}}, \bibinfo
  {author} {\bibfnamefont {D.~J.}\ \bibnamefont {Durian}},\ and\ \bibinfo
  {author} {\bibfnamefont {A.~J.}\ \bibnamefont {Liu}},\ }\bibfield  {title}
  {{\selectlanguage {English}\bibinfo {title} {Training self-learning circuits
  for power-efficient solutions}},\ }\href {https://doi.org/10.1063/5.0181382}
  {\bibfield  {journal} {\bibinfo  {journal} {APL Machine Learning}\ }\textbf
  {\bibinfo {volume} {2}},\ \bibinfo {pages} {016114} (\bibinfo {year}
  {2024})}\BibitemShut {NoStop}%
\bibitem [{\citenamefont {Kim}\ \emph {et~al.}(2020)\citenamefont {Kim},
  \citenamefont {Kim}, \citenamefont {Seo}, \citenamefont {Park}, \citenamefont
  {Song}, \citenamefont {Choi}, \citenamefont {Kim}, \citenamefont {Cha},
  \citenamefont {Park},\ and\ \citenamefont
  {Nam}}]{kim_nanoparticle-based_2020}%
  \BibitemOpen
  \bibfield  {author} {\bibinfo {author} {\bibfnamefont {S.}~\bibnamefont
  {Kim}}, \bibinfo {author} {\bibfnamefont {N.}~\bibnamefont {Kim}}, \bibinfo
  {author} {\bibfnamefont {J.}~\bibnamefont {Seo}}, \bibinfo {author}
  {\bibfnamefont {J.-E.}\ \bibnamefont {Park}}, \bibinfo {author}
  {\bibfnamefont {E.~H.}\ \bibnamefont {Song}}, \bibinfo {author}
  {\bibfnamefont {S.~Y.}\ \bibnamefont {Choi}}, \bibinfo {author}
  {\bibfnamefont {J.~E.}\ \bibnamefont {Kim}}, \bibinfo {author} {\bibfnamefont
  {S.}~\bibnamefont {Cha}}, \bibinfo {author} {\bibfnamefont {H.~H.}\
  \bibnamefont {Park}},\ and\ \bibinfo {author} {\bibfnamefont {J.-M.}\
  \bibnamefont {Nam}},\ }\bibfield  {title} {{\selectlanguage {English}\bibinfo
  {title} {Nanoparticle-based computing architecture for nanoparticle neural
  networks}},\ }\href {https://doi.org/10.1126/sciadv.abb3348} {\bibfield
  {journal} {\bibinfo  {journal} {Sci. Adv.}\ }\textbf {\bibinfo {volume}
  {6}},\ \bibinfo {pages} {eabb3348} (\bibinfo {year} {2020})}\BibitemShut
  {NoStop}%
\bibitem [{\citenamefont {Govern}\ and\ \citenamefont
  {Ten~Wolde}(2014)}]{govern_optimal_2014}%
  \BibitemOpen
  \bibfield  {author} {\bibinfo {author} {\bibfnamefont {C.~C.}\ \bibnamefont
  {Govern}}\ and\ \bibinfo {author} {\bibfnamefont {P.~R.}\ \bibnamefont
  {Ten~Wolde}},\ }\bibfield  {title} {{\selectlanguage {English}\bibinfo
  {title} {Optimal resource allocation in cellular sensing systems}},\ }\href
  {https://doi.org/10.1073/pnas.1411524111} {\bibfield  {journal} {\bibinfo
  {journal} {Proc. Natl. Acad. Sci. U.S.A.}\ }\textbf {\bibinfo {volume}
  {111}},\ \bibinfo {pages} {17486} (\bibinfo {year} {2014})}\BibitemShut
  {NoStop}%
\bibitem [{\citenamefont {O’Byrne}\ and\ \citenamefont
  {Jerbi}(2022)}]{obyrne_how_2022}%
  \BibitemOpen
  \bibfield  {author} {\bibinfo {author} {\bibfnamefont {J.}~\bibnamefont
  {O’Byrne}}\ and\ \bibinfo {author} {\bibfnamefont {K.}~\bibnamefont
  {Jerbi}},\ }\bibfield  {title} {\bibinfo {title} {How critical is brain
  criticality?},\ }\href {https://doi.org/10.1016/j.tins.2022.08.007}
  {\bibfield  {journal} {\bibinfo  {journal} {Trends in Neurosciences}\
  }\textbf {\bibinfo {volume} {45}},\ \bibinfo {pages} {820} (\bibinfo {year}
  {2022})}\BibitemShut {NoStop}%
\bibitem [{\citenamefont {Zhang}\ \emph {et~al.}(2021)\citenamefont {Zhang},
  \citenamefont {Feng}, \citenamefont {Chen},\ and\ \citenamefont
  {Lai}}]{zhang_edge_2021}%
  \BibitemOpen
  \bibfield  {author} {\bibinfo {author} {\bibfnamefont {L.}~\bibnamefont
  {Zhang}}, \bibinfo {author} {\bibfnamefont {L.}~\bibnamefont {Feng}},
  \bibinfo {author} {\bibfnamefont {K.}~\bibnamefont {Chen}},\ and\ \bibinfo
  {author} {\bibfnamefont {C.~H.}\ \bibnamefont {Lai}},\ }\href
  {https://doi.org/10.48550/arXiv.2107.09437} {\bibinfo {title} {Edge of chaos
  as a guiding principle for modern neural network training}} (\bibinfo {year}
  {2021}),\ \bibinfo {note} {arXiv:2107.09437 [cs]}\BibitemShut {NoStop}%
\bibitem [{\citenamefont
  {Levin}(2024{\natexlab{b}})}]{levin_self-improvising_2024}%
  \BibitemOpen
  \bibfield  {author} {\bibinfo {author} {\bibfnamefont {M.}~\bibnamefont
  {Levin}},\ }\bibfield  {title} {{\selectlanguage {English}\bibinfo {title}
  {Self-{Improvising} {Memory}: {A} {Perspective} on {Memories} as {Agential},
  {Dynamically} {Reinterpreting} {Cognitive} {Glue}}},\ }\href
  {https://doi.org/10.3390/e26060481} {\bibfield  {journal} {\bibinfo
  {journal} {Entropy}\ }\textbf {\bibinfo {volume} {26}},\ \bibinfo {pages}
  {481} (\bibinfo {year} {2024}{\natexlab{b}})},\ \bibinfo {note} {number: 6
  Publisher: Multidisciplinary Digital Publishing Institute}\BibitemShut
  {NoStop}%
\bibitem [{\citenamefont {Hadorn}\ \emph {et~al.}(2012)\citenamefont {Hadorn},
  \citenamefont {Boenzli}, \citenamefont {Sørensen}, \citenamefont
  {Fellermann}, \citenamefont {Eggenberger~Hotz},\ and\ \citenamefont
  {Hanczyc}}]{hadorn_specific_2012}%
  \BibitemOpen
  \bibfield  {author} {\bibinfo {author} {\bibfnamefont {M.}~\bibnamefont
  {Hadorn}}, \bibinfo {author} {\bibfnamefont {E.}~\bibnamefont {Boenzli}},
  \bibinfo {author} {\bibfnamefont {K.~T.}\ \bibnamefont {Sørensen}}, \bibinfo
  {author} {\bibfnamefont {H.}~\bibnamefont {Fellermann}}, \bibinfo {author}
  {\bibfnamefont {P.}~\bibnamefont {Eggenberger~Hotz}},\ and\ \bibinfo {author}
  {\bibfnamefont {M.~M.}\ \bibnamefont {Hanczyc}},\ }\bibfield  {title}
  {\bibinfo {title} {Specific and reversible {DNA}-directed self-assembly of
  oil-in-water emulsion droplets},\ }\href
  {https://doi.org/10.1073/pnas.1214386109} {\bibfield  {journal} {\bibinfo
  {journal} {Proceedings of the National Academy of Sciences}\ }\textbf
  {\bibinfo {volume} {109}},\ \bibinfo {pages} {20320} (\bibinfo {year}
  {2012})},\ \bibinfo {note} {publisher: Proceedings of the National Academy of
  Sciences}\BibitemShut {NoStop}%
\bibitem [{\citenamefont {Kimura}\ \emph {et~al.}(2021)\citenamefont {Kimura},
  \citenamefont {Sumida}, \citenamefont {Kurasaki}, \citenamefont {Imai},
  \citenamefont {Takishita},\ and\ \citenamefont
  {Nakashima}}]{kimura2021amorphous}%
  \BibitemOpen
  \bibfield  {author} {\bibinfo {author} {\bibfnamefont {M.}~\bibnamefont
  {Kimura}}, \bibinfo {author} {\bibfnamefont {R.}~\bibnamefont {Sumida}},
  \bibinfo {author} {\bibfnamefont {A.}~\bibnamefont {Kurasaki}}, \bibinfo
  {author} {\bibfnamefont {T.}~\bibnamefont {Imai}}, \bibinfo {author}
  {\bibfnamefont {Y.}~\bibnamefont {Takishita}},\ and\ \bibinfo {author}
  {\bibfnamefont {Y.}~\bibnamefont {Nakashima}},\ }\bibfield  {title} {\bibinfo
  {title} {Amorphous metal oxide semiconductor thin film, analog memristor, and
  autonomous local learning for neuromorphic systems},\ }\href@noop {}
  {\bibfield  {journal} {\bibinfo  {journal} {Scientific reports}\ }\textbf
  {\bibinfo {volume} {11}},\ \bibinfo {pages} {580} (\bibinfo {year}
  {2021})}\BibitemShut {NoStop}%
\bibitem [{\citenamefont {Boniface}\ \emph {et~al.}(2019)\citenamefont
  {Boniface}, \citenamefont {Cottin-Bizonne}, \citenamefont {Kervil},
  \citenamefont {Ybert},\ and\ \citenamefont
  {Detcheverry}}]{boniface_self-propulsion_2019}%
  \BibitemOpen
  \bibfield  {author} {\bibinfo {author} {\bibfnamefont {D.}~\bibnamefont
  {Boniface}}, \bibinfo {author} {\bibfnamefont {C.}~\bibnamefont
  {Cottin-Bizonne}}, \bibinfo {author} {\bibfnamefont {R.}~\bibnamefont
  {Kervil}}, \bibinfo {author} {\bibfnamefont {C.}~\bibnamefont {Ybert}},\ and\
  \bibinfo {author} {\bibfnamefont {F.}~\bibnamefont {Detcheverry}},\
  }\bibfield  {title} {{\selectlanguage {English}\bibinfo {title}
  {Self-propulsion of symmetric chemically active particles: {Point}-source
  model and experiments on camphor disks}},\ }\href
  {https://doi.org/10.1103/PhysRevE.99.062605} {\bibfield  {journal} {\bibinfo
  {journal} {Phys. Rev. E}\ }\textbf {\bibinfo {volume} {99}},\ \bibinfo
  {pages} {062605} (\bibinfo {year} {2019})}\BibitemShut {NoStop}%
\bibitem [{\citenamefont {Huber}\ \emph {et~al.}(2013)\citenamefont {Huber},
  \citenamefont {Schnauß}, \citenamefont {Rönicke}, \citenamefont {Rauch},
  \citenamefont {Müller}, \citenamefont {Fütterer},\ and\ \citenamefont
  {Käs}}]{huber_emergent_2013}%
  \BibitemOpen
  \bibfield  {author} {\bibinfo {author} {\bibfnamefont {F.}~\bibnamefont
  {Huber}}, \bibinfo {author} {\bibfnamefont {J.}~\bibnamefont {Schnauß}},
  \bibinfo {author} {\bibfnamefont {S.}~\bibnamefont {Rönicke}}, \bibinfo
  {author} {\bibfnamefont {P.}~\bibnamefont {Rauch}}, \bibinfo {author}
  {\bibfnamefont {K.}~\bibnamefont {Müller}}, \bibinfo {author} {\bibfnamefont
  {C.}~\bibnamefont {Fütterer}},\ and\ \bibinfo {author} {\bibfnamefont
  {J.}~\bibnamefont {Käs}},\ }\bibfield  {title} {\bibinfo {title} {Emergent
  complexity of the cytoskeleton: from single filaments to tissue},\ }\href
  {https://doi.org/10.1080/00018732.2013.771509} {\bibfield  {journal}
  {\bibinfo  {journal} {Adv Phys}\ }\textbf {\bibinfo {volume} {62}},\ \bibinfo
  {pages} {1} (\bibinfo {year} {2013})}\BibitemShut {NoStop}%
\bibitem [{\citenamefont {Feng}\ \emph {et~al.}(2020)\citenamefont {Feng},
  \citenamefont {Zhang},\ and\ \citenamefont {Lai}}]{feng_optimal_2020}%
  \BibitemOpen
  \bibfield  {author} {\bibinfo {author} {\bibfnamefont {L.}~\bibnamefont
  {Feng}}, \bibinfo {author} {\bibfnamefont {L.}~\bibnamefont {Zhang}},\ and\
  \bibinfo {author} {\bibfnamefont {C.~H.}\ \bibnamefont {Lai}},\ }\href
  {https://doi.org/10.48550/arXiv.1909.05176} {\bibinfo {title} {Optimal
  {Machine} {Intelligence} at the {Edge} of {Chaos}}} (\bibinfo {year}
  {2020}),\ \bibinfo {note} {arXiv:1909.05176 [cs]}\BibitemShut {NoStop}%
\bibitem [{\citenamefont {Cavagna}\ \emph {et~al.}(2010)\citenamefont
  {Cavagna}, \citenamefont {Cimarelli}, \citenamefont {Giardina}, \citenamefont
  {Parisi}, \citenamefont {Santagati}, \citenamefont {Stefanini},\ and\
  \citenamefont {Viale}}]{cavagna_scale-free_2010}%
  \BibitemOpen
  \bibfield  {author} {\bibinfo {author} {\bibfnamefont {A.}~\bibnamefont
  {Cavagna}}, \bibinfo {author} {\bibfnamefont {A.}~\bibnamefont {Cimarelli}},
  \bibinfo {author} {\bibfnamefont {I.}~\bibnamefont {Giardina}}, \bibinfo
  {author} {\bibfnamefont {G.}~\bibnamefont {Parisi}}, \bibinfo {author}
  {\bibfnamefont {R.}~\bibnamefont {Santagati}}, \bibinfo {author}
  {\bibfnamefont {F.}~\bibnamefont {Stefanini}},\ and\ \bibinfo {author}
  {\bibfnamefont {M.}~\bibnamefont {Viale}},\ }\bibfield  {title} {\bibinfo
  {title} {Scale-free correlations in starling flocks},\ }\href
  {https://doi.org/10.1073/pnas.1005766107} {\bibfield  {journal} {\bibinfo
  {journal} {Proceedings of the National Academy of Sciences}\ }\textbf
  {\bibinfo {volume} {107}},\ \bibinfo {pages} {11865} (\bibinfo {year}
  {2010})},\ \bibinfo {note} {publisher: Proceedings of the National Academy of
  Sciences}\BibitemShut {NoStop}%
\bibitem [{\citenamefont {Puy}\ \emph {et~al.}(2024)\citenamefont {Puy},
  \citenamefont {Gimeno}, \citenamefont {March-Pons}, \citenamefont {Miguel},\
  and\ \citenamefont {Pastor-Satorras}}]{puy_signatures_2024}%
  \BibitemOpen
  \bibfield  {author} {\bibinfo {author} {\bibfnamefont {A.}~\bibnamefont
  {Puy}}, \bibinfo {author} {\bibfnamefont {E.}~\bibnamefont {Gimeno}},
  \bibinfo {author} {\bibfnamefont {D.}~\bibnamefont {March-Pons}}, \bibinfo
  {author} {\bibfnamefont {M.~C.}\ \bibnamefont {Miguel}},\ and\ \bibinfo
  {author} {\bibfnamefont {R.}~\bibnamefont {Pastor-Satorras}},\ }\bibfield
  {title} {{\selectlanguage {English}\bibinfo {title} {Signatures of
  criticality in turning avalanches of schooling fish}},\ }\href
  {https://doi.org/10.1103/PhysRevResearch.6.033270} {\bibfield  {journal}
  {\bibinfo  {journal} {Phys. Rev. Research}\ }\textbf {\bibinfo {volume}
  {6}},\ \bibinfo {pages} {033270} (\bibinfo {year} {2024})}\BibitemShut
  {NoStop}%
\bibitem [{\citenamefont {Seifert}\ \emph {et~al.}(2024)\citenamefont
  {Seifert}, \citenamefont {Sealander}, \citenamefont {Marzen},\ and\
  \citenamefont {Levin}}]{seifert_reinforcement_2024}%
  \BibitemOpen
  \bibfield  {author} {\bibinfo {author} {\bibfnamefont {G.}~\bibnamefont
  {Seifert}}, \bibinfo {author} {\bibfnamefont {A.}~\bibnamefont {Sealander}},
  \bibinfo {author} {\bibfnamefont {S.}~\bibnamefont {Marzen}},\ and\ \bibinfo
  {author} {\bibfnamefont {M.}~\bibnamefont {Levin}},\ }\bibfield  {title}
  {\bibinfo {title} {From reinforcement learning to agency: {Frameworks} for
  understanding basal cognition},\ }\href
  {https://doi.org/10.1016/j.biosystems.2023.105107} {\bibfield  {journal}
  {\bibinfo  {journal} {BioSystems}\ }\textbf {\bibinfo {volume} {235}},\
  \bibinfo {pages} {105107} (\bibinfo {year} {2024})}\BibitemShut {NoStop}%
\bibitem [{\citenamefont {Friston}(2010)}]{friston_free-energy_2010}%
  \BibitemOpen
  \bibfield  {author} {\bibinfo {author} {\bibfnamefont {K.}~\bibnamefont
  {Friston}},\ }\bibfield  {title} {{\selectlanguage {English}\bibinfo {title}
  {The free-energy principle: a unified brain theory?}},\ }\href
  {https://doi.org/10.1038/nrn2787} {\bibfield  {journal} {\bibinfo  {journal}
  {Nat Rev Neurosci}\ }\textbf {\bibinfo {volume} {11}},\ \bibinfo {pages}
  {127} (\bibinfo {year} {2010})},\ \bibinfo {note} {publisher: Nature
  Publishing Group}\BibitemShut {NoStop}%
\bibitem [{\citenamefont {Parr}\ \emph {et~al.}(2022)\citenamefont {Parr},
  \citenamefont {Pezzulo},\ and\ \citenamefont {Friston}}]{parr_active_2022}%
  \BibitemOpen
  \bibfield  {author} {\bibinfo {author} {\bibfnamefont {T.}~\bibnamefont
  {Parr}}, \bibinfo {author} {\bibfnamefont {G.}~\bibnamefont {Pezzulo}},\ and\
  \bibinfo {author} {\bibfnamefont {K.~J.}\ \bibnamefont {Friston}},\ }\href
  {https://doi.org/10.7551/mitpress/12441.001.0001} {{\selectlanguage
  {English}\emph {\bibinfo {title} {Active {Inference}: {The} {Free} {Energy}
  {Principle} in {Mind}, {Brain}, and {Behavior}}}}}\ (\bibinfo  {publisher}
  {The MIT Press},\ \bibinfo {year} {2022})\BibitemShut {NoStop}%
\bibitem [{\citenamefont {Pezzulo}\ \emph {et~al.}(2024)\citenamefont
  {Pezzulo}, \citenamefont {Parr},\ and\ \citenamefont
  {Friston}}]{pezzulo_active_2024}%
  \BibitemOpen
  \bibfield  {author} {\bibinfo {author} {\bibfnamefont {G.}~\bibnamefont
  {Pezzulo}}, \bibinfo {author} {\bibfnamefont {T.}~\bibnamefont {Parr}},\ and\
  \bibinfo {author} {\bibfnamefont {K.}~\bibnamefont {Friston}},\ }\bibfield
  {title} {\bibinfo {title} {Active inference as a theory of sentient
  behavior},\ }\href {https://doi.org/10.1016/j.biopsycho.2023.108741}
  {\bibfield  {journal} {\bibinfo  {journal} {Biological Psychology}\ }\textbf
  {\bibinfo {volume} {186}},\ \bibinfo {pages} {108741} (\bibinfo {year}
  {2024})}\BibitemShut {NoStop}%
\bibitem [{\citenamefont {Huang}\ \emph {et~al.}(2022)\citenamefont {Huang},
  \citenamefont {Gu},\ and\ \citenamefont {Nelson}}]{huang_increasingly_2022}%
  \BibitemOpen
  \bibfield  {author} {\bibinfo {author} {\bibfnamefont {T.-Y.}\ \bibnamefont
  {Huang}}, \bibinfo {author} {\bibfnamefont {H.}~\bibnamefont {Gu}},\ and\
  \bibinfo {author} {\bibfnamefont {B.~J.}\ \bibnamefont {Nelson}},\ }\bibfield
   {title} {{\selectlanguage {English}\bibinfo {title} {Increasingly
  {Intelligent} {Micromachines}}},\ }\href
  {https://doi.org/10.1146/annurev-control-042920-013322} {\bibfield  {journal}
  {\bibinfo  {journal} {Annual Review of Control, Robotics, and Autonomous
  Systems}\ }\textbf {\bibinfo {volume} {5}},\ \bibinfo {pages} {279} (\bibinfo
  {year} {2022})},\ \bibinfo {note} {publisher: Annual Reviews}\BibitemShut
  {NoStop}%
\bibitem [{\citenamefont {Falk}\ \emph {et~al.}(2023)\citenamefont {Falk},
  \citenamefont {Wu}, \citenamefont {Matthews}, \citenamefont {Sachdeva},
  \citenamefont {Pashine}, \citenamefont {Gardel}, \citenamefont {Nagel},\ and\
  \citenamefont {Murugan}}]{falk_learning_2023}%
  \BibitemOpen
  \bibfield  {author} {\bibinfo {author} {\bibfnamefont {M.~J.}\ \bibnamefont
  {Falk}}, \bibinfo {author} {\bibfnamefont {J.}~\bibnamefont {Wu}}, \bibinfo
  {author} {\bibfnamefont {A.}~\bibnamefont {Matthews}}, \bibinfo {author}
  {\bibfnamefont {V.}~\bibnamefont {Sachdeva}}, \bibinfo {author}
  {\bibfnamefont {N.}~\bibnamefont {Pashine}}, \bibinfo {author} {\bibfnamefont
  {M.~L.}\ \bibnamefont {Gardel}}, \bibinfo {author} {\bibfnamefont {S.~R.}\
  \bibnamefont {Nagel}},\ and\ \bibinfo {author} {\bibfnamefont
  {A.}~\bibnamefont {Murugan}},\ }\bibfield  {title} {\bibinfo {title}
  {Learning to learn by using nonequilibrium training protocols for adaptable
  materials},\ }\href {https://doi.org/10.1073/pnas.2219558120} {\bibfield
  {journal} {\bibinfo  {journal} {Proceedings of the National Academy of
  Sciences}\ }\textbf {\bibinfo {volume} {120}},\ \bibinfo {pages}
  {e2219558120} (\bibinfo {year} {2023})},\ \bibinfo {note} {publisher:
  Proceedings of the National Academy of Sciences}\BibitemShut {NoStop}%
\bibitem [{\citenamefont {O'Brien}\ \emph {et~al.}(2024)\citenamefont
  {O'Brien}, \citenamefont {Stremmel}, \citenamefont {Pio-Lopez}, \citenamefont
  {McMillen}, \citenamefont {Rasmussen-Ivey},\ and\ \citenamefont
  {Levin}}]{obrien_machine_2024}%
  \BibitemOpen
  \bibfield  {author} {\bibinfo {author} {\bibfnamefont {T.}~\bibnamefont
  {O'Brien}}, \bibinfo {author} {\bibfnamefont {J.}~\bibnamefont {Stremmel}},
  \bibinfo {author} {\bibfnamefont {L.}~\bibnamefont {Pio-Lopez}}, \bibinfo
  {author} {\bibfnamefont {P.}~\bibnamefont {McMillen}}, \bibinfo {author}
  {\bibfnamefont {C.}~\bibnamefont {Rasmussen-Ivey}},\ and\ \bibinfo {author}
  {\bibfnamefont {M.}~\bibnamefont {Levin}},\ }\bibfield  {title}
  {{\selectlanguage {english}\bibinfo {title} {Machine learning for hypothesis
  generation in biology and medicine: exploring the latent space of
  neuroscience and developmental bioelectricity}},\ }\href
  {https://doi.org/10.1039/D3DD00185G} {\bibfield  {journal} {\bibinfo
  {journal} {Digital Discovery}\ }\textbf {\bibinfo {volume} {3}},\ \bibinfo
  {pages} {249} (\bibinfo {year} {2024})},\ \bibinfo {note} {publisher:
  RSC}\BibitemShut {NoStop}%
\bibitem [{\citenamefont {Chen}\ \emph {et~al.}(2023)\citenamefont {Chen},
  \citenamefont {Hu}, \citenamefont {Zhu},\ and\ \citenamefont
  {Li}}]{chen_metamaterials_2023}%
  \BibitemOpen
  \bibfield  {author} {\bibinfo {author} {\bibfnamefont {J.}~\bibnamefont
  {Chen}}, \bibinfo {author} {\bibfnamefont {S.}~\bibnamefont {Hu}}, \bibinfo
  {author} {\bibfnamefont {S.}~\bibnamefont {Zhu}},\ and\ \bibinfo {author}
  {\bibfnamefont {T.}~\bibnamefont {Li}},\ }\bibfield  {title}
  {{\selectlanguage {English}\bibinfo {title} {Metamaterials: {From}
  fundamental physics to intelligent design}},\ }\href
  {https://doi.org/10.1002/idm2.12049} {\bibfield  {journal} {\bibinfo
  {journal} {Interdisciplinary Materials}\ }\textbf {\bibinfo {volume} {2}},\
  \bibinfo {pages} {5} (\bibinfo {year} {2023})},\ \bibinfo {note} {\_eprint:
  https://onlinelibrary.wiley.com/doi/pdf/10.1002/idm2.12049}\BibitemShut
  {NoStop}%
\bibitem [{\citenamefont {Paixão}\ \emph {et~al.}(2022)\citenamefont
  {Paixão}, \citenamefont {Sadoulet-Reboul}, \citenamefont {Foltête},
  \citenamefont {Chevallier},\ and\ \citenamefont
  {Cogan}}]{paixao_leveraging_2022}%
  \BibitemOpen
  \bibfield  {author} {\bibinfo {author} {\bibfnamefont {J.}~\bibnamefont
  {Paixão}}, \bibinfo {author} {\bibfnamefont {E.}~\bibnamefont
  {Sadoulet-Reboul}}, \bibinfo {author} {\bibfnamefont {E.}~\bibnamefont
  {Foltête}}, \bibinfo {author} {\bibfnamefont {G.}~\bibnamefont
  {Chevallier}},\ and\ \bibinfo {author} {\bibfnamefont {S.}~\bibnamefont
  {Cogan}},\ }\bibfield  {title} {{\selectlanguage {English}\bibinfo {title}
  {Leveraging physical intelligence for the self-design of high performance
  engineering structures}},\ }\href
  {https://doi.org/10.1038/s41598-022-15229-z} {\bibfield  {journal} {\bibinfo
  {journal} {Sci Rep}\ }\textbf {\bibinfo {volume} {12}},\ \bibinfo {pages}
  {11640} (\bibinfo {year} {2022})},\ \bibinfo {note} {publisher: Nature
  Publishing Group}\BibitemShut {NoStop}%
\bibitem [{\citenamefont {Allaire}\ \emph {et~al.}(2012)\citenamefont
  {Allaire}, \citenamefont {He}, \citenamefont {Deyst},\ and\ \citenamefont
  {Willcox}}]{allaire_information-theoretic_2012}%
  \BibitemOpen
  \bibfield  {author} {\bibinfo {author} {\bibfnamefont {D.}~\bibnamefont
  {Allaire}}, \bibinfo {author} {\bibfnamefont {Q.}~\bibnamefont {He}},
  \bibinfo {author} {\bibfnamefont {J.}~\bibnamefont {Deyst}},\ and\ \bibinfo
  {author} {\bibfnamefont {K.}~\bibnamefont {Willcox}},\ }\bibfield  {title}
  {{\selectlanguage {English}\bibinfo {title} {An {Information}-{Theoretic}
  {Metric} of {System} {Complexity} {With} {Application} to {Engineering}
  {System} {Design}}},\ }\href {https://doi.org/10.1115/1.4007587} {\bibfield
  {journal} {\bibinfo  {journal} {Journal of Mechanical Design}\ }\textbf
  {\bibinfo {volume} {134}},\ \bibinfo {pages} {100906} (\bibinfo {year}
  {2012})}\BibitemShut {NoStop}%
\bibitem [{\citenamefont {Thorsen}\ \emph {et~al.}(2002)\citenamefont
  {Thorsen}, \citenamefont {Maerkl},\ and\ \citenamefont
  {Quake}}]{thorsen_microfluidic_2002}%
  \BibitemOpen
  \bibfield  {author} {\bibinfo {author} {\bibfnamefont {T.}~\bibnamefont
  {Thorsen}}, \bibinfo {author} {\bibfnamefont {S.~J.}\ \bibnamefont
  {Maerkl}},\ and\ \bibinfo {author} {\bibfnamefont {S.~R.}\ \bibnamefont
  {Quake}},\ }\bibfield  {title} {\bibinfo {title} {Microfluidic
  {Large}-{Scale} {Integration}},\ }\href
  {https://doi.org/10.1126/science.1076996} {\bibfield  {journal} {\bibinfo
  {journal} {Science}\ }\textbf {\bibinfo {volume} {298}},\ \bibinfo {pages}
  {580} (\bibinfo {year} {2002})},\ \bibinfo {note} {publisher: American
  Association for the Advancement of Science}\BibitemShut {NoStop}%
\bibitem [{\citenamefont {Aguilera}\ and\ \citenamefont
  {Bedia}(2018)}]{aguilera_exploring_2018}%
  \BibitemOpen
  \bibfield  {author} {\bibinfo {author} {\bibfnamefont {M.}~\bibnamefont
  {Aguilera}}\ and\ \bibinfo {author} {\bibfnamefont {M.~G.}\ \bibnamefont
  {Bedia}},\ }\bibfield  {title} {{\selectlanguage {English}\bibinfo {title}
  {Exploring {Criticality} as a {Generic} {Adaptive} {Mechanism}}},\ }\bibfield
   {journal} {\bibinfo  {journal} {Front. Neurorobot.}\ }\textbf {\bibinfo
  {volume} {12}},\ \href {https://doi.org/10.3389/fnbot.2018.00055}
  {10.3389/fnbot.2018.00055} (\bibinfo {year} {2018}),\ \bibinfo {note}
  {publisher: Frontiers}\BibitemShut {NoStop}%
\bibitem [{\citenamefont {Yang}\ \emph {et~al.}(2019)\citenamefont {Yang},
  \citenamefont {Joglekar}, \citenamefont {Song}, \citenamefont {Newsome},\
  and\ \citenamefont {Wang}}]{yang_task_2019}%
  \BibitemOpen
  \bibfield  {author} {\bibinfo {author} {\bibfnamefont {G.~R.}\ \bibnamefont
  {Yang}}, \bibinfo {author} {\bibfnamefont {M.~R.}\ \bibnamefont {Joglekar}},
  \bibinfo {author} {\bibfnamefont {H.~F.}\ \bibnamefont {Song}}, \bibinfo
  {author} {\bibfnamefont {W.~T.}\ \bibnamefont {Newsome}},\ and\ \bibinfo
  {author} {\bibfnamefont {X.-J.}\ \bibnamefont {Wang}},\ }\bibfield  {title}
  {{\selectlanguage {English}\bibinfo {title} {Task representations in neural
  networks trained to perform many cognitive tasks}},\ }\href
  {https://doi.org/10.1038/s41593-018-0310-2} {\bibfield  {journal} {\bibinfo
  {journal} {Nat Neurosci}\ }\textbf {\bibinfo {volume} {22}},\ \bibinfo
  {pages} {297} (\bibinfo {year} {2019})},\ \bibinfo {note} {publisher: Nature
  Publishing Group}\BibitemShut {NoStop}%
\bibitem [{\citenamefont {Beck}\ and\ \citenamefont
  {Ramstead}(2025)}]{beck_dynamic_2025}%
  \BibitemOpen
  \bibfield  {author} {\bibinfo {author} {\bibfnamefont {J.}~\bibnamefont
  {Beck}}\ and\ \bibinfo {author} {\bibfnamefont {M.~J.~D.}\ \bibnamefont
  {Ramstead}},\ }\href {https://doi.org/10.48550/arXiv.2502.21217} {\bibinfo
  {title} {Dynamic {Markov} {Blanket} {Detection} for {Macroscopic} {Physics}
  {Discovery}}} (\bibinfo {year} {2025}),\ \bibinfo {note} {arXiv:2502.21217
  [q-bio]}\BibitemShut {NoStop}%
\bibitem [{\citenamefont {Man}\ and\ \citenamefont
  {Damasio}(2019)}]{man_homeostasis_2019}%
  \BibitemOpen
  \bibfield  {author} {\bibinfo {author} {\bibfnamefont {K.}~\bibnamefont
  {Man}}\ and\ \bibinfo {author} {\bibfnamefont {A.}~\bibnamefont {Damasio}},\
  }\bibfield  {title} {{\selectlanguage {English}\bibinfo {title} {Homeostasis
  and soft robotics in the design of feeling machines}},\ }\href
  {https://doi.org/10.1038/s42256-019-0103-7} {\bibfield  {journal} {\bibinfo
  {journal} {Nat Mach Intell}\ }\textbf {\bibinfo {volume} {1}},\ \bibinfo
  {pages} {446} (\bibinfo {year} {2019})},\ \bibinfo {note} {publisher: Nature
  Publishing Group}\BibitemShut {NoStop}%
\bibitem [{\citenamefont {Friston}\ \emph {et~al.}(2023)\citenamefont
  {Friston}, \citenamefont {Da~Costa}, \citenamefont {Sakthivadivel},
  \citenamefont {Heins}, \citenamefont {Pavliotis}, \citenamefont {Ramstead},\
  and\ \citenamefont {Parr}}]{friston_path_2023}%
  \BibitemOpen
  \bibfield  {author} {\bibinfo {author} {\bibfnamefont {K.}~\bibnamefont
  {Friston}}, \bibinfo {author} {\bibfnamefont {L.}~\bibnamefont {Da~Costa}},
  \bibinfo {author} {\bibfnamefont {D.~A.~R.}\ \bibnamefont {Sakthivadivel}},
  \bibinfo {author} {\bibfnamefont {C.}~\bibnamefont {Heins}}, \bibinfo
  {author} {\bibfnamefont {G.~A.}\ \bibnamefont {Pavliotis}}, \bibinfo {author}
  {\bibfnamefont {M.}~\bibnamefont {Ramstead}},\ and\ \bibinfo {author}
  {\bibfnamefont {T.}~\bibnamefont {Parr}},\ }\bibfield  {title} {\bibinfo
  {title} {Path integrals, particular kinds, and strange things},\ }\href
  {https://doi.org/10.1016/j.plrev.2023.08.016} {\bibfield  {journal} {\bibinfo
   {journal} {Physics of Life Reviews}\ }\textbf {\bibinfo {volume} {47}},\
  \bibinfo {pages} {35} (\bibinfo {year} {2023})}\BibitemShut {NoStop}%
\bibitem [{\citenamefont {Sharma}\ \emph {et~al.}(2023)\citenamefont {Sharma},
  \citenamefont {Czégel}, \citenamefont {Lachmann}, \citenamefont {Kempes},
  \citenamefont {Walker},\ and\ \citenamefont {Cronin}}]{sharma_assembly_2023}%
  \BibitemOpen
  \bibfield  {author} {\bibinfo {author} {\bibfnamefont {A.}~\bibnamefont
  {Sharma}}, \bibinfo {author} {\bibfnamefont {D.}~\bibnamefont {Czégel}},
  \bibinfo {author} {\bibfnamefont {M.}~\bibnamefont {Lachmann}}, \bibinfo
  {author} {\bibfnamefont {C.~P.}\ \bibnamefont {Kempes}}, \bibinfo {author}
  {\bibfnamefont {S.~I.}\ \bibnamefont {Walker}},\ and\ \bibinfo {author}
  {\bibfnamefont {L.}~\bibnamefont {Cronin}},\ }\bibfield  {title}
  {{\selectlanguage {English}\bibinfo {title} {Assembly theory explains and
  quantifies selection and evolution}},\ }\href
  {https://doi.org/10.1038/s41586-023-06600-9} {\bibfield  {journal} {\bibinfo
  {journal} {Nature}\ }\textbf {\bibinfo {volume} {622}},\ \bibinfo {pages}
  {321} (\bibinfo {year} {2023})},\ \bibinfo {note} {publisher: Nature
  Publishing Group}\BibitemShut {NoStop}%
\bibitem [{\citenamefont {Howard}\ \emph {et~al.}(2019)\citenamefont {Howard},
  \citenamefont {Eiben}, \citenamefont {Kennedy}, \citenamefont {Mouret},
  \citenamefont {Valencia},\ and\ \citenamefont
  {Winkler}}]{howard_evolving_2019}%
  \BibitemOpen
  \bibfield  {author} {\bibinfo {author} {\bibfnamefont {D.}~\bibnamefont
  {Howard}}, \bibinfo {author} {\bibfnamefont {A.~E.}\ \bibnamefont {Eiben}},
  \bibinfo {author} {\bibfnamefont {D.~F.}\ \bibnamefont {Kennedy}}, \bibinfo
  {author} {\bibfnamefont {J.-B.}\ \bibnamefont {Mouret}}, \bibinfo {author}
  {\bibfnamefont {P.}~\bibnamefont {Valencia}},\ and\ \bibinfo {author}
  {\bibfnamefont {D.}~\bibnamefont {Winkler}},\ }\bibfield  {title}
  {{\selectlanguage {English}\bibinfo {title} {Evolving embodied intelligence
  from materials to machines}},\ }\href
  {https://doi.org/10.1038/s42256-018-0009-9} {\bibfield  {journal} {\bibinfo
  {journal} {Nat Mach Intell}\ }\textbf {\bibinfo {volume} {1}},\ \bibinfo
  {pages} {12} (\bibinfo {year} {2019})},\ \bibinfo {note} {publisher: Nature
  Publishing Group}\BibitemShut {NoStop}%
\bibitem [{\citenamefont {Hughes}\ \emph {et~al.}(2022)\citenamefont {Hughes},
  \citenamefont {Abdulali}, \citenamefont {Hashem},\ and\ \citenamefont
  {Iida}}]{hughes_embodied_2022}%
  \BibitemOpen
  \bibfield  {author} {\bibinfo {author} {\bibfnamefont {J.}~\bibnamefont
  {Hughes}}, \bibinfo {author} {\bibfnamefont {A.}~\bibnamefont {Abdulali}},
  \bibinfo {author} {\bibfnamefont {R.}~\bibnamefont {Hashem}},\ and\ \bibinfo
  {author} {\bibfnamefont {F.}~\bibnamefont {Iida}},\ }\bibfield  {title}
  {\bibinfo {title} {Embodied {Artificial} {Intelligence}: {Enabling} the
  {Next} {Intelligence} {Revolution}},\ }\href
  {https://doi.org/10.1088/1757-899X/1261/1/012001} {\bibfield  {journal}
  {\bibinfo  {journal} {IOP Conf. Ser.: Mater. Sci. Eng.}\ }\textbf {\bibinfo
  {volume} {1261}},\ \bibinfo {pages} {012001} (\bibinfo {year}
  {2022})}\BibitemShut {NoStop}%
\bibitem [{\citenamefont {Yang}\ \emph
  {et~al.}(2022{\natexlab{b}})\citenamefont {Yang}, \citenamefont {Liu},
  \citenamefont {Berrueta}, \citenamefont {Zhang}, \citenamefont {Brooks},
  \citenamefont {Koman}, \citenamefont {Yang}, \citenamefont {Gong},
  \citenamefont {Murphey},\ and\ \citenamefont {Strano}}]{yang_memristor_2022}%
  \BibitemOpen
  \bibfield  {author} {\bibinfo {author} {\bibfnamefont {J.~F.}\ \bibnamefont
  {Yang}}, \bibinfo {author} {\bibfnamefont {A.~T.}\ \bibnamefont {Liu}},
  \bibinfo {author} {\bibfnamefont {T.~A.}\ \bibnamefont {Berrueta}}, \bibinfo
  {author} {\bibfnamefont {G.}~\bibnamefont {Zhang}}, \bibinfo {author}
  {\bibfnamefont {A.~M.}\ \bibnamefont {Brooks}}, \bibinfo {author}
  {\bibfnamefont {V.~B.}\ \bibnamefont {Koman}}, \bibinfo {author}
  {\bibfnamefont {S.}~\bibnamefont {Yang}}, \bibinfo {author} {\bibfnamefont
  {X.}~\bibnamefont {Gong}}, \bibinfo {author} {\bibfnamefont {T.~D.}\
  \bibnamefont {Murphey}},\ and\ \bibinfo {author} {\bibfnamefont {M.~S.}\
  \bibnamefont {Strano}},\ }\bibfield  {title} {{\selectlanguage
  {English}\bibinfo {title} {Memristor {Circuits} for {Colloidal} {Robotics}:
  {Temporal} {Access} to {Memory}, {Sensing}, and {Actuation}}},\ }\href
  {https://doi.org/10.1002/aisy.202100205} {\bibfield  {journal} {\bibinfo
  {journal} {Advanced Intelligent Systems}\ }\textbf {\bibinfo {volume} {4}},\
  \bibinfo {pages} {2100205} (\bibinfo {year}
  {2022}{\natexlab{b}})}\BibitemShut {NoStop}%
\bibitem [{\citenamefont {Gumuskaya}\ \emph {et~al.}(2024)\citenamefont
  {Gumuskaya}, \citenamefont {Srivastava}, \citenamefont {Cooper},
  \citenamefont {Lesser}, \citenamefont {Semegran}, \citenamefont {Garnier},\
  and\ \citenamefont {Levin}}]{gumuskaya_motile_2024}%
  \BibitemOpen
  \bibfield  {author} {\bibinfo {author} {\bibfnamefont {G.}~\bibnamefont
  {Gumuskaya}}, \bibinfo {author} {\bibfnamefont {P.}~\bibnamefont
  {Srivastava}}, \bibinfo {author} {\bibfnamefont {B.~G.}\ \bibnamefont
  {Cooper}}, \bibinfo {author} {\bibfnamefont {H.}~\bibnamefont {Lesser}},
  \bibinfo {author} {\bibfnamefont {B.}~\bibnamefont {Semegran}}, \bibinfo
  {author} {\bibfnamefont {S.}~\bibnamefont {Garnier}},\ and\ \bibinfo {author}
  {\bibfnamefont {M.}~\bibnamefont {Levin}},\ }\bibfield  {title}
  {{\selectlanguage {English}\bibinfo {title} {Motile {Living} {Biobots}
  {Self}‐{Construct} from {Adult} {Human} {Somatic} {Progenitor} {Seed}
  {Cells}}},\ }\href {https://doi.org/10.1002/advs.202303575} {\bibfield
  {journal} {\bibinfo  {journal} {Advanced Science}\ }\textbf {\bibinfo
  {volume} {11}},\ \bibinfo {pages} {2303575} (\bibinfo {year}
  {2024})}\BibitemShut {NoStop}%
\bibitem [{\citenamefont {Rouleau}\ and\ \citenamefont
  {Levin}(2023)}]{rouleau_multiple_2023}%
  \BibitemOpen
  \bibfield  {author} {\bibinfo {author} {\bibfnamefont {N.}~\bibnamefont
  {Rouleau}}\ and\ \bibinfo {author} {\bibfnamefont {M.}~\bibnamefont
  {Levin}},\ }\bibfield  {title} {{\selectlanguage {English}\bibinfo {title}
  {The {Multiple} {Realizability} of {Sentience} in {Living} {Systems} and
  {Beyond}}},\ }\bibfield  {journal} {\bibinfo  {journal} {eNeuro}\ }\textbf
  {\bibinfo {volume} {10}},\ \href
  {https://doi.org/10.1523/ENEURO.0375-23.2023} {10.1523/ENEURO.0375-23.2023}
  (\bibinfo {year} {2023}),\ \bibinfo {note} {publisher: Society for
  Neuroscience Section: Commentary}\BibitemShut {NoStop}%
\end{thebibliography}


\end{document}